\documentclass[12pt]{iopart}

\usepackage{iopams}
\usepackage{setstack}
\usepackage{subfig}
\usepackage{graphicx}
\usepackage{float}
\usepackage{cite}
\usepackage[labelfont=bf, textfont=it]{caption}
\usepackage{hyperref}
\usepackage{xcolor}
\usepackage{tabularx}
\usepackage{colortbl}
\usepackage{multirow}
\usepackage{bm}
\usepackage{booktabs}
\usepackage{amssymb}
\usepackage{enumitem}

\newcommand{\dd}{\mathrm d}

\newcommand{\p}{_{\parallel}}

\newcommand{\secref}[1]{\S\ref{#1}}

\definecolor{Gray}{gray}{0.85}
\definecolor{LightCyan}{rgb}{0.88,1,1}

\newcolumntype{a}{>{\columncolor{Gray}}c}

\begin{document}

\title{Electromagnetic gyrokinetic instabilities in STEP}
% subject to change! 

\author{D. Kennedy$^1$, M. Giacomin$^2$, F. J. Casson$^1$,  D. Dickinson$^2$, W. A. Hornsby$^1$, B. S. Patel$^1,$ and C. M. Roach$^1$}

\address{$^1$Culham Centre for Fusion Energy, Abingdon OX14 3DB, United Kingdom}
\address{$^2$York Plasma Institute, University of York, York, YO10 5DD, United Kingdom}
\ead{daniel.kennedy@ukaea.uk}

\begin{abstract}

We present herein the results of a linear gyrokinetic analysis of electromagnetic microinstabilites in the conceptual high$-\beta,$ reactor-scale, tight-aspect-ratio tokamak STEP (Spherical Tokamak for Energy Production, \url{https://step.ukaea.uk}). We examine a range of flux surfaces between the deep core and the pedestal top for two candidate flat-top operating points of the prototype device. Local linear gyrokinetic analysis is performed to determine the type of microinstabilities that arise under these reactor-relevant conditions. We find that the equilibria are dominated at ion binormal scales by a hybrid version of the Kinetic Ballooning Mode (KBM) instability {that has significant linear drive contributions from the ion temperature gradient and from trapped electrons}, while collisional Microtearing Modes (MTMs) are sub-dominantly also unstable at similar binormal scales.  The \textit{hybrid}-KBM and MTM exhibit very different radial scales. We study the sensitivity of these instabilities to physics parameters, and discuss potential mechanisms for mitigating them. The results of this investigation are compared to a small set of similar conceptual reactor designs in the literature. A detailed benchmark of the linear results is performed using three gyrokinetic codes; alongside extensive resolution testing and sensitivity to numerical parameters providing confidence in the results of our calculations, and paving the way for detailed nonlinear studies in a companion article. 

\end{abstract}

\section{Introduction}
\label{sec:intro}
Magnetically confined fusion is promising as a future power source. However, the viability of fusion power plants is strongly influenced by how well the thermal energy can be confined in the plasma. Often, the dominant process governing confinement is microinstability-driven plasma turbulence. The beneficial impacts of equilibrium geometry on the microstability properties of spherical tokamaks (STs) \cite{peng1986} were uncovered in early studies motivated by START, MAST and NSTX \cite{rewoldt1996,roach1995,applegate2004,bourdelle2003}; favorable magnetic drifts, allied with higher radial pressure gradients in STs, were found capable of suppressing some of the drift-wave instabilities that drive anomalous transport in other devices \cite{kotschenreuther2000}. Furthermore, in experiments with tangential NBI, the compact nature of the ST leads to high toroidal flows {(see\cite{Field_2011})} that can act to suppress turbulence, especially at ion Larmor scales (see recent review of transport and confinement in STs \cite{kaye2021} and references therein), though it is anticipated that an ST power plant will have minimal momentum input and only modest externally driven flow. On the other hand, the higher trapping fraction in STs contributes to an increased drive for trapped electron modes (TEMs) at high density gradients, though this drive is mitigated if the magnetic drifts are favourable \cite{roach1995}. In addition, the high $\beta$ (the ratio of thermal pressure to magnetic pressure) accessible in STs (e.g., \cite{kaye_2005,kaye2007}) can drive electromagnetic modes unstable, which can significantly increase the core turbulent transport. 

It is precisely these aforementioned electromagnetic instabilities which we expect to dominate transport in high-$\beta,$ reactor-scale, tight-aspect-ratio tokamaks such as STEP (Spherical Tokamak for Energy Production) \cite{meyer2022}. To be economically competitive, ST power plant designs such as STEP require a high $\beta,$ which further necessitates a high $\beta^\prime$ (the radial gradient of pressure) in a compact device. As a result, we also require sufficiently low turbulent transport in order to sustain these steep gradients and thus to maximise the self-driven bootstrap current and reduce the need for external current drive in a steady state device. In plasmas such as STEP where $\beta$ and $\beta^\prime$ are sufficiently high, the curvature of the confining magnetic field and the plasma kinetic gradients can excite electromagnetic instabilities such as kinetic ballooning modes (KBMs) and microtearing modes (MTMs). The KBM is driven by electrons and ions at binormal-scales approaching the ion Larmor radius ($k_{y}\rho_{i} \lesssim 1$), propagates in the ion diamagnetic direction, and is closely related to the ideal ballooning mode of magnetohydrodynamics (MHD)  \cite{dickinson2011}. MTMs excite radially localised current layers on rational
surfaces, are primarily driven unstable by the electron temperature gradient, and propagate in the electron diamagnetic direction. They generate magnetic islands on rational surfaces that tear the confining equilibrium flux surfaces and enhance electron heat transport through magnetic field line stochasticisation~\cite{applegate2007,guttenfelder2011,nevins2011,giacomin2023a,Kotschenreuther_2019}. 
In devices where $\beta$ exceeds a certain threshold value, electromagnetic instabilities can become the fastest growing instabilities in the system and the dominant sources of transport in the plasma core \cite{patel2021}. 

These two instabilities, and the nonlinear interactions between them, will likely play a crucial role in setting the transport levels in the core of devices such as STEP, and dictate the confinement times attainable in next-generation STs such as STEP. Fully understanding the transport impacts of these modes is one of the major physics questions which must be
answered to build confidence in the feasibility of designs of future ST power plants. 

In this work, the first of two related papers, our contributions are: (a) to report on the main results of the gyrokinetic linear analysis of two candidate STEP equilibria, STEP-EC-HD-v5\footnote{SimDB UUID: 2bb77572-d832-11ec-b2e3-679f5f37cafe, Alias: smars/jetto/step/88888/apr2922/seq-1} (hereinafter STEP-EC-HD) and STEP-EB-HD-v4\footnote{SimDB UUID: 8ea23452-dc00-11ec-9736-332ed6514f8d, Alias: twilson/jetto/step/88888/may2422/seq-1} (hereinafter STEP-EB-HD), at various surfaces between the core and the pedestal top; (b) to identify the dominant and sub-dominant instabilities and elucidate the nature of these modes; and
(c) to explore the resolution requirements for nonlinear simulations, thus paving the way for the companion work \cite{giacomin2023b} (hereinafter referred to as Paper (II)), in which we will present the first local nonlinear turbulence simulations for a STEP conceptual design.

We begin in \secref{sec:equilibria} by introducing the STEP equilibria\footnote{We remark that the STEP plasma design has not been finalised and these equilibria are thus subject to change.} and the associated plasma parameters, providing some motivation of the design choices and a brief discussion of how the equilibria compare to similar ST design points \cite{patel2021,tdotp}. In \secref{sec:overview of simulation results}, we present the main results of the gyrokinetic linear analysis of the STEP-EC-HD equilibrium at four flux-surfaces between the core and the pedestal top. This analysis reveals the importance of two particular electromagnetic instabilities, a \textit{hybrid}-KBM and a collisional MTM; in \secref{sec:hybrid TEM/KBM instability}, \secref{sec:stabilising the hybrid KBM}, and \secref{sec:subdominant MTM instability}, we explore the salient features of these modes {(primarily focusing on one mid-radius surface in STEP-EC-HD and parameter scans around this flux surface)}. In \secref{sec:code_comparison}, we present the results of a three-code microstability comparison {for two surfaces in STEP-EC-HD and one surface in STEP-EB-HD.} Finally, we present our conclusions and outlook in \secref{sec:conclusions}.

\section{The STEP-EC-HD and STEP-EB-HD equilibria}
\label{sec:equilibria}

STEP is a UK programme that aims to demonstrate the ability to generate net electricity from fusion. STEP is planned to be a compact prototype power plant (based on the ST concept) designed to deliver net electric power ${P} > 100$ MW to the  national grid \cite{wilson2020,STEP}. The first phase of this ambitious programme is to develop a conceptual design of a STEP Prototype Plant (SPP) and STEP Plasma Reference (SPR) equilibria for preferred flat-top operating points. The ST concept maximises fusion power $\mathrm{P}_{\mathrm{fus}} \propto (\kappa\beta_{N}B_{t})^{4}/A$ \cite{menard2016} and bootstrap current fraction $f_{\mathrm{BS}} = I_{\mathrm{BS}}/I_{p}$ in a compact device at relatively low toroidal field by allowing operation at high normalised pressure $\beta_{N} \simeq 4 - 5$ and high elongation $\kappa > 2.8.$ However, alongside these advantages, the ST concept also poses unique challenges, not only in terms of plasma microstability (the focus of this work) but also in terms of the engineering constraints; the compactness restricts significantly the available space for a solenoid, so the required plasma current of $I_{p} \simeq 20$ MA has to be driven, ramped up and ramped down non-inductively. 

STEP plasma concepts (with the global parameters given in Table~\ref{tab:sprs}) have mainly been designed \cite{meyer2022} using the integrated modelling suite JINTRAC \cite{ROMANELLI2014}, to model transport and sources self-consistently in the core plasma with prescribed boundary conditions: simplified models are used for the pedestal boundary conditions, pellet fuelling, heating and current drive, and core transport uses an empirical Bohm-gyro-Bohm (BgB) model which has been tuned both to give dominant electron heat transport as observed experimentally in MAST, and also to give a desired $\beta_N$. In the present design of the operating points, plasma confinement is largely assumed, with the confinement enhancement (or H-factor) over an energy confinement scaling law indicating the level of confinement required to achieve a particular non-inductive operating point satisfying a prescribed set of additional constraints. The primary drivers that constrain the confinement needed for a viable operating point include a specified fusion gain $Q > 11$ (a proxy for net electricity generation), a specified fusion power $P_{\mathrm{fus}} > 1.5$ GW, current drive efficiency validated against full wave modelling of either Electron Cyclotron (EC) or Electron Bernstein Wave (EBW) systems, $P_{\mathrm{aux}} < 160$ MW, a current profile consistent with MHD, vertical stability and divertor shaping constraints (see \cite{tholerus2023} for further details). Importantly, the STEP parameter regime is outside the range of validity of the most advanced reduced core transport models available, typically developed for present-day conventional tokamaks, which often do not capture the electromagnetic (EM) transport expected to prevail in STEP, as such, it is important to test the assumptions of BgB transport using linear and nonlinear GK simulations; a key thrust of this current work. 

Here, we focus on two steady-state, non-inductive flat-top operating points, STEP-EC-HD and STEP-EB-HD, both of which are designed to deliver a fusion power $P_{\mathrm{fus}} \sim 1.8$ GW. These two designs both use RF heating instead of neutral beams to generate the current drive, in order to maximise the wall area available for Tritium breeding and minimise the recirculating power fraction \cite{meyer2022}. There are modest differences between these equilibria because they use different RF current drive schemes:

\begin{itemize}
    \item \textbf{STEP-EC-HD} utilises only Electron Cyclotron Current Drive (ECCD) heating. 
    \item \textbf{STEP-EB-HD} utilises a mixture of ECCD and Electron Bernstein Wave (EBW) heating.
\end{itemize}

\begin{figure}
    \centering
    \subfloat[]{\includegraphics[height=0.27\textheight]{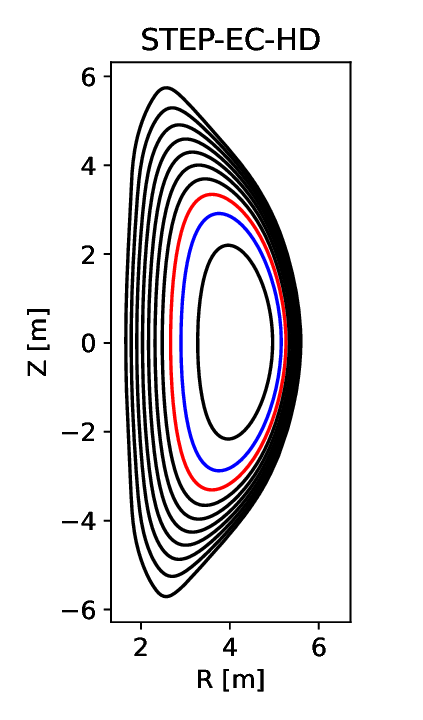}}
    \subfloat[]{\includegraphics[height=0.27\textheight]{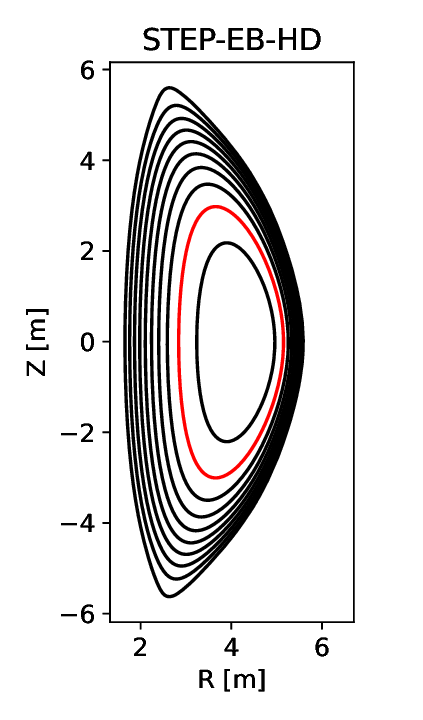}}
    \subfloat[]{\includegraphics[height=0.255\textheight]{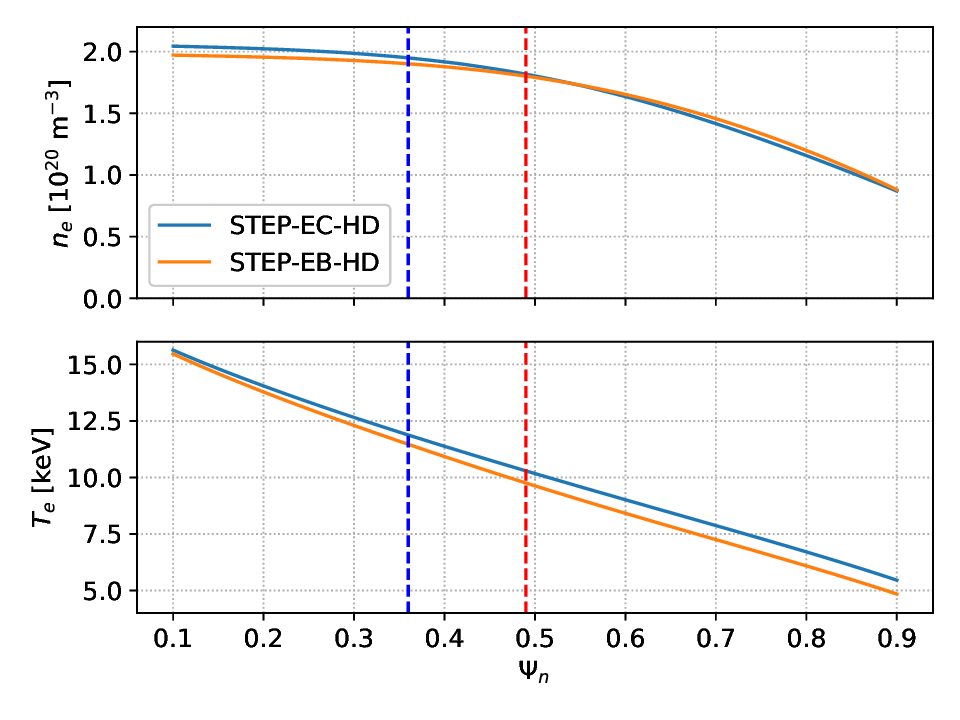}}
    \caption{Magnetic flux surfaces of the of the STEP-EC-HD (a) and STEP-EB-HD (b) equilibria. The red line denotes the flux surface corresponding to $q=3.5$, while the blue line refers to the flux surface at $q=3.0$ (considered only in STEP-EC-HD). (c) Electron density and electron temperature radial profiles of STEP-EC-HD and STEP-EB-HD. {The coloured STEP-EC-HD surfaces in (a) and the vertical dashed lines in (c) denote the q=3 and q=3.5 surfaces in STEP-EC-HD, used for the bulk of our analysis in Sections~\secref{sec:hybrid TEM/KBM instability} to \secref{sec:subdominant MTM instability}.  These surfaces together with the coloured STEP-EB-HD surface in (b) are used in three-code comparisons in Section~\secref{sec:code_comparison}.}}
    \label{fig:equilibria}
\end{figure}

Key global parameters of the preferred flat top operating points are shown in the highlighted columns of Table \ref{tab:sprs}, and a contour plot of the magnetic flux surfaces in these two design points is shown in Figure~\ref{fig:equilibria}, alongside the corresponding electron density and electron temperature radial profiles as functions of the normalised poloidal flux $\Psi_{n}.$ { In Section~\secref{sec:overview of simulation results} we perform linear microstability analysis on the surfaces $\Psi_n=0.36,0.49,0.58,0.71$ in STEP-EC-HD. Our primary focus in Sections~\secref{sec:hybrid TEM/KBM instability} to \secref{sec:subdominant MTM instability} will be on the $q = 3.5$ surface ($\Psi_n=0.49$) of STEP-EC-HD.  Two surfaces from STEP-EC-HD and one surface from STEP-EB-HD are used in the three-code microstability comparisons reported in Section~\secref{sec:code_comparison}.}

\begin{table}
    \centering
    \begin{tabular}{|c|a|a|c|c|}
     \hline
      & {\textbf{STEP-EC-HD}} & \textbf{STEP-EB-HD} & \textbf{TDoTP-high-$q_{0}$} \cite{tdotp} & \textbf{BurST} \cite{patel2021} \\
    \hline
    $R_{\mathrm{geo}}$  & 3.60 & 3.60 & 2.5 & 2.5 \\
    \hline
    $A$  & 1.8 & 1.8 & 1.67 & 1.67  \\
     \hline
    $B_{T} \, (R_{\mathrm{geo}})$ [T]  & 3.2 & 3.2 & 2.25 & 2.4 \\
    \hline
     $I_{p}$ [MA]  & 20.9 & 22.0 & 16.5 & 21.0 \\
    \hline 
    $n_{e0}$ [$10^{20}$ m$^{-3}$]  & 2.05 & 1.98 & 2.15 & 1.72  \\
    \hline 
    $T_{e0}$ [keV]  & 18.0 & 18.0 & 17.5 & 28.0 \\
    \hline 
    $\kappa$  & 2.93 & 2.93 & 2.80 & 2.80\\
    \hline 
    $\delta$  & 0.59 & 0.50 & 0.54 & 0.55 \\
    \hline
        $P_{\mathrm{fus}}$ [GW]  & 1.76 & 1.77 & 0.81 & 1.10 \\
    \hline
        $P_{\mathrm{heat}}$ [MW]  & 150 & 154 & 60 & 94 \\
    \hline 
        $P_{\mathrm{ECCD}}$ [MW] & 150 & 55.40 & - & - \\
    \hline
        $P_{\mathrm{EBW}}$ [MW] & 0 & 98.60 & - & - \\
    \hline
        $P_{\mathrm{rad}}$ [MW] & 338 & 341 & 220 & 250 \\
    \hline
            $Q$  & 11.8 & 11.5 & 13.5 & 11.7 \\
    \hline
            $\beta_{N}$ & 4.4 & 4.1 & 5.5 & 5.5\\
    \hline
     H98 & 1.60 & 1.48 & - & - \\
     \hline 
          H98* & 1.10 & 1.02 & - & - \\
    \hline
            $f_\mathrm{BS}$ & 0.88 & 0.78 & 0.67 & 0.61 \\
    \hline
    HCD technique         & ECCD  & ECCD / EBW  & - & NBI \\
    \hline
    \end{tabular}
    \caption{Basic global plasma parameters including the tokamak major radius, $R_\mathrm{geo}$, the aspect-ratio, $A$, the toroidal magnetic field at the tokamak magnetic axis, $B_T$, the plasma current, $I_p$, the electron density and temperature values at the magnetic axis, $n_{e0}$ and $T_{e0}$, the fusion power, $P_\mathrm{fus}$, the total heating power, $P_\mathrm{tot}$, the ECCD heating power, $P_{\mathrm{ECCD}},$ the EBW heating power, $P_{\mathrm{EBW}},$ the radiated power, $P_\mathrm{rad}$, the fusion gain, $Q$, the normalised $\beta$, $\beta_N=\beta aB_T/I_p$, the energy confinement times normalized to the multi-machine-based ITER-H98(y,2) scaling law, H98 (with radiation) and H98* (without radiation), the bootstrap fraction, $f_\mathrm{BS},$ {and the technique used for heating and current drive, HCD,} for the two baseline operating points examined in this work (STEP-EC-HD and STEP-EB-HD) are shown alongside parameters for comparable ST design points; the TDoTP high$-q_{0}$ baseline \cite{tdotp}; and an earlier prototype burning ST design BurST \cite{patel2021}. Also shown are the plasma elongation $\kappa$ and triangularity $\delta$ at the last closed flux surface.}
    \label{tab:sprs}
\end{table}

Table \ref{tab:sprs} also provides key global equilibrium parameters for two other recently developed conceptual burning ST plasma equilibria: the TDoTP high $q_{0}$ case \cite{tdotp} and an earlier concept BurST \cite{patel2021}.

For each flux surface considered, a Miller parameterisation \cite{miller1998} was used to model the local plasma equilibrium. Miller parameters were fitted to the surface using \texttt{pyrokinetics} \cite{pyrokinetics}, a python library aiming to standardise gyrokinetic analysis between different GK codes and conventions. \texttt{Pyrokinetics} was also used throughout to facilitate the conversion of input files between the different GK codes used in this work (see the three code comparison reported in \secref{sec:code_comparison}). Table~\ref{tab:equilibrium} reports the local value of the normalised poloidal magnetic flux $\Psi_n$,  magnetic shear, $\hat{s}=(\rho/q)\mathrm{d}q/\mathrm{d}\rho$, radial position, $\rho=r/a$, elongation and its radial derivative, $\kappa$ and $\kappa'$, triangularity and its radial derivative (the symbol $'$ denotes derivative with respect to $\rho$), $\delta$ and $\delta'$, the radial derivative of the Shafranov shift, $\Delta'$, the electron $\beta$, $\beta_e = 2\mu_0 n_e T_e/B_T^2$, and electron and deuterium density and temperature gradients at different flux surfaces corresponding to low order rational values of the safety factor $q$. For each surface, we report the value of the binormal wavenumber $k_{y}\rho_{s}$ corresponding to the toroidal mode number $n=1$, with $\rho_{s} = c_{sD}/\Omega_D$ where $c_{sD} = \sqrt{T_e/m_D}$, $\Omega_D = eB/m_D$ and $m_D$ the deuterium mass.

Nominally, fives species (electron, deuterium, tritium, thermalised helium ash, and a heavy impurity species) are included in the integrated modelling of the STEP-EC-HD and STEP-EB-HD equilibria considered in this paper. The simulations performed in this paper are carried out with 3 kinetic species (electron, deuterium, and tritium) unless explicitly stated otherwise and a future work will explore the influence of fast $\alpha$ particles which are completely neglected in this analysis. 

\begin{table}
    \centering
    \begin{tabular}{|c|c|c|c|c|c|}
     \hline
      & \multicolumn{4}{c|}{\textbf{STEP-EC-HD}} & \textbf{STEP-EB-HD}\\
      \hline
    $q$         & 3.0  & 3.5  & 4.0  & 5.0  &  3.5\\
    \hline
    $\hat{s}$   & 0.60 & 1.20 & 1.56 & 2.24 & 0.87\\
    \hline
    $\Psi_n$    & 0.36 & 0.49 & 0.58 & 0.71 & 0.35\\
    \hline
    { $\rho=r/a$}    & 0.54 & 0.64 & 0.70 & 0.79 & 0.55\\
    \hline
    $\kappa$    & 2.57 & 2.56 & 2.57 & 2.60 & 2.57\\
    \hline 
    $\kappa'$   &-0.09 & 0.06 & 0.19 & 0.43 & 0.32\\
    \hline
    $\delta$    & 0.23 & 0.29 & 0.32 & 0.40 & 0.28\\ 
    \hline
    $\delta'$   & 0.36 & 0.46 & 0.54 & 0.70 & 0.60\\
    \hline
    $\Delta'$   &-0.34 & -0.40& -0.44& -0.49& -0.30\\
    \hline
    $\beta_e$     & 0.12 & 0.09 & 0.07 & 0.05 & 0.11\\
    \hline
    $\beta'$    &-0.45 & -0.48& -0.47& -0.44& -0.40\\
    \hline
    $a/L_{n_e}$ & 0.45 & 1.06 & 1.54 & 2.58 & 0.30\\
    \hline
    $a/L_{T_e}$ & 1.32 & 1.58 & 1.77 & 2.15 & 1.40\\
    \hline
    $a/L_{n_D}$ & 0.48 & 1.06 & 1.61 & 2.61 & 0.33\\
    \hline
    $a/L_{T_D}$ & 1.67 & 1.82 & 1.96 & 2.41 & 1.74\\
    \hline
    $a/L_{n_T}$ & 0.41 & 0.99 & 1.54 & 2.54 & 0.24\\
    \hline
    $a/L_{T_T}$ & 1.67 & 1.82 & 1.96 & 2.41 & 1.74\\
    \hline
    $k_y^{n=1}\rho_{s}$ & 0.0056 & 0.0047 & 0.0044 & 0.0039 & 0.0061\\
    \hline
    \end{tabular}
    \caption{Local parameters of all flux surfaces considered in this work, which include the surface at $\Psi_n = 0.36$ ($q=3.0$), $\Psi_n = 0.49$ ($q=3.5$), $\Psi_n = 0.58$ ($q=4.0$) and $\Psi_n = 0.71$ ($q=5.0$) of STEP-EC-HD and at $\Psi_n = 0.36$ ($q=3.5$) of STEP-EB-HD. Included also is the binormal wavenumber $k_y^{n=1}\rho_{s}$ corresponding to the toroidal mode number $n=1$.}
    \label{tab:equilibrium}
\end{table}

\section{Overview of simulation results for various surfaces in STEP-EC-HD}
\label{sec:overview of simulation results}

We begin by finding the dominant linearly unstable modes at an initial ballooning angle of $\theta_{0} = 0,$ i.e., those modes centered on the outboard midplane. For now, we focus our attention on STEP-EC-HD, with the results for STEP-EB-HD reported in \secref{sec:code_comparison}. The linear simulations presented here are carried out with the gyrokinetic code GS2~\cite{gs2} (commit \texttt{675f0870}). Later (in \secref{sec:code_comparison}) we will verify the fidelity by comparing the main results obtained with GS2 against CGYRO~\cite{candy2016} (commit \texttt{399deb4c}) and GENE~\cite{gene} (commit \texttt{de99981}) in a detailed three-code benchmark \secref{sec:code_comparison}. Table~\ref{tab:resolution_spr45} indicates grid parameters used in each code (see highlighted columns for GS2) for calculations that include ($f_{B} = 1$) or neglect ($f_{B} = 0$) the compressional magnetic perturbation $\delta B_\parallel$. We find that neglecting $\delta B_\parallel$ is sufficient to suppress the dominant instability in our simulations, see \secref{sec:subdominant MTM instability}. Otherwise, the physics included in $f_{B} = 0$ is the same as that in $f_{B} = 1$ simulations, evolving three kinetic species (electrons, deuterium and tritium). The linearized Fokker-Planck collision model of \cite{barnes2009} is used to model collisions in the system. In this Section, we will report primarily on simulations of the dominant instability (using parameters in the $f_{B} = 1$ column), and a thorough discussion of simulations of the subdominant instability (using parameters in the $f_{B} = 0$ column) will be deferred to \secref{sec:subdominant MTM instability}.

\begin{table}[ht]
    \centering
    \begin{tabular}{|c|a|a|c|c|c|c|}
    \hline
    \multirow{2}{*}{\textbf{Grid parameter}} & \multicolumn{2}{c|}{\textbf{GS2}} & \multicolumn{2}{c|}{\textbf{GENE}} & \multicolumn{2}{c|}{\textbf{CGYRO}}\\
    \cline{2-7}
                & $f_{B} = 1$ & $f_{B} = 0$ & $f_{B} = 1$ & $f_{B} = 0$ & $f_{B} = 1$ & $f_{B} = 0$ \\
    \hline
    $n_\theta$                  & 64 & 32 & 64 & 96 & 32 & 64\\
    \hline
    $n_r$       & 5  & 32 & 64  & 64  & 16 & 64\\
    \hline
    $n_\lambda$, $n_{v_{\parallel}}$, $n_\xi$   & 41  & 25  & 32  & 64 &  24 & 96\\
    \hline
    $n_\epsilon$, $n_\mu$          & 16 & 16 & 24  & 32 &  12 & 12\\
    \hline
    \end{tabular}
    \caption{Numerical resolution used in CGYRO, GENE and GS2 linear simulations of the full model ($f_{B} = 1$) and the model without $\delta B_\parallel$ ($f_{B} = 0$) at $\Psi_n = 0.49$ ($q=3.5$) of STEP-EC-HD.
    In CGYRO, $n_\xi$ is the number of Legendre pseudospectral meshpoints in the pitch-angle space and $n_\epsilon$ is the number of generalized-Laguerre pseudospectral meshpoints.
    In GENE, $n_{v_\parallel}$ and $n_\mu$ are the number of grid points in the $v_\parallel$ and $\mu$ direction, respectively. 
    In GS2, $n_\epsilon$ is the number of energy grid points and $n_\lambda$ is the number of pitch-angles.}
    \label{tab:resolution_spr45}
\end{table}

We begin by performing linear initial value calculations to find the dominant unstable modes (i.e., the fastest growing unstable mode) across a range of different binormal wavenumbers. In Figure~\ref{fig:psin}, we plot growth rate, $\gamma$, and mode frequency, $\omega$, (both normalised to the ion sound frequency) as functions of the normalised perpendicular binormal wavenumber, $k_y \rho_{s}$, at various radial locations corresponding to low $q$ rational surfaces: $\Psi_n = 0.36$ ($q=3.0$); $\Psi_n = 0.49$ ($q=3.5$); $\Psi_n = 0.58$ ($q=4.0$); and $\Psi_n = 0.71$ ($q=5.0$).

\subsection{Electron Larmor radius scale modes}

\begin{figure}
    \centering
    \subfloat[]{\includegraphics[height=0.23\textheight]{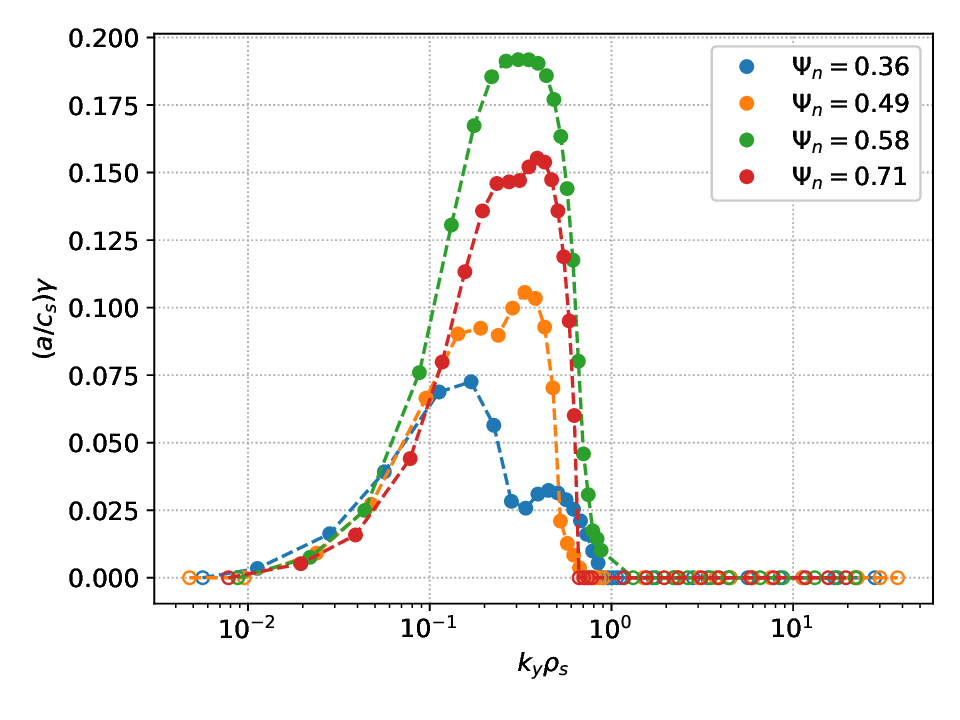}}\quad
    \subfloat[]{\includegraphics[height=0.23\textheight]{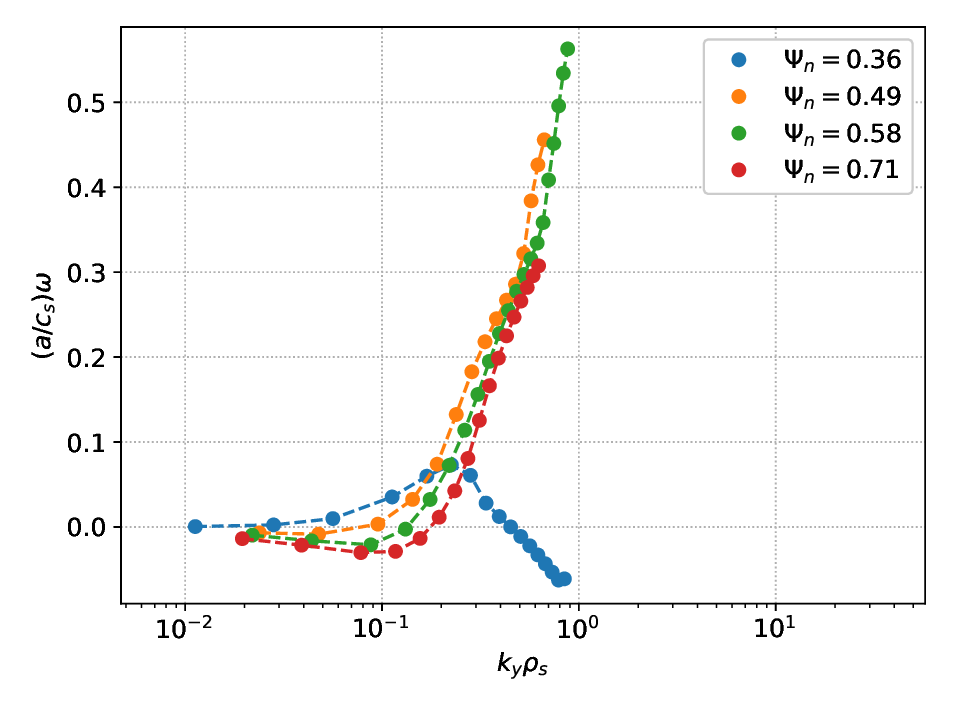}}
    \caption{Growth rate (a) and mode frequency (b) as functions of $k_y \rho_s$ from GS2 linear simulations of STEP-EC-HD at various radial surfaces corresponding to low $q$ rational surfaces. The considered $k_y$ values cover a range corresponding to toroidal mode numbers between $n=2$ and $n=5000$. Frequency values are shown only for unstable modes with $n<200$. Solid and open markers refer to unstable and stable modes, respectively. {The growth rate of stable modes is set to zero.} No unstable modes are found below $n=5$.}
    \label{fig:psin}
\end{figure}

We begin by noting that there is no purely electron scale instability in the system at any of the core flux surfaces considered (see Fig.~\ref{fig:psin}), a result which is due to the large $\beta$ compared to conventional tokamaks. We also remark here that we see a distinct absence of the collisionless MTM which tends to dominate the instability spectrum at intermediate scales $k_{y}\rho_{s} = \mathcal{O}(1)$ in similar ST equilibria \cite{patel2021,dickinson2022};  the absence of the collisionless MTM is due to the larger value of the density gradient owing to pellet fuelling (which strongly stabilises the collisionless MTM). 

\subsection{Ion Larmor radius scale modes}

Approaching the binormal Deuterium Larmor radius scale $k_{y}\rho_{s} \lesssim 1,$ the stability landscape is somewhat more complicated. For clarity, we group the flux surfaces by spectra structure.

\subsubsection{$q = 3.5$ $(\Psi_n = 0.49)$} \hfill \\

\noindent From Figure~\ref{fig:psin}, we note that the maximum growth rate occurs approximately at $k_y\rho_s \simeq 0.4$ for the surface at $\Psi_n = 0.49.$ Interestingly, the growth rate as a function of $k_{y}\rho_s$ has two local maxima, similar to that seen in similar ST designs \cite{dickinson2022,patel2021}. The mode frequency is positive (i.e., the most unstable mode is propagating in the ion diamagnetic direction) for all unstable $k_y \rho_s$ modes (i.e., those modes where $\gamma>0$). The mode is stable as we approach the sub-ion Larmor radius scale $(k_{y}\rho_{s} > 0.65, \, n > 130)$ but is unstable down to very long wavelengths $(k_{y}\rho_{s} = 0.023, \, n = 5).$ 

\subsubsection{$q = 4.0$ $(\Psi_n = 0.58)$ and $q = 4.5$ $(\Psi_n = 0.71)$} \hfill\\

\noindent Similarly to $\Psi_n = 0.49,$ the maximum growth rate occurs on both surfaces approximately at $k_y\rho_s \simeq 0.4.$ For $\Psi_n = 0.71,$ the growth rate spectrum once again has two local maxima. For $\Psi_n = 0.58,$ there is a single local maxima but we observe a similar plateau structure in the growth rate spectrum. One key difference with respect to the $q = 3.5$ ($\Psi_n = 0.49$) surface is that the longest wavelength unstable modes have weakly negative mode frequency (i.e., the mode is propagating in the electron diamagnetic direction). We note however, that the growth rate and frequency vary smoothly as $k_{y}\rho_{s}$ decreases (c.f., the abrupt change of sign in the real frequency between the unstable modes $k_{y}\rho_s < 0.65$ and the stable modes $k_{y}\rho_s > 0.65$ on the $\Psi_n = 0.49$ surface) suggesting that this is perhaps not a discrete mode transition but instead is a change in the nature of the dominant instability (see \secref{sec:hybrid TEM/KBM instability} for further discussion).

\subsubsection{$q = 3.0$ $({\Psi_n = 0.36})$} \hfill \\

\noindent The maximum growth rate moves to slightly longer wavelengths $k_y\rho_s \simeq 0.2$ at $\Psi_n = 0.36,$ though it once again possess two local maxima. Again, we observe a slightly different dependence of the mode frequency on $k_y \rho_s$ at $\Psi_n \simeq 0.36$ compared to the $q = 3.5$ ($\Psi_n = 0.49$) surface: the frequency increases at low $k_y$, reaching a maximum around $k_y\rho_s \simeq 0.3$, the frequency then decreases and changes sign at $k_y\rho_s \simeq 0.4.$ Once again, we note that this change in frequency occurs smoothly. 

The remainder of this manuscript is largely devoted to studying the linear instabilities identified in the STEP-EC-HD equilibrium, focusing in particular on the $q = 3.5$ flux surface $(\Psi_n = 0.49)$, unless otherwise explicitly indicated.  

\section{Hybrid-KBM instability}
\label{sec:hybrid TEM/KBM instability}

Based on previous results (see e.g., \cite{kaye2021} and references therein), we might expect in these high-$\beta$ plasmas that some of the instabilities in Figure~\ref{fig:psin} propagating in the ion diamagnetic direction are electromagnetic KBMs, especially where the local equilibrium profiles do not access second stability \cite{Hong1989, davies2022} This section is dedicated to studying the physics of the dominant instability identified in Figure~\ref{fig:psin}. 

\subsection{Is the mode electromagnetic or electrostatic?}

\begin{figure}
    \centering
    \subfloat[]{\includegraphics[width=0.32\textwidth]{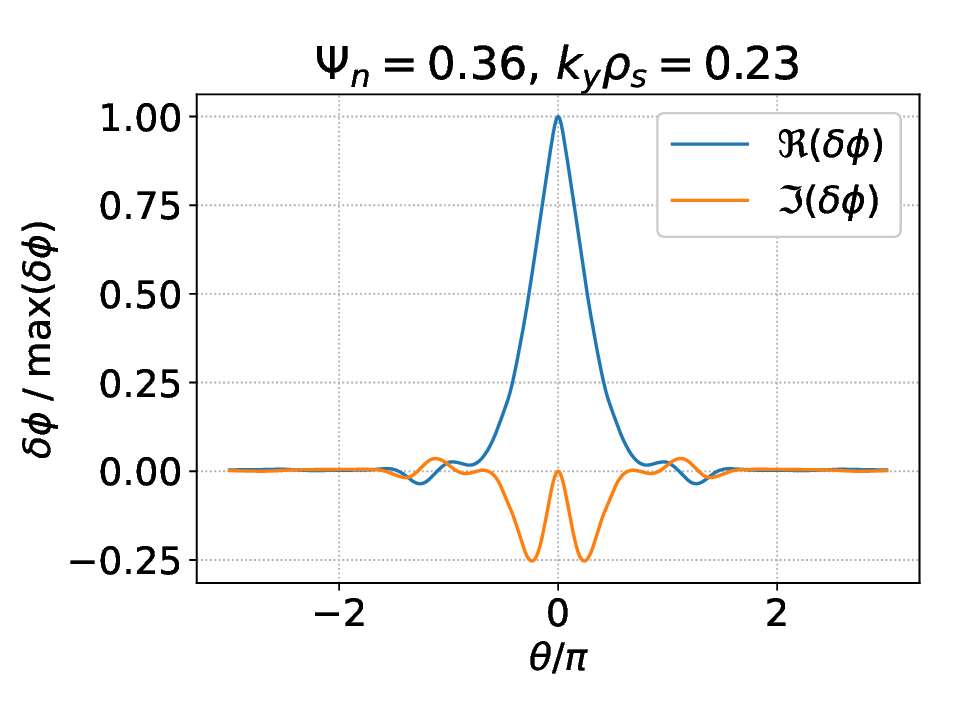}}\,
    \subfloat[]{\includegraphics[width=0.32\textwidth]{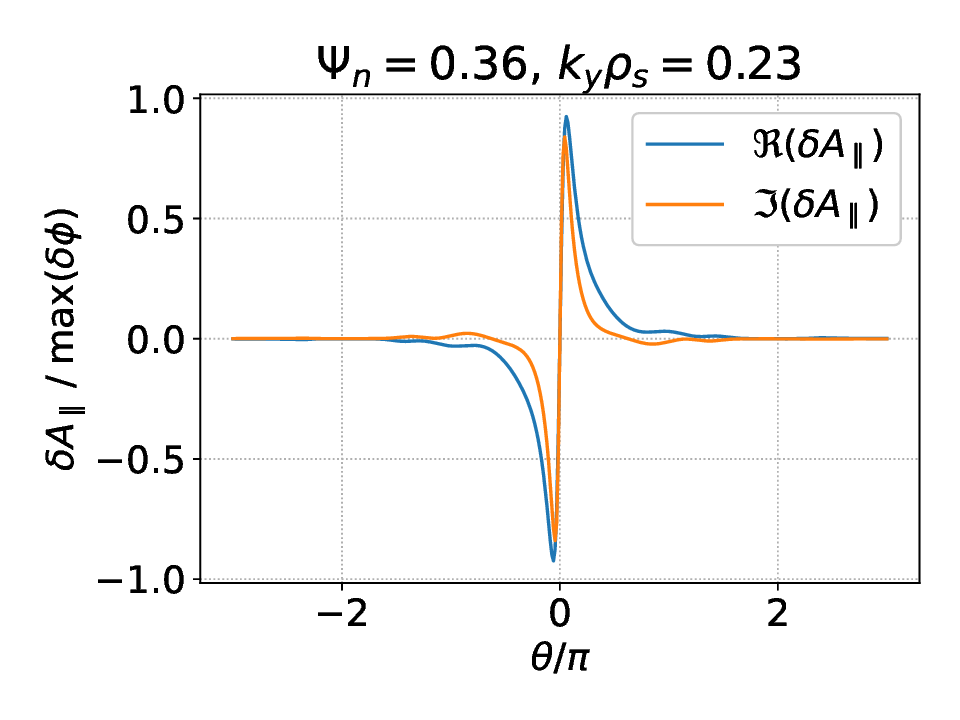}}\, 
    \subfloat[]{\includegraphics[width=0.32\textwidth]{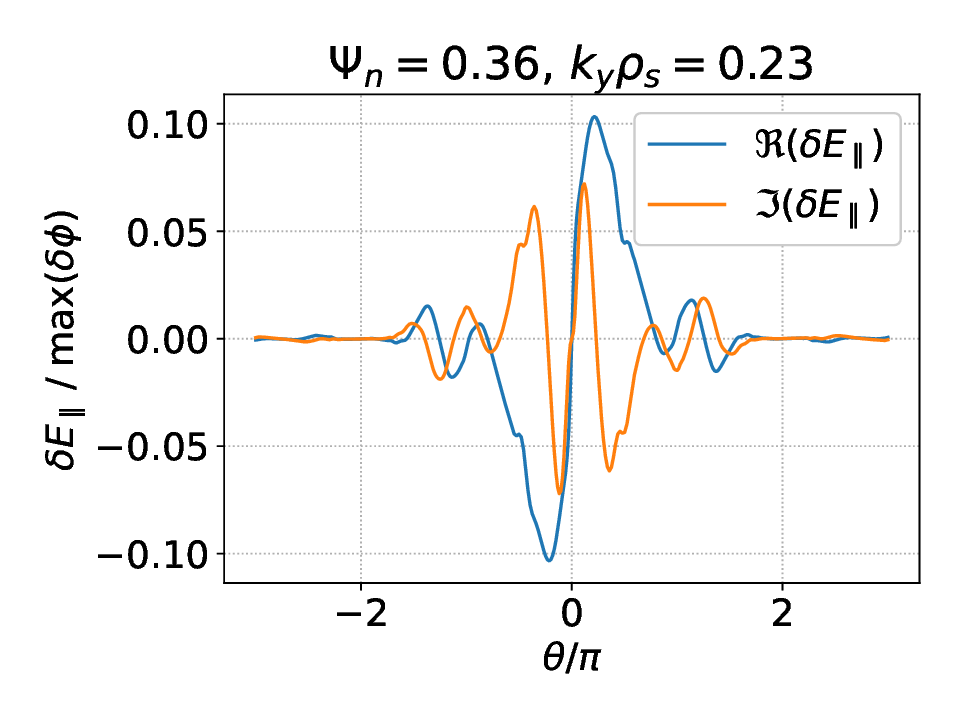}}\\
    \subfloat[]{\includegraphics[width=0.32\textwidth]{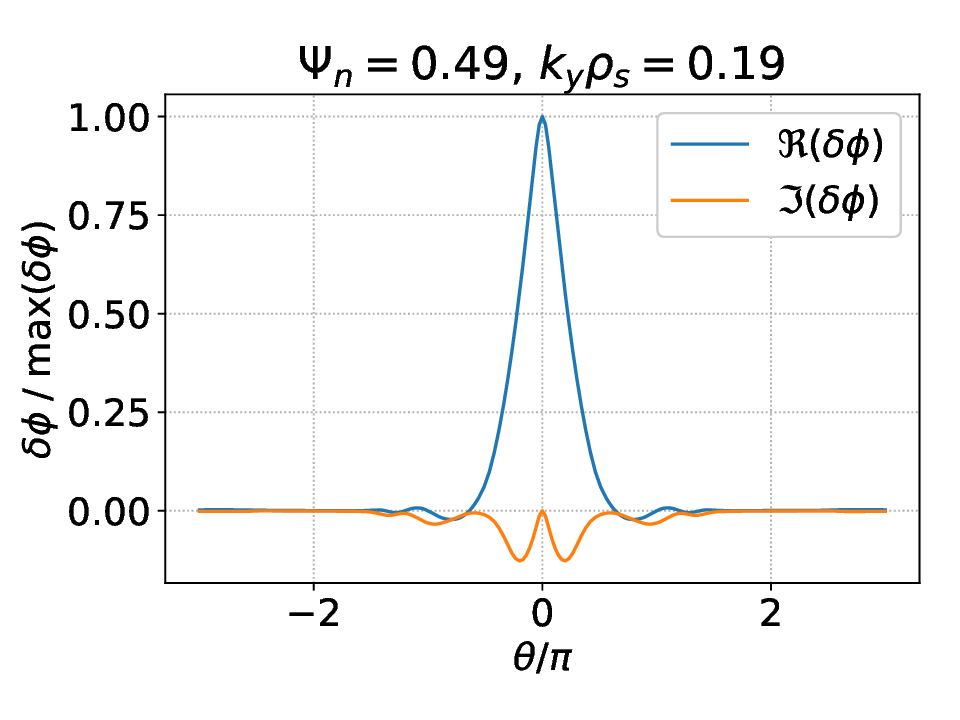}}\,
    \subfloat[]{\includegraphics[width=0.32\textwidth]{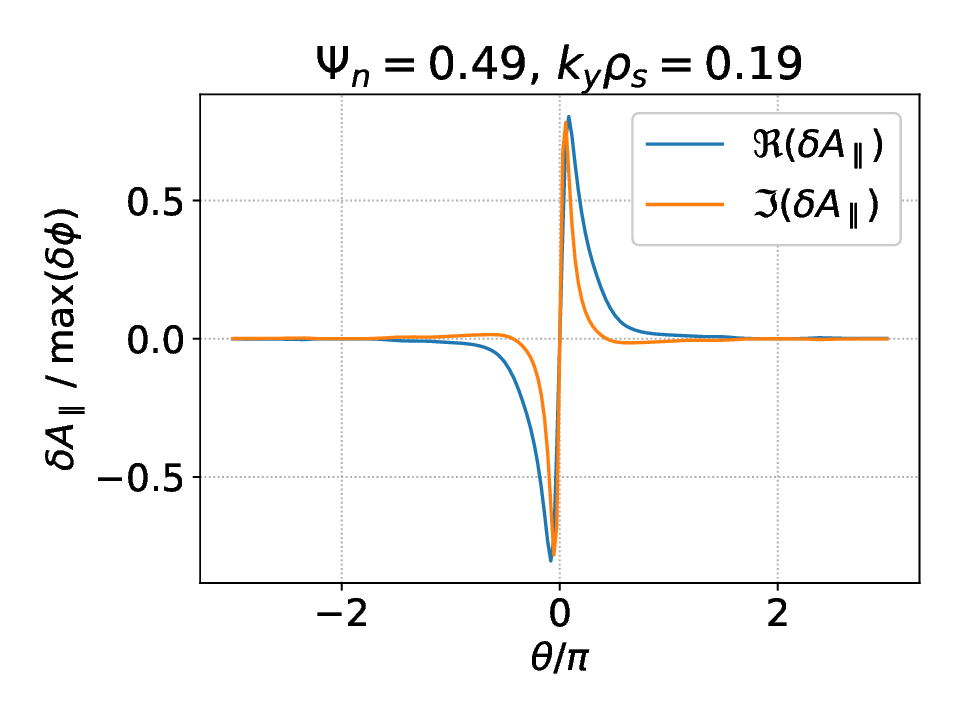}}\,
    \subfloat[]{\includegraphics[width=0.32\textwidth]{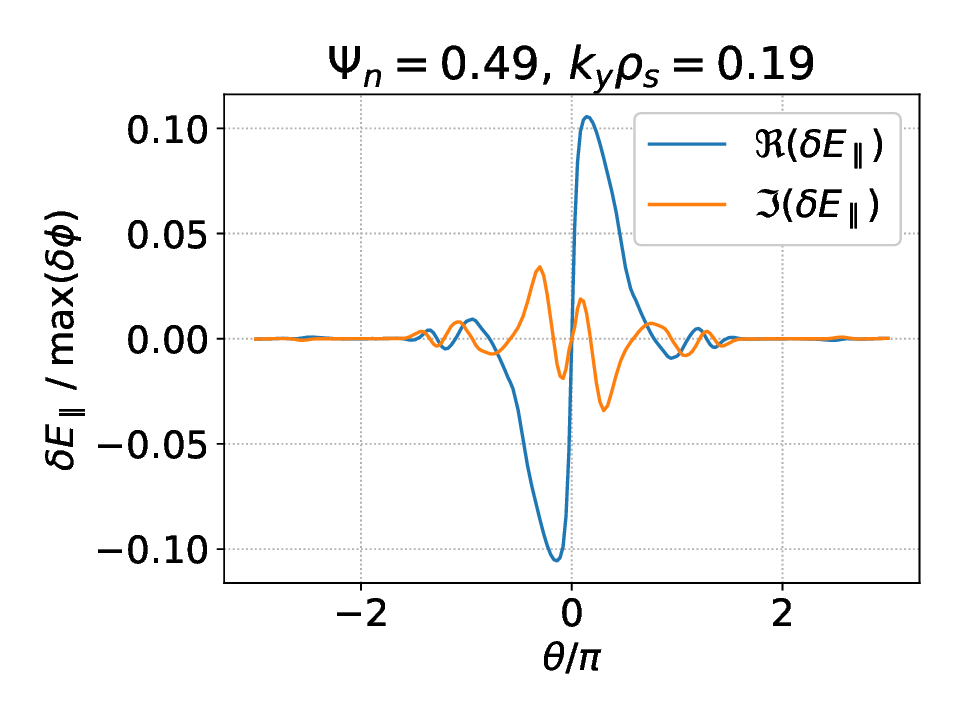}}
    \caption{Real and imaginary part of $\delta\phi/\max(\delta\phi)$ [(a) and (d)], $\delta A_\parallel/\max(\delta \phi)$ [(b) and (e)] and $\delta E_\parallel$ [(c) and (f)] at $\Psi_n=0.49$ and $\Psi_n=0.36$ of STEP-EC-HD for the $k_y\rho_s\simeq 0.2$ mode.}
    \label{fig:eig_kbm_02}
\end{figure}

\begin{figure}
    \centering
    \subfloat[]{\includegraphics[width=0.32\textwidth]{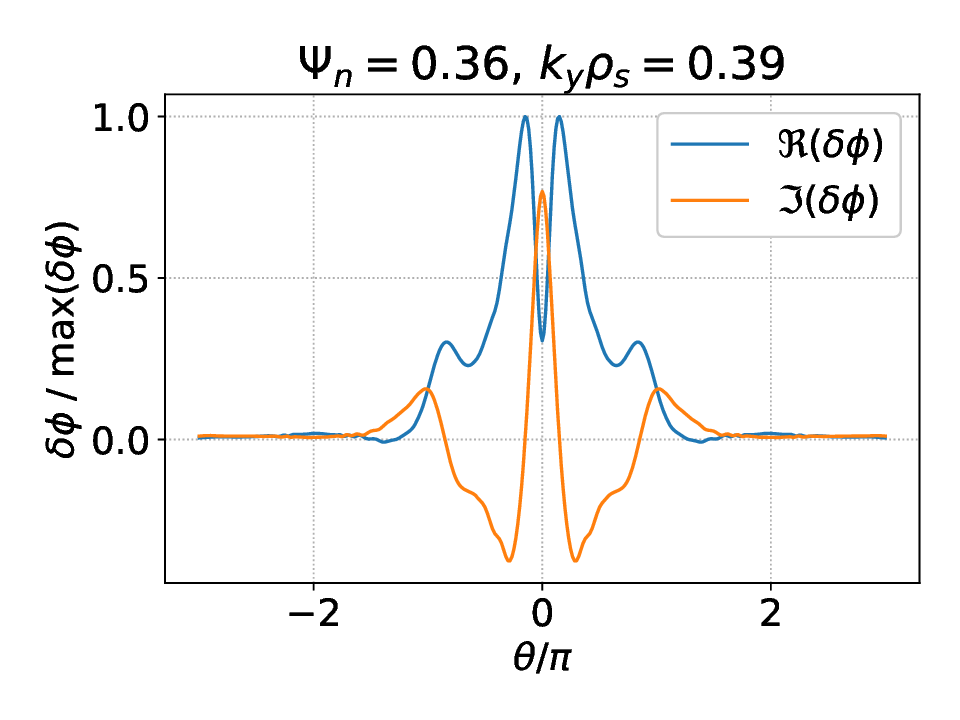}}\,
    \subfloat[]{\includegraphics[width=0.32\textwidth]{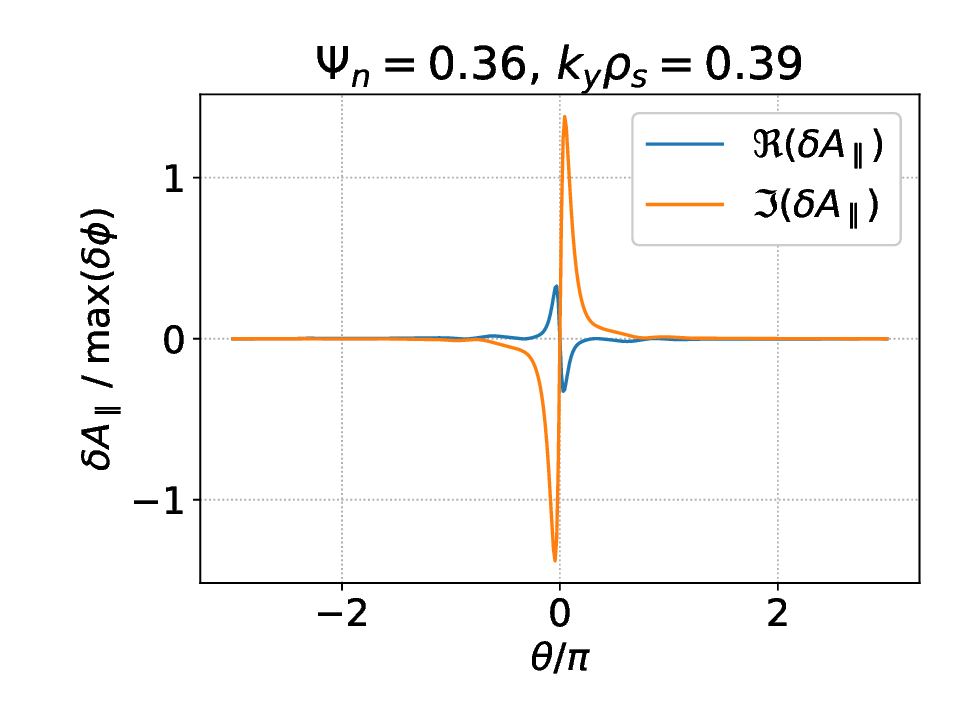}}\,
    \subfloat[]{\includegraphics[width=0.32\textwidth]{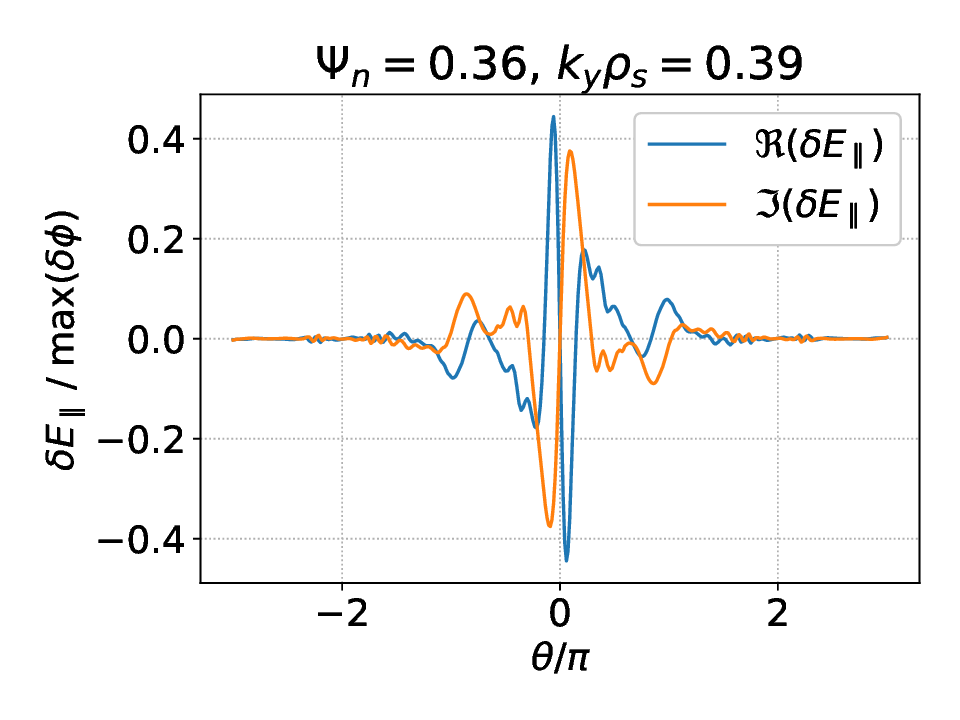}}\\
    \subfloat[]{\includegraphics[width=0.32\textwidth]{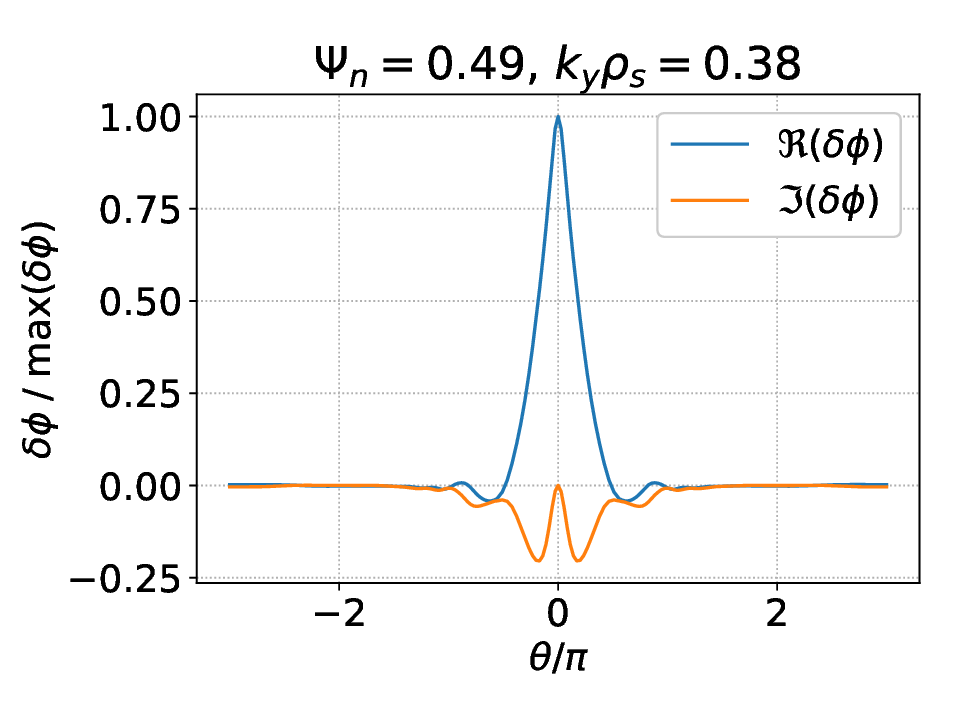}}\,
    \subfloat[]{\includegraphics[width=0.32\textwidth]{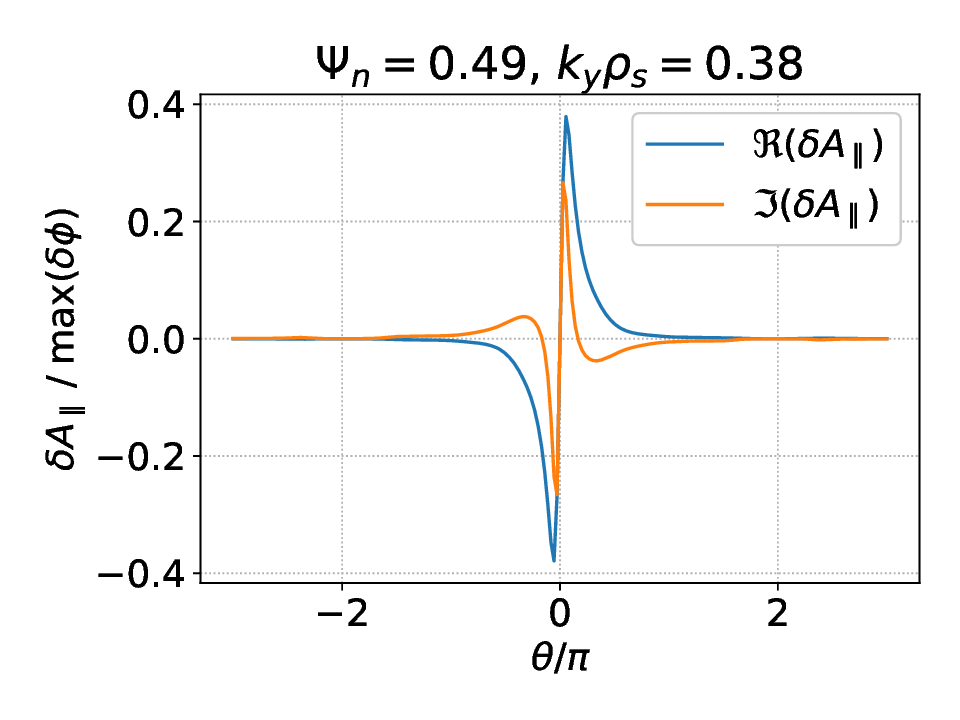}}\,
    \subfloat[]{\includegraphics[width=0.32\textwidth]{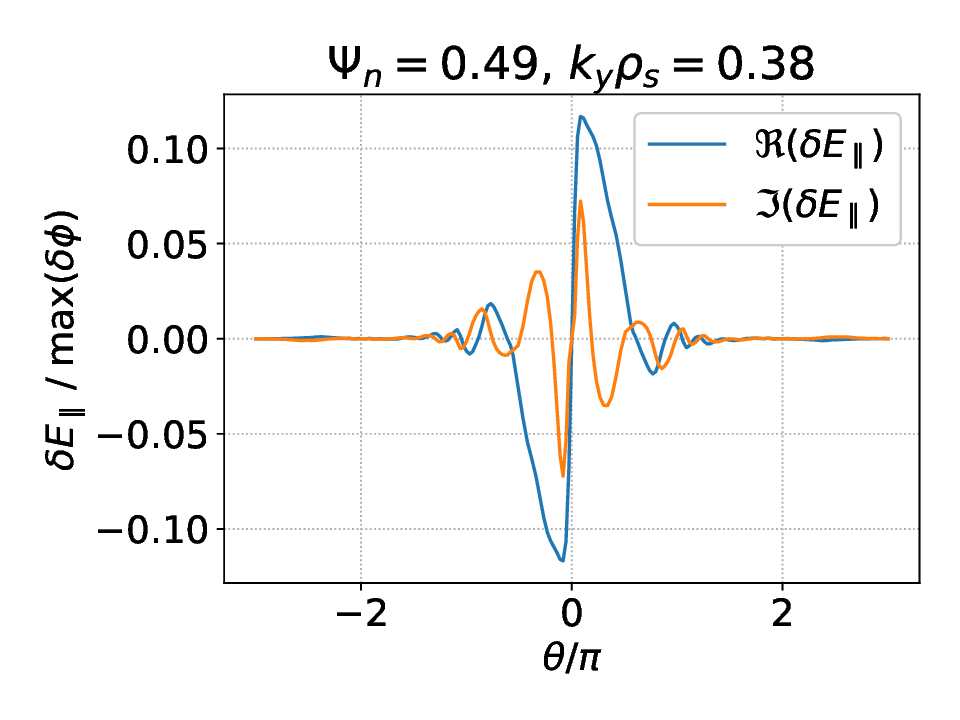}}
    \caption{Real and imaginary part of $\delta\phi/\max(\delta\phi)$ [(a) and (d)], $\delta A_\parallel/\max(\delta\phi)$ [(b) and (e)] and $\delta E_\parallel/\max(\delta\phi)$ [(c) and (f)] at $\Psi_n=0.49$ and $\Psi_n=0.36$ of STEP-EC-HD for the $k_y\rho_s\simeq 0.4$ mode.}
    \label{fig:eig_kbm_04}
\end{figure}

A sensible first step towards classifying and understanding this instability is to examine whether the mode is predominantly electrostatic or electromagnetic, this can be done by examining the eigenfunctions of the dominant instability identified in \secref{sec:overview of simulation results}. In Figures~\ref{fig:eig_kbm_02}-\ref{fig:eig_kbm_04}, we plot the $\delta\phi$ and $\delta A_{\parallel}$ eigenmode structures (both normalised to the maximum value of $\delta \phi$) as functions of ballooning angle $\theta,$ at $k_y\rho_s\simeq0.2$ (Figure~\ref{fig:eig_kbm_02}) and at $k_y\rho_s\simeq0.4$ (Figure~\ref{fig:eig_kbm_04}) for the flux surfaces with $\Psi_n = 0.36$ and $\Psi_n = 0.49$ respectively. We note that the amplitudes of $\delta\phi$ and $\delta A_\parallel$ are comparable, thus suggesting that

\begin{enumerate}[label=\textbf{P.\arabic*}]
    \item\label{item1}{the mode is predominantly electromagnetic.} 
\end{enumerate}

Electrostatic instabilities are typically characterised by $|\delta A_{\parallel}| \ll |\delta \phi|$. 
At both radial locations, the mode is strongly peaked around $\theta = 0,$ with even parity in $\phi$ and odd parity in $A_{\parallel}.$ Conventionally, even parity $\phi$ modes are called ‘twisting parity’ and odd parity $\phi$ modes are called ‘tearing parity’. Therefore, 

\begin{enumerate}[label=\textbf{P.\arabic*},resume]
    \item{the mode has twisting parity.} \label{item2}
\end{enumerate}


Based on properties \ref{item1} and \ref{item2}, and in agreement with previous results in \cite{dickinson2022,patel2021} with a similar parameter regime, the dominant instability may be associated with a KBM. We can investigate this further by examining whether the mode is indeed active where the equilibrium profiles do not access second stability. 

\subsection{Is the mode a KBM?}

A general description of KBMs was presented by \cite{antonsen1980} and \cite{tang1980}, in which the linear electromagnetic gyrokinetic equation is solved for the gyrokinetic distribution function in terms of the perturbed fields $\delta \phi$, $\delta A_{\parallel},$ and $\delta B_{\parallel},$ and the expression for the gyrokinetic distribution function is inserted into the field equations. This results in three coupled, linear, integro-differential equations which may then be solved in certain limits. In theory, one could analyze this system of equations to determine which design choices (e.g., shaping) are beneficial for KBM stability. However, the complexity of these equations make it difficult to assess whether kinetic effects have a net stabilising or net destabilising effect beyond simple limits \cite{aleynikova2018,aleynikova2022}. In a complex physical system such as an ST, accurately describing KBMs thus typically requires gyrokinetic simulations to explore the sensitivity of these modes. As a first step, we investigate the stability with respect to the ideal ballooning boundary, which is often used a simple proxy for KBM stability.

\subsubsection{The Ideal Ballooning Mode}

The KBM instability is often associated with the MHD ideal ballooning mode (IBM) in the limit of $n \to \infty$, which is derived from ideal MHD and thereby neglecting kinetic effects such as the finite Larmor radius and the effect of trapped particles. Despite making considerable simplifications to the physics, the IBM still describes the basic physics of the pure KBM instability; a competition between the stabilising effect of magnetic field line bending and the destabilising effect of a plasma pressure gradient combined with “bad” magnetic curvature. Moreover, IBM stability is much more easily assessed for a given plasma, and is sometimes used as a proxy for KBM stability in models such as the predictive pedestal model EPED (see e.g., \cite{snyder2009}) and a good correlation is generally found in the pedestal of conventional tokamaks between the region where KBMs dominate and the region that is unstable to $n \to \infty$ ideal ballooning modes (see e.g., \cite{dickinson2011,davies2022} and discussion therein).

An approach pioneered by \cite{connor1970} allows one to calculate stability quickly and easily by integrating a one-dimensional differential equation for a given field line. This has been numerically implemented in GS2’s module \texttt{ideal\_ball}. Moreover, for some fixed set of geometric parameters, \texttt{ideal\_ball} can be used to scan the normalised pressure gradient $\alpha = -R q^2 \mathrm{d}\beta/\mathrm{d}r$, and magnetic shear $\hat{s} \equiv \partial q / \partial \psi$ to evaluate IBM stability for a given flux surface as a function of $(\hat{s},\alpha).$ 
We therefore investigate where STEP-EC-HD and STEP-EB-HD are located with respect to the region of IBM stability and whether this is consistent with our suspicion that these equilibria are KBM dominated.
\begin{figure}
    \centering
    \includegraphics[scale=0.6]{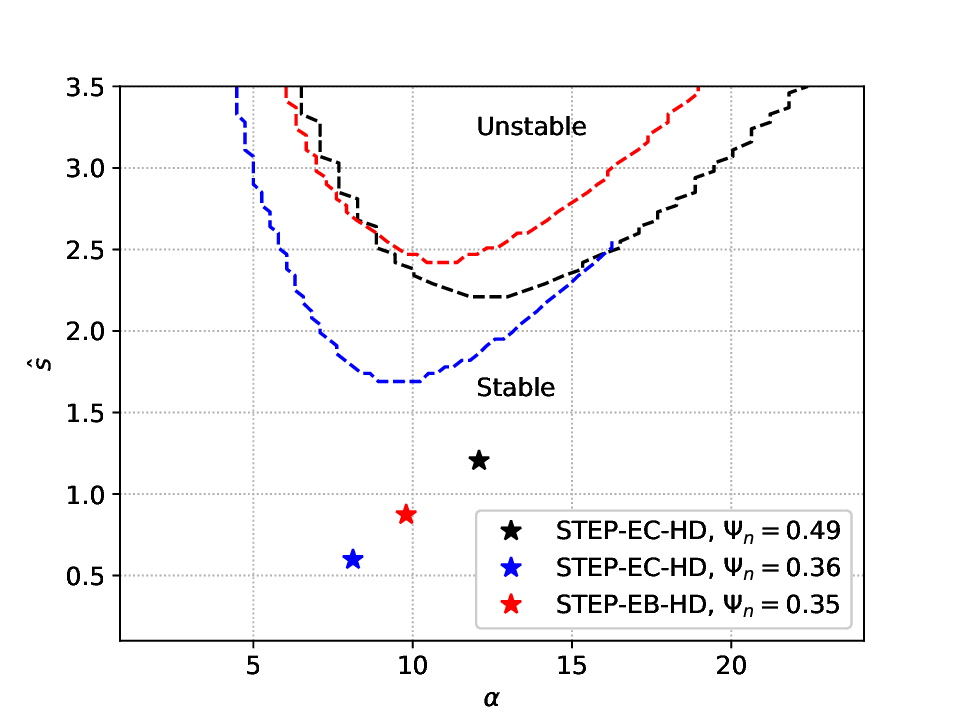}
    \caption{Ideal ballooning stability boundary in the $\hat{s}-\alpha$ plane of STEP-EC-HD  at $\Psi_n=0.49$ (black line) and at $\Psi_n=0.36$ (blue line) and of STEP-EB-HD at $\Psi_n=0.35$ (red line). The makers denote the equilibrium value of $\hat{s}$ and $\alpha$ for each surface.}
    \label{fig:ideal_ball}
\end{figure}

The results of these calculations are shown in Figure~\ref{fig:ideal_ball}, where we see that all of the flux surfaces we have considered are well outside the unstable region. As such, we expect STEP plasma operating in these regions of parameter space to be stable to IBMs, a sensible proxy for KBM stability. This is an important piece of information about the dominant mode. 
\begin{enumerate}[label=\textbf{P.\arabic*},resume]
    \item{The dominant mode can be unstable in the region where the IBM is stable.}\label{item3}
\end{enumerate}
{ It is well known that kinetic effects can make the KBM unstable in the IBM-stable region, e.g. a finite ion temperature gradient was found to make KBM unstable below the beta threshold of the IBM \cite{cheng1982}. Therefore, we emphasise that \ref{item3} is insufficient to exclude the dominant mode from being labelled as a KBM, though it supports the need to explore broader mode properties, which will be a major focus of this paper.} 

\subsection{Mode fingerprinting}
\label{subsec:fingerprinting}

Statements \ref{item1} and \ref{item2} indicate that the dominant mode has clear features consistent with the KBM instability, while \ref{item3} seems to suggest a mode with different instability characteristics, i.e. this mode might be KBM-like, but it may also be coupling to some other modes in our system in order to be driven unstable. We can examine whether this might be the case by fingerprinting the mode. Mode ‘fingerprints' \cite{Kotschenreuther_2019} to identify the instabilities that cause transport losses in modern experiments from among widely posited candidates such as the KBM and others. The key idea underpinning mode fingerprinting is that analysis of both the gyrokinetic-Maxwell equations and gyrokinetic simulations of experiments reveals that each mode type produces characteristic ratios of transport in the density and heat channels. Thus, by examining the electron to ion heat and particle flux ratios, we might shed light on the nature of our instability. The important quantities for fingerprinting analysis are the particle and heat diffusivity, $D_\alpha = \Gamma_\alpha / (\dd n_\alpha / \dd r),$ and  $\chi_\alpha = [Q_\alpha - (3/2)T_\alpha\Gamma_\alpha] / (n_\alpha \dd T_\alpha / \dd r)$, where $\alpha$ is the species label and $\Gamma$ and $Q$ denote the particle and heat flux respectively. 
\begin{table}
    \centering
    \begin{tabular}{|c|c|}
    \hline
    \multicolumn{2}{|c|}{\textbf{Mode fingerprint}}\\
    \hline
    $\chi_e/\chi_i$  & 0.86 \\
    \hline
    $D_e/\chi_e$  & 0.68\\
     \hline
    \end{tabular}
    \caption{Electron to ion heat diffusion coefficient ratio, $\chi_e/\chi_i$, and electron particle to heat diffusion coefficient ratio, $D_e/\chi_e$, for the $k_y\rho_s=0.2$ mode of $q = 3.5 (\Psi_n=0.49)$. The  heat and particle diffusion coefficients are computed as $\chi_s=Q_s/(\partial p_s/\partial r)$ and $D_s = \Gamma_s/(\partial n_s/\partial r)$, respectively, where $Q_s$ and $\Gamma_s$ are the heat and particle fluxes of species $s$. }
    \label{tab:fingerprint}
\end{table}
The mode fingerprints identified in \cite{Kotschenreuther_2019} are reported in Table~\ref{tab:fingerprint} for the $k_y\rho_s = 0.2$ mode of $q = 3.5$ $(\Psi_n = 0.49)$. Comparing our simulation results with the fingerprint identifiers given in Table 1 of \cite{Kotschenreuther_2019} we see that our dominant instability does indeed have features in common with MHD-like modes (including the KBM) which cause very comparable diffusivities in all channels, and are characterised by $|\delta E_\parallel| \ll |\delta \phi|$ (see Figures~\ref{fig:eig_kbm_02} and \ref{fig:eig_kbm_04}). However, we remark that this fingerprint may also be also compatible with the ion-temperature gradient mode (ITG) and trapped electron mode (TEM). 

\begin{enumerate}[label=\textbf{P.\arabic*},resume]
    \item {The mode can be fingerprinted as a KBM or ITG/TEM.} \label{item4}
\end{enumerate}

Observation \ref{item4} provides us with a way to reconcile \ref{item3} with \ref{item1} and \ref{item2}, the dominant mode is likely a hybrid instability. We now wish to study this hybrid instability. 

\subsection{Sensitivity to different physics parameters and hybridisation of the KBM}

Based on our fingerprinting analysis, we have deduced that the dominant instability is likely a hybrid mode which shares the features of the KBM and of something reminiscent of an ITG or TEM. In the following, we attempt to further characterise the main instability by analysing the dependence on the inclusion of collisions, the numbers of species, the local gradients, the magnetic shear and safety factor, and $\beta_e$.

\subsubsection{Pressure gradient}

\begin{figure}
    \centering
    \subfloat[]{\includegraphics[width=0.31\textwidth]{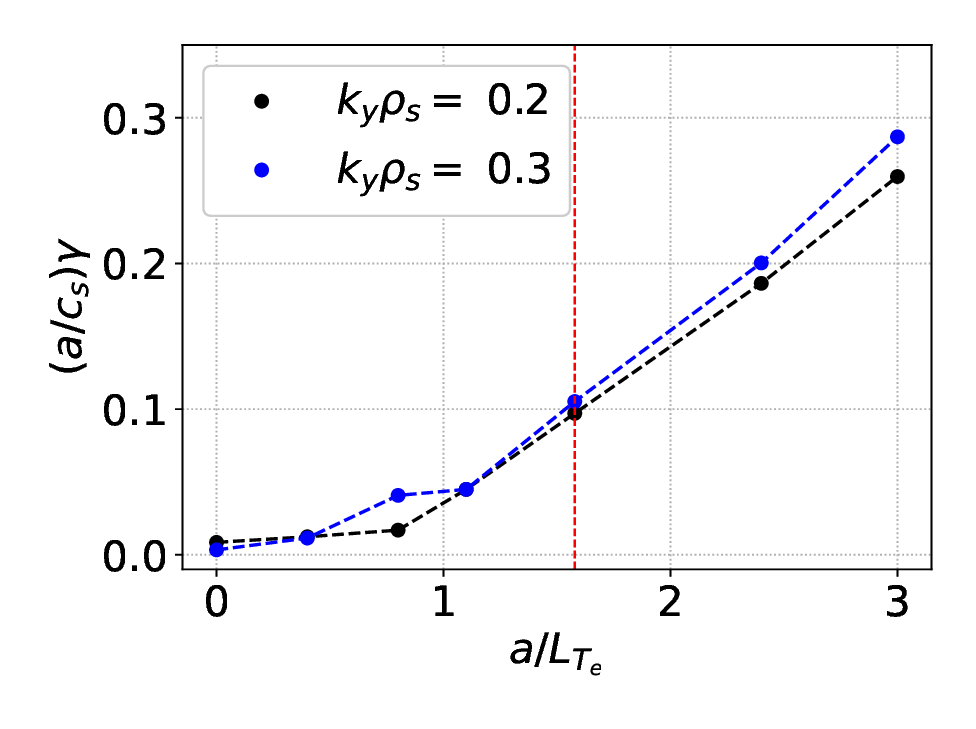}}
    \subfloat[]{\includegraphics[width=0.31\textwidth]{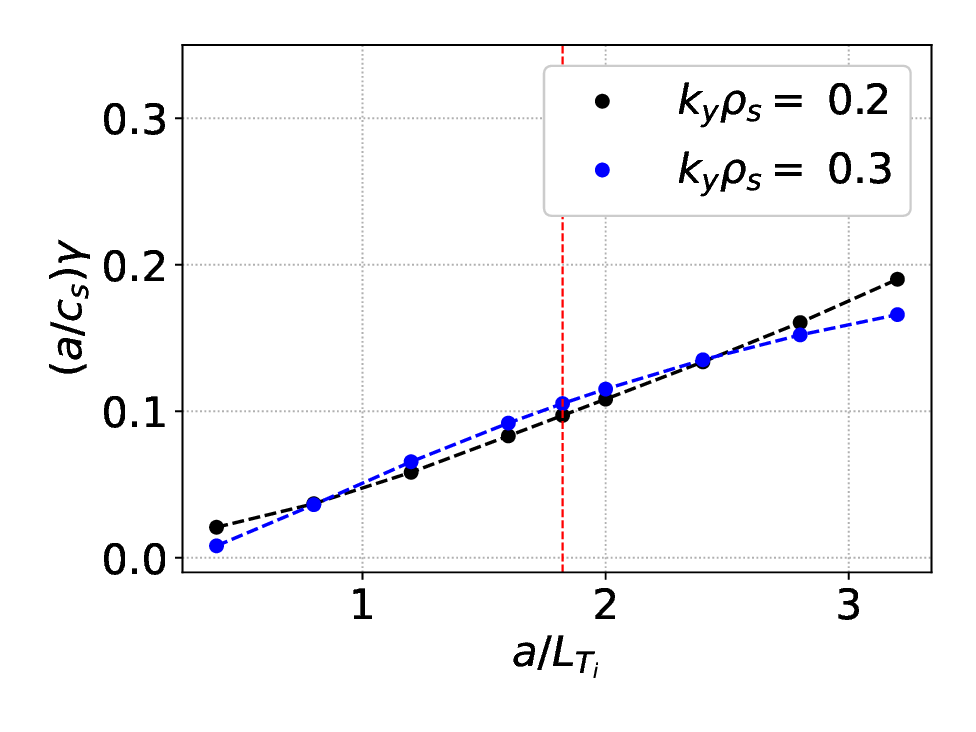}}
    \subfloat[]{\includegraphics[width=0.31\textwidth]{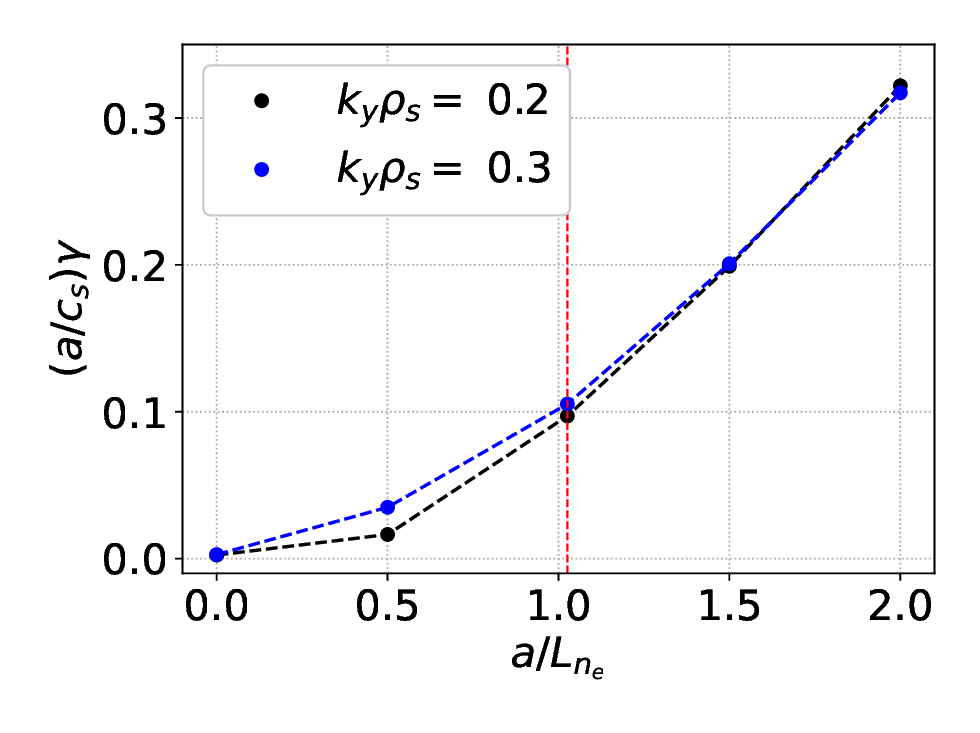}}
    \caption{Growth rate from GS2 linear simulations as a function of $a/L_{T_e}$ (a), $a/L_{T_i}$ (b) and $a/L_n$ (c) at $k_y\rho_s=0.2$ (black line) and $k_y\rho_s=0.3$ (blue line). The red dashed line represents the reference value. All the other parameters except $a/L_{T_e}$ in (a), $a/L_{T_i}$ in (b) and $a/L_n$ in (c) are kept fixed. Results at the surface $\Psi_n=0.49$ of STEP-EC-HD.}
    \label{fig:scan}
\end{figure}

\begin{figure}
    \centering
    \subfloat[]{\includegraphics[height=0.23\textheight]{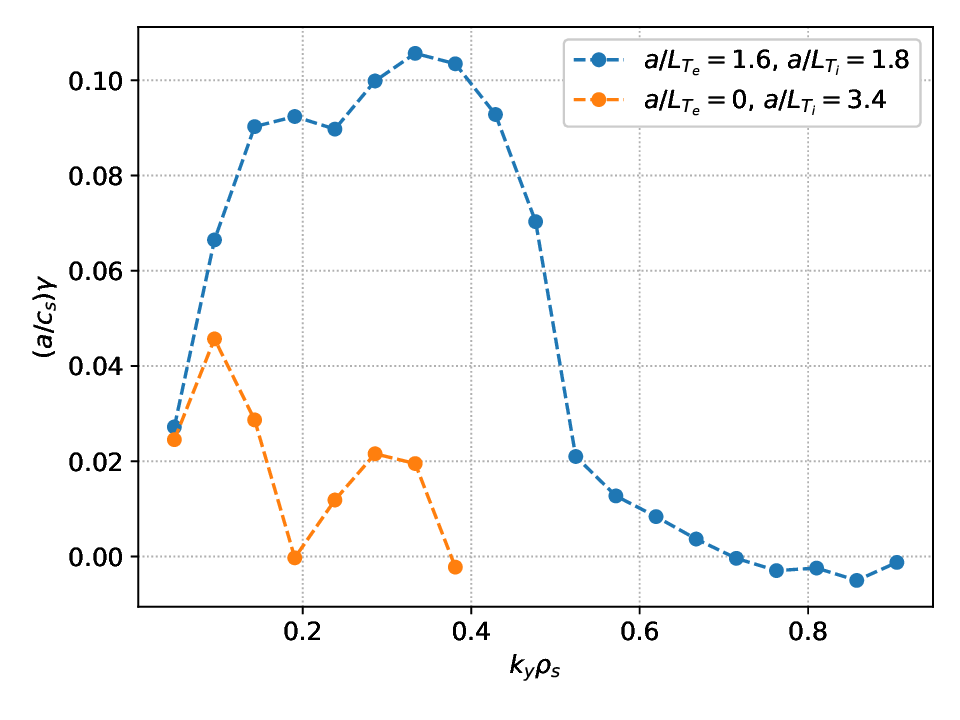}}\quad
    \subfloat[]{\includegraphics[height=0.23\textheight]{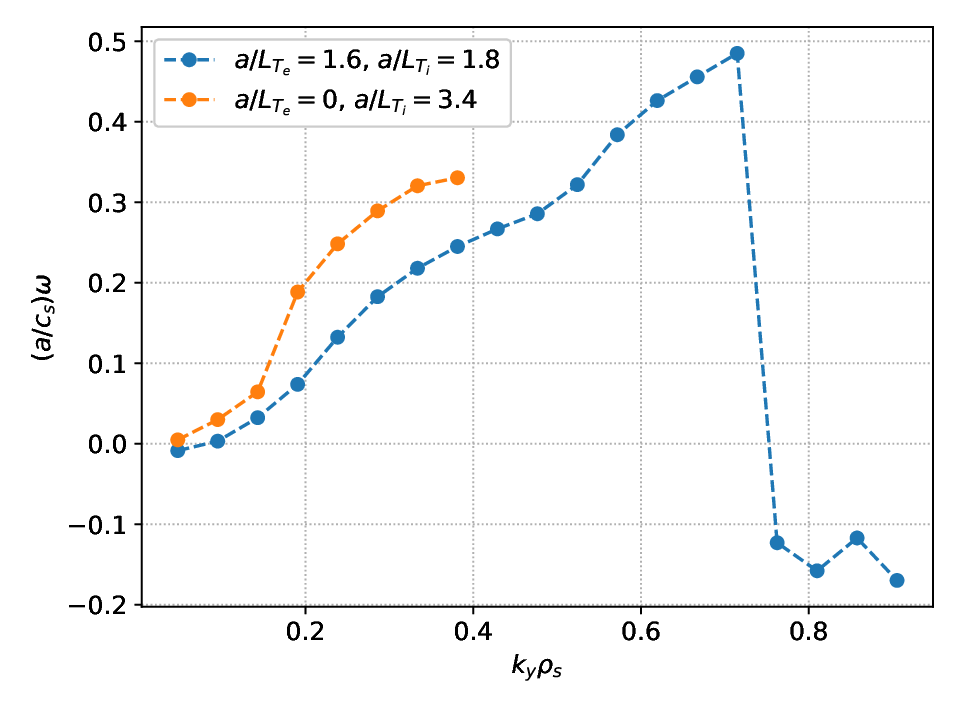}}
    \caption{Growth rate (a) and mode frequency (b) as functions of $k_y \rho_s$ from GS2 linear simulations with nominal (blue markers) and zero (orange markers) electron temperature gradient and same pressure gradient. Results at $\Psi_n=0.49$ of STEP-EC-HD.}
    \label{fig:alte0}
\end{figure}

When the local geometry is held fixed, the growth rate of a pure KBM increases with the total pressure gradient and $\beta_e$. Thus, assessing the sensitivity of the dominant mode to these parameters allows us to test to what extent the mode is KBM-like. In Figure~\ref{fig:scan} we explore the dependence of the mode on the electron and ion temperature gradients, $a/L_{Te}$ and $a/L_{Ti}$, as well as on the density gradient\footnote{Note that we do not vary the electron and ion density gradients independently since quasineutrality requires $n_{e}=n_{i}$ which in turn demands $a/L_{n_e} = a/L_{n_i}$ globally).} $a/L_n$, whilst all other parameters are held constant. Here, we note that both ion species temperatures are changed together. Figure~\ref{fig:scan} reveals another important characteristic of the hybrid mode,
\begin{enumerate}[label=\textbf{P.\arabic*},resume]
    \item{the growth rate is sensitive to changes in the pressure gradient (a KBM-like feature).}\label{item5}
\end{enumerate}
From comparing Figures~\ref{fig:scan}~(a) and (b) we note that the growth rate appears to depend more strongly on the electron temperature gradient than on the ion temperature gradient for both of the binormal wavenumbers considered. In Figure~\ref{fig:alte0}, we show a comparison of the linear spectrum for two linear simulations with the same total kinetic pressure gradient but different electron temperature gradients; one with the nominal electron temperature gradient (blue markers) and one with zero electron temperature gradient (orange markers). In the second simulation, the total pressure gradient is kept fixed by putting the electron temperature gradient contribution into the ion temperature gradient. 
\begin{enumerate}[label=\textbf{P.\arabic*},resume]
    \item{The growth rate is sensitive to how the pressure gradient is varied - i.e. the mode is sensitive to the partitioning of the pressure gradient into electron and ion contributions.}\label{item6}
\end{enumerate}
If the instability was the MHD-like KBM, the growth rate would be the same and the two curves would be coincident\footnote{Strictly speaking, this is only true if $T_{i} = T_{e}.$ In our simulations $T_{i}/T_{e} = 1.03$ and thus any deviation from MHD-like KBM behaviour might be expected to have more sensitivity to the ions - the opposite of what we observe.} . However, what we observe is that the growth rate is much smaller in the case with zero electron temperature gradient across all scales. 
When understood alongside Figure~\ref{fig:scan} we thus deduce that this mode is indeed much more sensitive to changes in the electron temperature gradient than the ion temperature gradient. This { asymmetry will have an impact on the transport properties associated with the mode, and} suggests the KBM is hybridising with an electron instability (see e.g., \cite{belli2010}). Combining \ref{item4}, \ref{item5} and \ref{item6} suggests that while the mode is KBM-like, it deviates significantly from the pure KBM through kinetic effects.

\begin{figure}
    \centering
    \subfloat[]{\includegraphics[height=0.23\textheight]{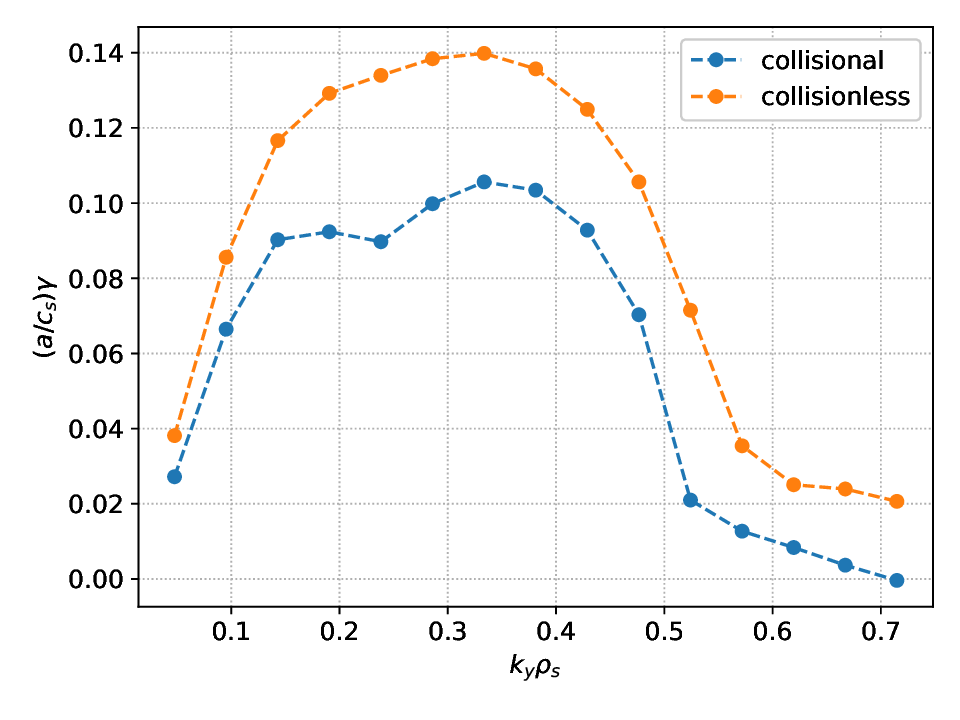}}\quad
    \subfloat[]{\includegraphics[height=0.23\textheight]{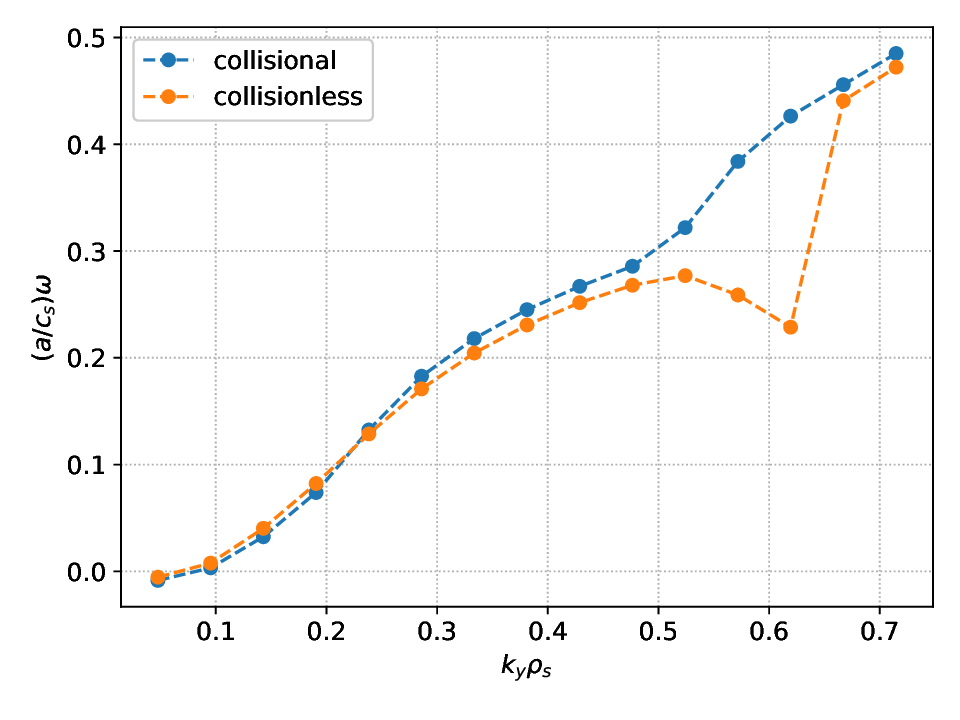}}
    \caption{Growth rate (a) and mode frequency (b) as functions of $k_y\rho_s$ with and without collisions. Results from GS2 simulations at $\Psi_n=0.49$ of STEP-EC-HD.}
    \label{fig:coll}
\end{figure}

{
\subsubsection{Collisions} The impact of collisions on the growth rate of the \textit{hybrid}-KBM is analysed in Figure~\ref{fig:coll} which shows the growth rate and mode frequency from GS2 linear simulations with and without collisions. We note that the growth rate values in the collisional case are lower than in the collisionless case, while the mode frequency is largely unaffected. That is, at the level of collisionality in this local equilibrium, collisions have a weakly stabilising effect on this dominant mode.}

% This result is consistent with our fingerprinting analysis \secref{subsec:fingerprinting} which highlighted the TEM and the ITG modes as potential culprits for hybridising with the KBM. Both the ITG and TEM modes are destabilised by the trapped electron population, and it is therefore not surprising that both are stabilised by collisions since these cause detrapping of the trapped population \cite{belli2010}. The KBM on the other hand, tends to be weakly destabilised by collision \cite{belli2017}. 

\subsubsection{Species}

\begin{figure}
    \centering
    \subfloat[]{\includegraphics[height=0.23\textheight]{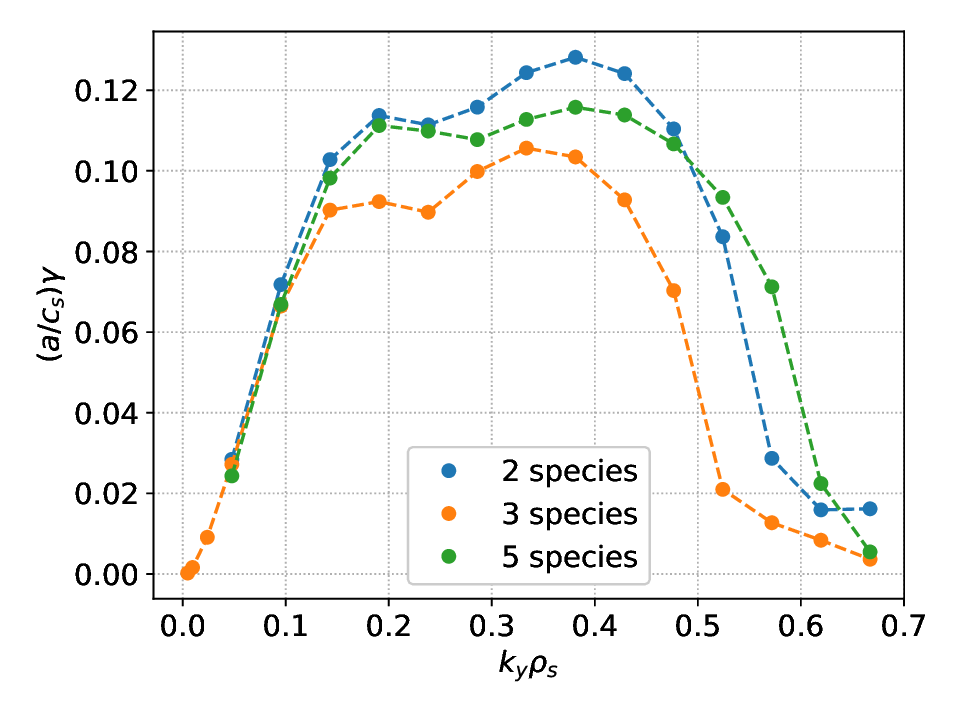}}\quad
    \subfloat[]{\includegraphics[height=0.23\textheight]{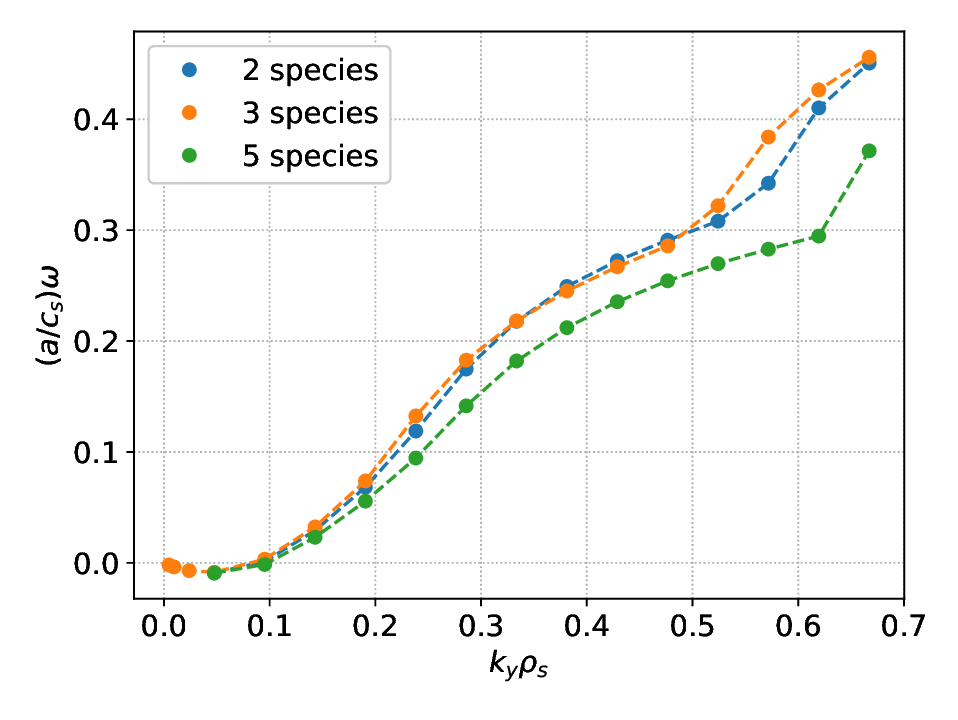}}
    \caption{Growth rate (a) and mode frequency (b) as functions of $k_y \rho_s$ in GS2 linear simulations with two, three and five species at $\Psi_n = 0.49$ of STEP-EC-HD.}
    \label{fig:species}
\end{figure}

{One can also explore the role of the different kinetic species in the simulation. Figure~\ref{fig:species} shows the growth rate and mode frequency from GS2 simulations with two (electron and deuterium), three (electron, deuterium, and tritium) and five species (electron, deuterium, tritium, thermalised helium ash, and a heavy impurity). In each instance, we ensure that the quasineutrality constraint is satisfied by making small adjustments to the value of the electron density gradient. Although there are some small quantitative differences between the three simulations, there is no large change to the linear properties of the \textit{hybrid}-KBM observed when varying the species number\footnote{We note that the linear spectrum is slightly more sensitive to the plasma composition when a different collision model is used, although there is no qualitative change in the main instability. See \ref{app:collision_model} for details.}.}

%This indifference to the plasma ion composition once again points towards the KBM (which are typically sensitive to the total pressure gradient rather than to the contribution of individual species) coupling with an electron instability

{In this paper we primarily consider linear simulations with three species (electrons, deuterium and tritium). The results of Figure~\ref{fig:species} support this choice and have motivated using only two species in the nonlinear calculations reported in \cite{giacomin2023b}.}

\subsubsection{Trapped particles}

\begin{figure}
    \centering
    \subfloat[]{\includegraphics[height=0.23\textheight]{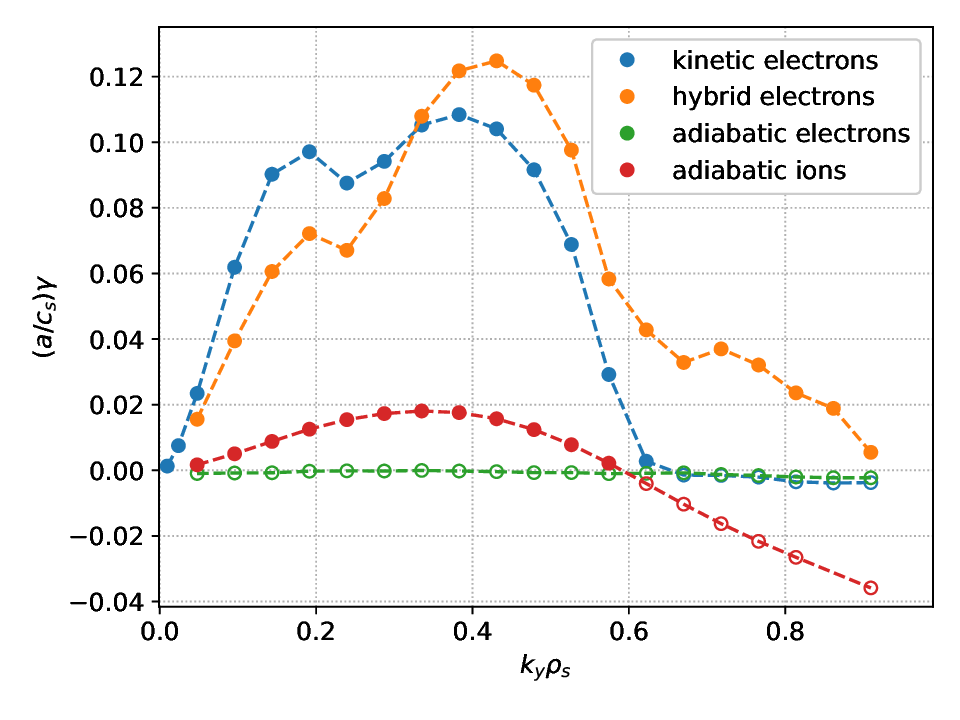}}\quad
    \subfloat[]{\includegraphics[height=0.23\textheight]{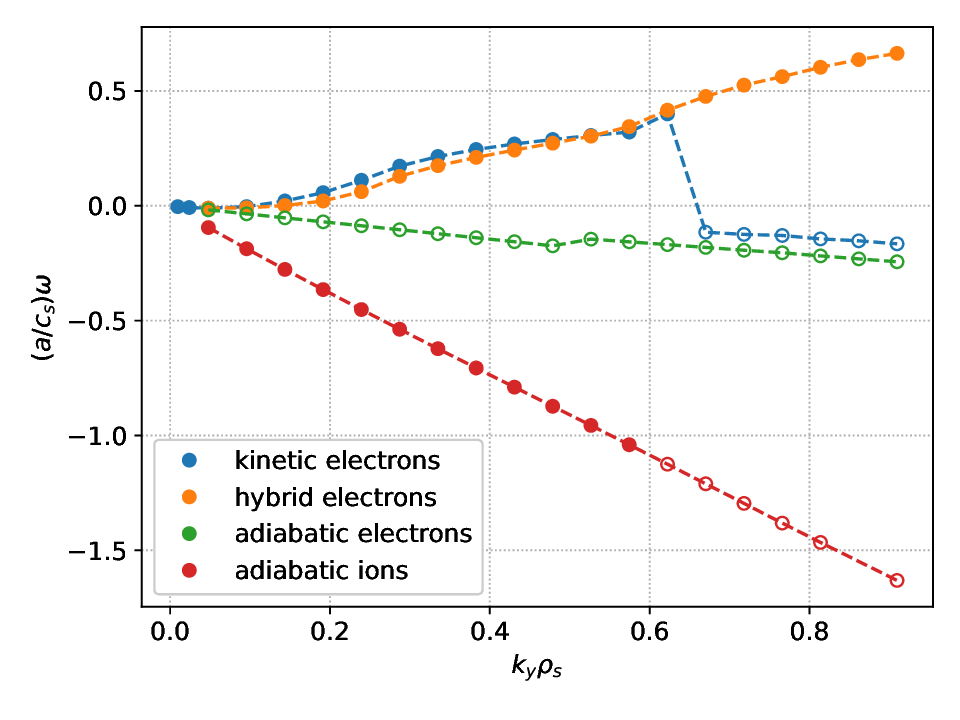}}
    \caption{Growth rate (a) and mode frequency (b) as functions of $k_y$ from GS2 linear simulations with kinetic ions and electrons (blue makers), and hybrid electrons (orange markers), adiabatic electrons (green markers) and adiabatic ions (red markers). Solid and open markers refer to unstable and stable modes, respectively. The simulation with adiabatic electrons is electrostatic and there are no unstable modes.}
    \label{fig:hybrid}
\end{figure}

Thus far, our sensitivity study has shown once again that our dominant mode has many properties in common with the KBM, but also has some non-KBM-like properties. The fingerprinting analysis in \secref{subsec:fingerprinting} suggested that the non-KBM-like properties might be due to coupling to a ITG or TEM (\ref{item4}).
A further investigation of the dominant mode is presented in Figure~\ref{fig:hybrid}, which shows the growth rate and mode frequency as  functions of $k_y\rho_s$ from four GS2 linear simulations: (i) the nominal simulation, (ii) a simulation with hybrid electrons, where the passing electrons are treated adiabatically (i.e., the passing particles have a Maxwellian response to $\delta\phi$ perturbations), while trapped electrons are treated kinetically, (iii) a simulation with adiabatic electrons, and (iv) a simulation with adiabatic ions.   
We note that the hybrid and kinetic electron curves (i.e., (i) and (ii)) follow each other closely, although the hybrid electron curve is not suddenly stabilised at $k_{y}\rho_{s} > 0.6$ (note also there is no sudden change in mode frequency). We thus determine that; 
\begin{enumerate}[label=\textbf{P.\arabic*},resume]
    \item{the dominant instability has a substantial drive from trapped electrons.}\label{item7}
\end{enumerate}
 Furthermore, we note that the simulation with adiabatic electrons is marginally stable at all binormal scales, once again highlighting the importance of kinetic electrons. The ion dynamics also provide an important drive for this instability, this can be seen by noting that the growth rate is strongly reduced in the simulation with adiabatic ions.

\subsection{Labelling the dominant mode}

{The careful analysis presented here has revealed that the dominant instability is KBM-like, but also has properties that suggest this mode is hybridising with other modes. For example, sensitivity scans in $\beta_e$ and $\beta'$ (discussed later in \secref{subsec:beta}) show that this instability can also be tracked to the electrostatic limit where it connects to a ion temperature gradient (ITG) mode. Furthermore Figure~\ref{fig:hybrid} clearly highlights the importance of ion and trapped electron dynamics, indicating hybridisation of the KBM with ITG and TEM drive mechanisms. Henceforth, based on the properties uncovered above, we will choose to refer to this mode as a \textit{hybrid-}KBM.}

% We also note that a similar hybrid TEM/KBM instability has been also observed in \cite{belli2010} in NSTX gyrokinetic simulations. However, sensitivity scans in $\beta_e$ and $\beta'$ (discussed later in \secref{subsec:beta}) show that this instability can also be tracked to the electrostatic limit where it connects to a ion temperature gradient (ITG) mode, and Figure~\ref{fig:hybrid} clearly highlights the importance of ion dynamics. This may imply an additional coupling with the ITG instability (see e.g., \cite{patelthesis}).

{It is important to highlight that ultimately the name \textit{hybrid}-KBM is just a convenient label to refer to the properties stated below.}  
 \begin{itemize}
     \item The mode generally propagates in the ion-diamagnetic direction (Figure~\ref{fig:psin}).
     \item The mode is electromagnetic (\ref{item1}).
     \item The mode eigenfunction is strongly peaked in ballooning space and the mode has {twisting} parity (\ref{item2}).
     \item The mode is driven by the pressure gradient (\ref{item5}) and by $\beta$ (see discussion in \secref{subsec:beta}). 
    \item The mode is unstable in a regime well below the ideal $n=\infty$ MHD limit. 
    \item The mode is sensitive to how the pressure gradient is varied e.g., it varies more strongly with the electron temperature gradient than with the ion temperature gradient (\ref{item6}).
    \item The mode is driven by trapped electrons (\ref{item7}).
    \item The mode couples smoothly to an electrostatic instability (see discussion in \secref{subsec:beta}).
    \item The mode requires access to $\delta B_{\parallel}$ drive in order to be unstable (see discussion in \secref{sec:subdominant MTM instability}). 
\end{itemize}

\section{Stabilising the hybrid KBM}
\label{sec:stabilising the hybrid KBM}
It should be emphasized here that understanding the nature of this mode {(rather than simply naming it)} is of the utmost importance for studying the high performance phase in conceptual ST reactors similar to STEP. Earlier studies for similar high beta conceptual burning ST plasmas \cite{patel2021} found MTMs dominating over several ranges in $k_y \rho_s$, and inferred that MTMs could cause substantial transport. Here, however, for STEP STEP-EC-HD and STEP-EB-HD (see \secref{sec:code_comparison} for STEP-EB-HD results) we find that these \textit{hybrid}-KBMs dominate at all scales across a range of equilibria at various surfaces between the deep core and pedestal top. 
Paper (II) shows  that this \textit{hybrid}-KBM instability is responsible for driving most of the heat and particle transport in the STEP plasmas considered here. It is thus important to understand the nature of this mode and find  strategies to mitigate it. 

\subsection{Sensitivity to $\theta_0$}
\label{subsec:stabilising the hybrid KBM theta}

\begin{figure}
    \centering
    \subfloat[]{\includegraphics[height=0.23\textheight]{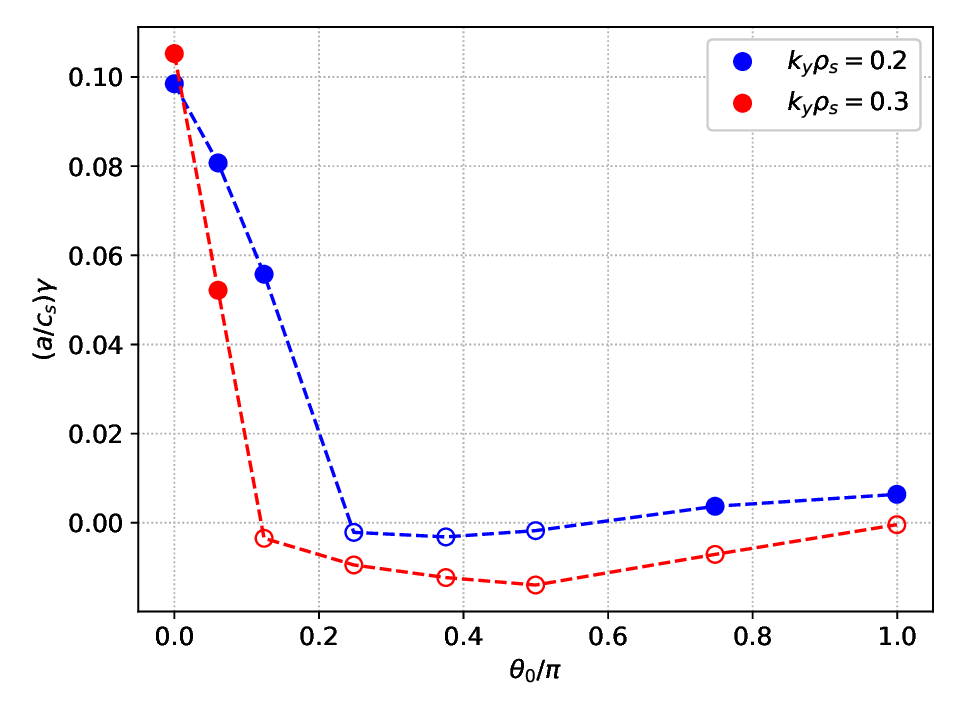}}\quad
    \subfloat[]{\includegraphics[height=0.23\textheight]{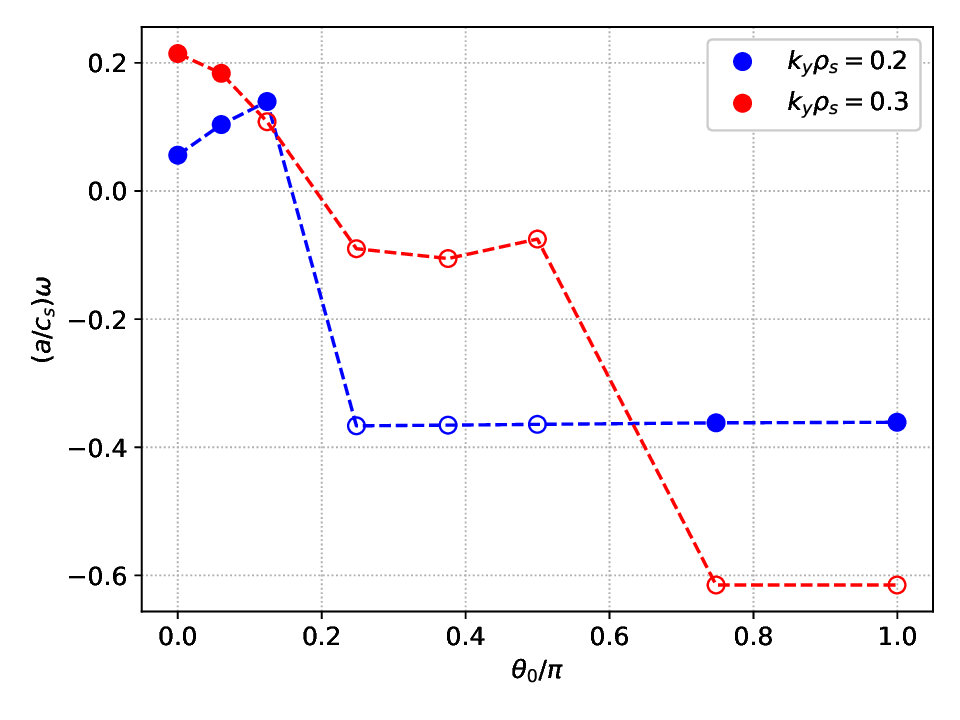}}
    \caption{Growth rate (a) and mode frequency (b) as functions of $\theta_0$ at $k_y\rho_s=0.2$ (blue line) and $k_y\rho_s=0.3$ (red line) for the surface at $\Psi_n = 0.49$ of STEP-EC-HD. Unstable and stable modes are represented by filled and open markers, respectively.}
    \label{fig:theta0}
\end{figure}

Shear $E \times B$ flows are known to play a stabilising role in gyrokinetics; a result which has been established theoretically \cite{terry2000} and borne out experimentally \cite{burrell1999}. Sheared $E \times B$ flows in linear local gyrokinetic simulations can be modelled by introducing a time-dependence into the radial wavenumber $k_{x},$ which corresponds to the ballooning parameter, $\theta_{0}=k_{x,0}/k_{y}\hat{s}$. For a mode in ballooning space at a given $k_y$, the dependence of growth rate on $\theta_0$ is a useful indicator of the mode's susceptibility to flow shear stabilisation \cite{roach2009}. If the mode is stable at some $\theta_{0},$ then when flow shear advects the mode it can be moved into a stabilising region, reducing its effective growth rate. 
Figure~\ref{fig:theta0} shows the growth rate and frequency of the dominant mode at $k_y\rho_s=0.2$ and $k_y\rho_s=0.3$ as functions of $\theta_0$ for the surface at $\Psi_n=0.49$ of STEP-EC-HD. At $k_y\rho_s=0.3$, the growth rate is strongly suppressed as $\theta_0$ increases, with the \textit{hybrid}-KBM instability being stable already at $\theta_0 \geq \pi/8$. The growth rate at $k_y\rho_s=0.2$ decreases as $\theta_0$ increases from 0 to $\pi/4$. At $\theta_0 \simeq \pi/4$, the mode is stable and remains close to the marginal stability until $\theta_0 \simeq 3\pi/4$. 
At $k_y \rho_s=0.2$, a different instability propagating in the electron drift direction (actually the previously subdominant MTM which has no such simple dependence on $\theta_0$ \cite{patel2023a}) appears at $\theta_0 \simeq \pi$.

The high sensitivity of the \textit{hybrid}-KBM instability to $\theta_0$ suggests a possible important effect of flow shear, a relationship which is explored further in Paper (II). We note that a strong dependence of a KBM-like instability on $\theta_0$ was also observed in a similar STEP conceptual design~\cite{patel2021}, where it was noted that in the local equilibrium studied flow shear effects may largely suppress transport from KBMs\footnote{It is important to remark that the stiffness of the pure KBM counters this argument by suggesting that even a small increase in drive could compensate for any stabilisation. It is also worth mentioning that in the pedestal of conventional aspect ratio tokamaks, it has been noted that owing to the stiffness of KBM transport, the KBM may still play a role in limiting gradients close to the critical value even when the mode is marginally stable \cite{Kotschenreuther_2019}}. 

{All simulations in this paper in this paper are performed at $\theta_0 = 0$ unless explicitly stated otherwise.} 

\subsection{Sensitivity to $\beta_{e}$ and $\beta^{\prime}$}
\label{subsec:beta}

Motivated by the {electromagnetic nature} of the dominant instability, we study the sensitivity of the mode with respect to $\beta_{e}.$ Since the mode has many features in common with the KBM, we would expect the mode to be stable below some finite value of $\beta_e,$ and for the growth rate to scale with $\beta_e$ above this threshold. 

\subsubsection{Varying $\beta_e$ with $\beta^\prime$ fixed}

\begin{figure}
    \centering
    \subfloat[]{\includegraphics[height=0.23\textheight]{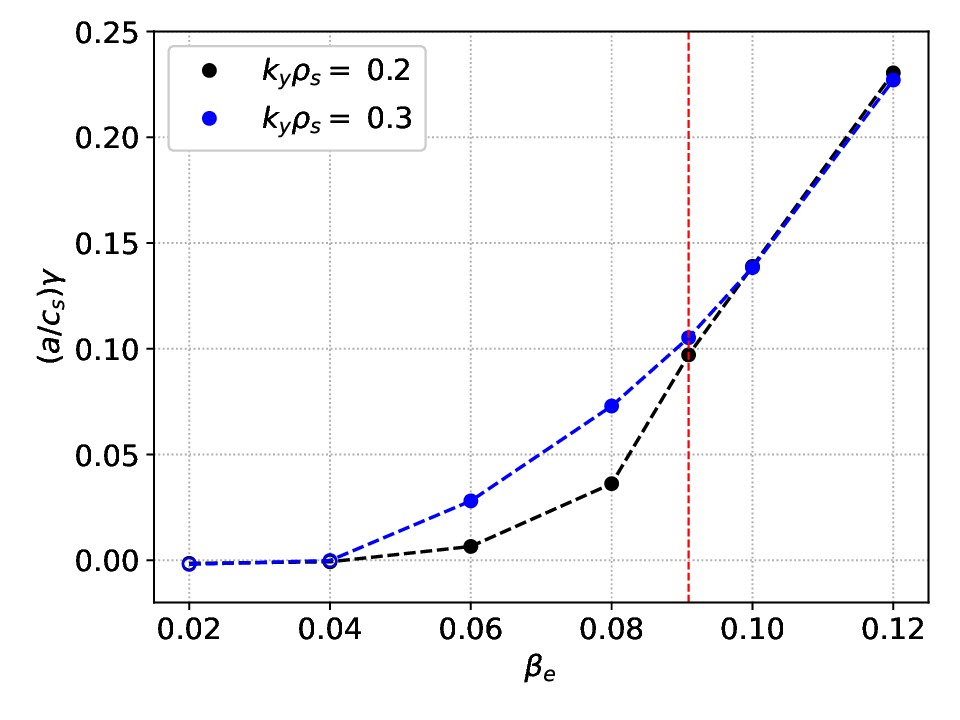}}\quad
    \subfloat[]{\includegraphics[height=0.23\textheight]{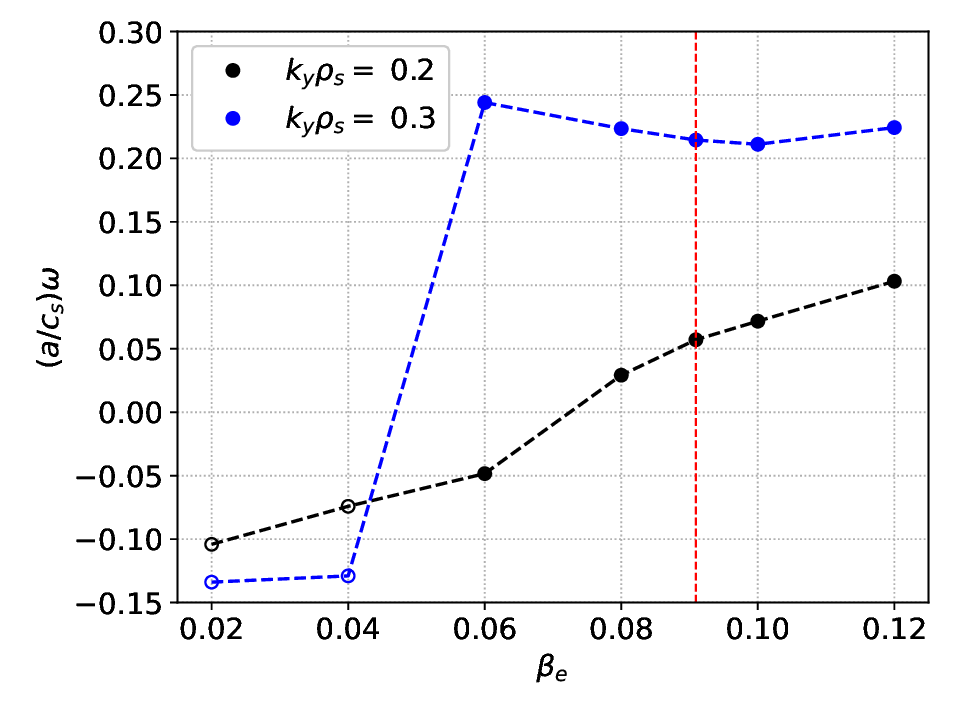}}
    \caption{Growth rate (a) and mode frequency (b) from GS2 linear simulations as functions of $\beta_e$ at $k_y\rho_s=0.2$ (black line) and $k_y\rho_s=0.3$ (blue line). The red dashed line represents the reference value. The value of $\beta'$ is kept constant while varying $\beta_e$. Filled and open markers refer to unstable and stable modes, respectively. Results at $\Psi_n=0.49$ of STEP-EC-HD.}
    \label{fig:beta}
\end{figure}

In Figure~\ref{fig:beta}, we study the impact of varying $\beta_e$ whilst the other parameters (notably $\beta^\prime$) are held fixed (the nominal value is denoted by a red vertical line). We note that in this case the mode is stabilised as $\beta_e$ is dropped (and the growth rate increases when $\beta_e$ is increased) which indicates that the dominant mode is accessing the electromagnetic component of the drive terms (the electromagnetic component of the drive is reduced at lower $\beta$, while the stabilising effect of $\beta'$ is retained). 

Later, in \secref{sec:subdominant MTM instability}, we will see that the \textit{hybrid}-KBM necessitates the inclusion of parallel magnetic fluctuations $\delta B_\parallel$ in order to access the electromagnetic drive. This result is in line with earlier works \cite{zocco2015,belli2010} that find parallel magnetic fluctuations act to destabilise the KBM  and an absence of $\delta B_\parallel$ effects lead to a decrease of the KBM growth rate up to a factor of 6 \cite{aleynikova2018}. However, this is even more severe in simulation of the \textit{hybrid}-KBM, which we find to be everywhere stable when $\delta B_\parallel$ is neglected.

\subsubsection{Varying $\beta_e$ and $\beta^\prime$ consistently}

\begin{figure}
    \centering
    \subfloat[]{\includegraphics[height=0.23\textheight]{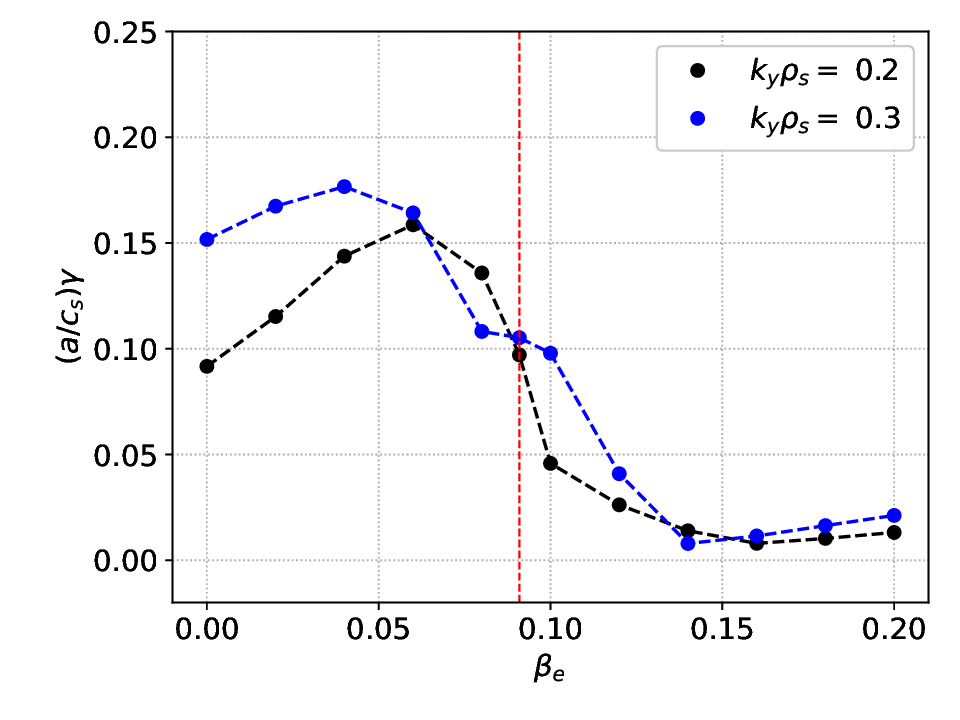}}\quad
    \subfloat[]{\includegraphics[height=0.23\textheight]{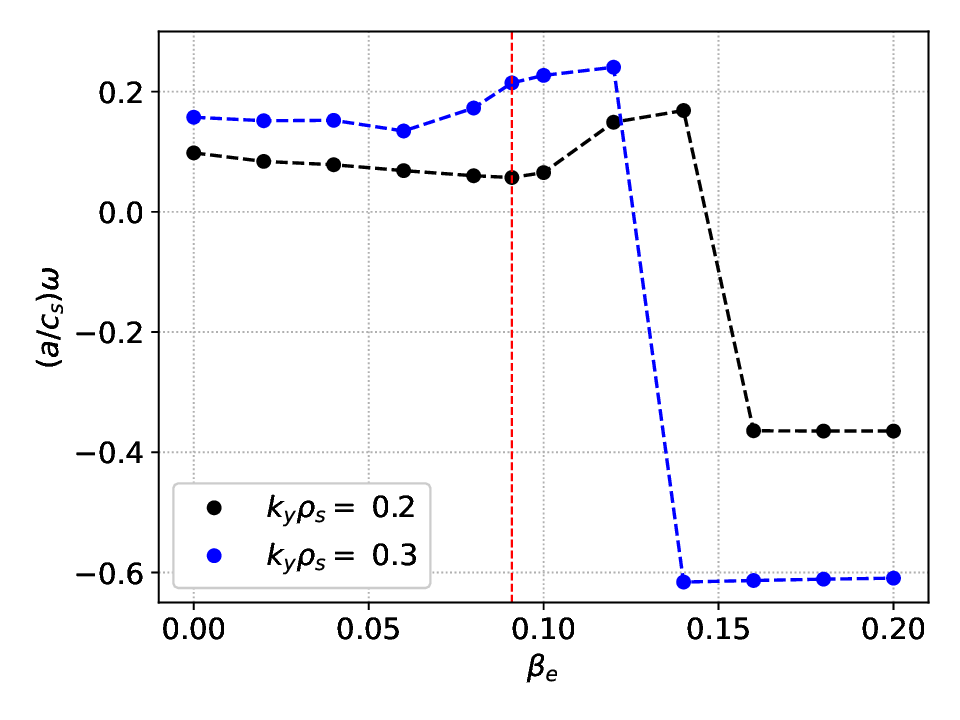}}
    \caption{Growth rate (a) and mode frequency (b) from GS2 linear simulations as functions of $\beta_e$ at $k_y\rho_s=0.2$ (black line) and $k_y\rho_s=0.3$ (blue line). The red dashed line represents the reference value. The value of $\beta'$ is consistently varied with $\beta_e$. Results at $\Psi_n=0.49$ of STEP-EC-HD.}
    \label{fig:betap}
\end{figure}

In Figure~\ref{fig:betap} we study the effect of varying $\beta_e$ whilst also varying $\beta^\prime$ consistently with the local equilibrium, following an approach similar to that outlined in \cite{dickinson2022}. Figure~\ref{fig:betap} reveals that this \textit{hybrid}-KBM mode remains unstable even in electrostatic limit ($\beta_e = 0, \, \beta^\prime = 0$). The smooth variation of the growth rate (and the real frequency) with $\beta_e$ indicates that the mode is coupling to some electrostatic instability which prevents stabilisation. This is another feature of the \textit{hybrid}-KBM which markedly distinguishes it from the simple KBM (which would be stable in an electrostatic simulation). We remark that the behaviour seen here is consistent with coupled KBM-ITG, and similar
behaviour has been seen in both theory \cite{dickinson2022} and experiment \cite{Bowman_2018}. 

Figure~\ref{fig:beta} also shows that the growth rate is reduced at higher $\beta_e$ due to $\beta'$ stabilisation, noting that a mode transition occurs as $\beta_e$ passes through $\beta_e \sim 0.13$ (note the abrupt change of sign of frequency and the further reduction of the growth rate). At these values of $\beta_e$ and $\beta'$, the \textit{hybrid}-KBM is fully stabilised, revealing the underlying subdominant MTM instability (see \secref{sec:subdominant MTM instability}). {As discussed in Section 5.1, \textit{hybrid}-KBMs are strongly ballooning with growth rates that peak strongly at $\theta_0=0$, while MTM growth rates depend very differently on $\theta_0$, with a typically weak dependence at low $k_y$ \cite{patel2021}.  For the reference equilibrium the \textit{hybrid}-KBM growth rate is much more unstable at $\theta_0=0$ than the highest MTM growth rate at any $\theta_0$.} Figure~\ref{fig:beta} indicates that increasing $\beta_e$ together with $\beta^\prime$ results in stabilisation of the dominant \textit{hybrid}-KBM.

\begin{figure}
    \centering
    \subfloat[]{\includegraphics[height=0.23\textheight]{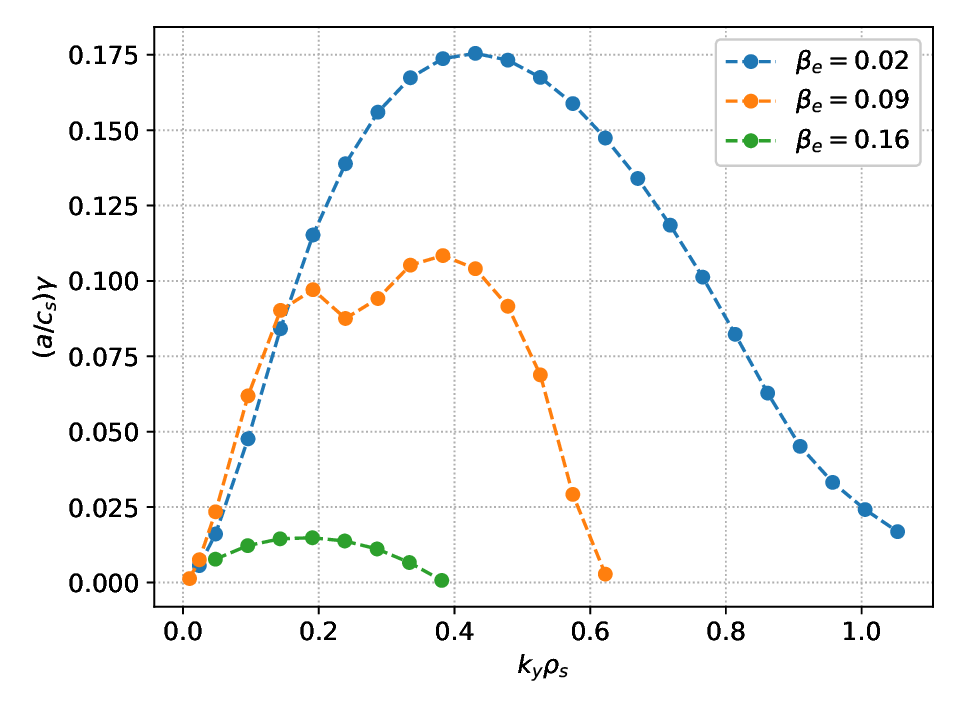}}\quad
    \subfloat[]{\includegraphics[height=0.23\textheight]{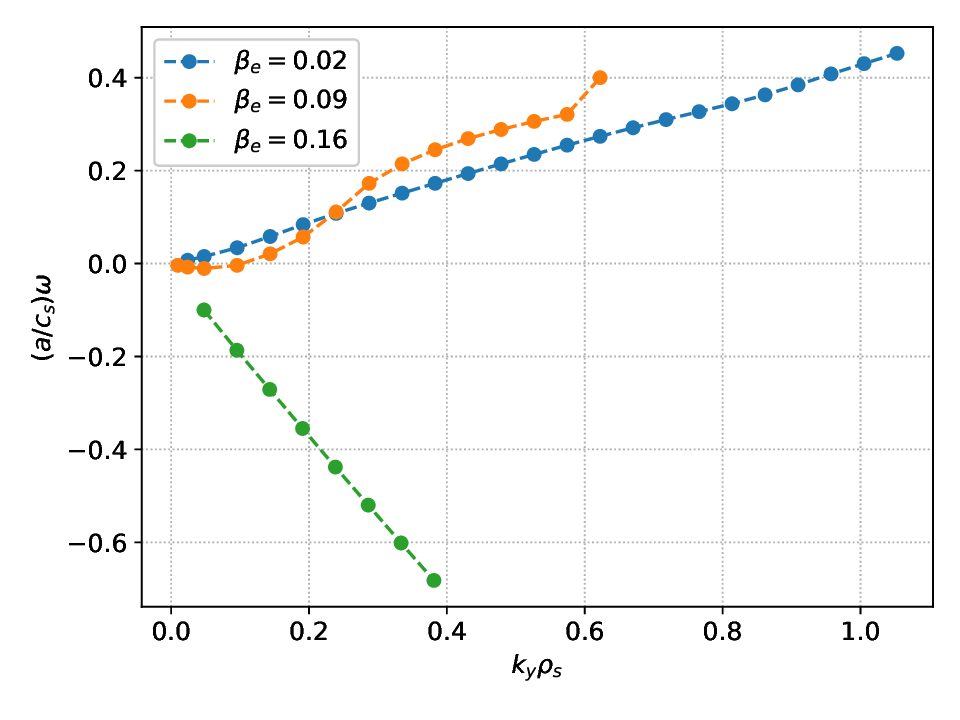}}
    \caption{Growth rate (a) and mode frequency (b) as functions of $k_y$ from GS2 linear simulations with lower (blue markers), nominal (orange markers) and higher (green markers) $\beta_e$ values. The $\beta'$ value is varied consistently. Results at $\Psi_n=0.49$ of STEP-EC-HD.}
    \label{fig:beta_full}
\end{figure}

As a complement, Figure~\ref{fig:beta_full} shows the linear growth rate spectrum for three different values of $\beta_e$ with consistently varied $\beta^\prime$ - the nominal case (orange markers) and a lower and higher $\beta_e$ case (blue and green markers). In the case with higher $\beta_e$, the \textit{hybrid}-KBM instability vanishes and the MTM instability (see \secref{sec:subdominant MTM instability}) becomes the most unstable mode in the system. In the case of lower $\beta_e$, the ITG instability drives an unstable mode with a higher growth rate than the most unstable hybrid mode at the nominal $\beta$ value.

\subsubsection{Implications for the current ramp}

We remark here that, as mentioned in \secref{sec:equilibria}, one of the major challenges for STEP is the need to generate the required plasma current of $I_{p} \simeq 20$ MA. Although Figures~\ref{fig:betap} and \ref{fig:beta_full} demonstrate that a high beta regime free of the hybrid-mode exists, it is less clear how the hybrid mode could be avoided on the approach to such a flat-top during the $I_p$ ramp. During the current ramp, the plasma equilibria will evolve continuously from a $\beta_e = 0$ state, where it will be dominated by electrostatic instabilities, up to the reference $\beta_e$ where it is dominated by the \textit{hybrid}-KBM. However, we have seen in Figures~\ref{fig:betap} and \ref{fig:beta_full} that the \textit{hybrid}-KBM becomes active at much smaller $\beta_e$ than that which we are aiming to achieve in the flat top. Thus, in getting to this equilibria one must first pass through a region where the \textit{hybrid}-KBM is active, and this could shut down the evolution of the plasma e.g., if the turbulent transport were too large to sustain the profiles (see Paper (II) for further discussion). It is currently not clear whether it's possible to avoid the onset of the \textit{hybrid}-KBM completely or how much heating and fueling would be required burn through it should it appear earlier in the current ramp. It is also worth remarking that the $\beta_e = 0.16$ case will probably exceed the resistive wall mode control limit and thus is likely not viable for other reasons. 

\subsubsection{Safety factor and magnetic shear}

\begin{figure}
    \centering
    \subfloat[]{\includegraphics[height=0.23\textheight]{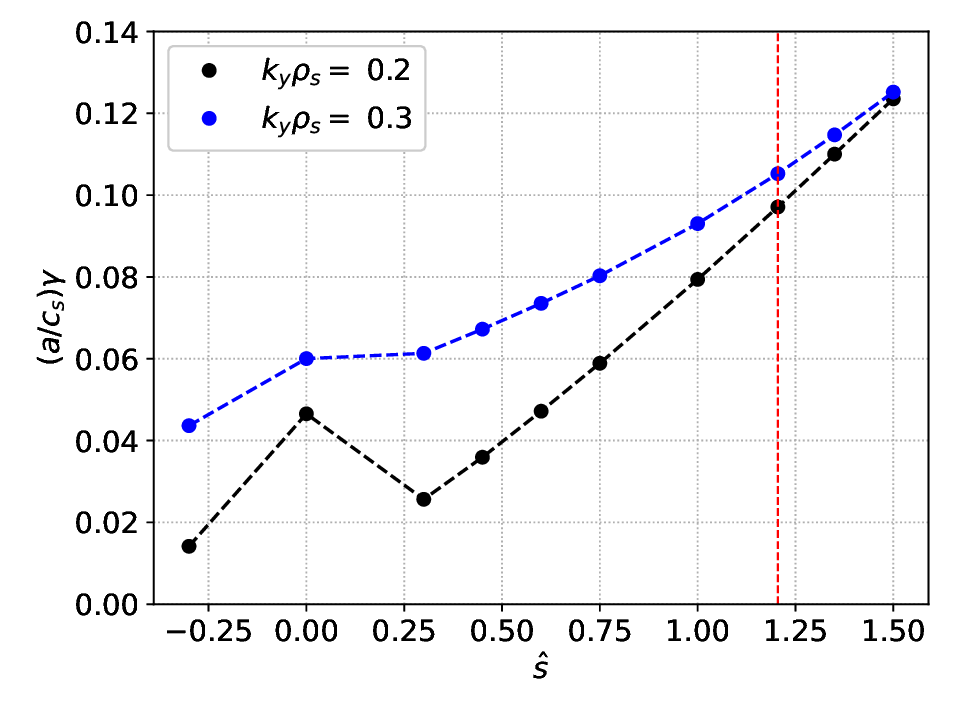}}\quad
    \subfloat[]{\includegraphics[height=0.23\textheight]{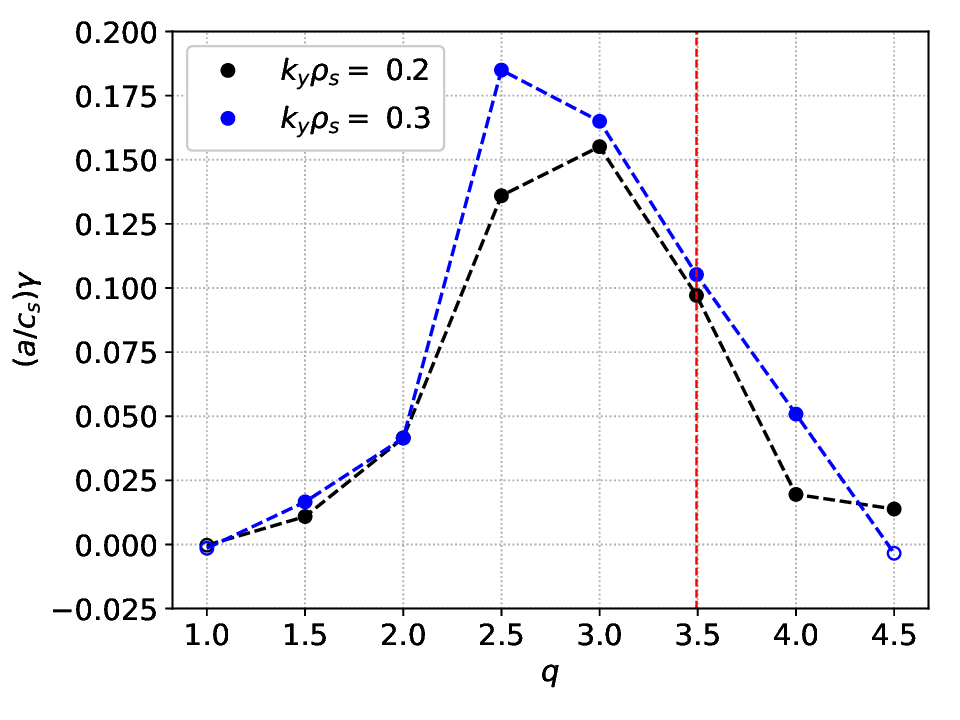}}
    \caption{Growth rate from GS2 linear simulations as a function of $\hat{s}$ (a) and $q$ (b) at $k_y\rho_s=0.2$ (black line) and $k_y\rho_s=0.3$ (blue line). The red dashed line represents the reference value. Results at $\Psi_n=0.49$ of STEP-EC-HD.}
    \label{fig:scan_sq}
\end{figure}

{The \textit{hybrid}-KBM also shows some sensitivity to changes to the local equilibrium parameters such as the magnetic $\hat{s}$ and the safety factor $q,$ (see Figure~\ref{fig:scan_sq}). Studying the sensitivity of the \textit{hybrid}-KBM to $\hat{s}$ and $q$ reveals that, although we have placed great emphasis on the properties of this mode that distinguish it from the IBM, some of the physical intuition we can develop from the ideal theory is still useful for these \textit{hybrid}-KBMs. Panel (a) of Figure~\ref{fig:scan_sq} shows how the dominant mode is destabilised by increasing $\hat{s},$ consistent with the ideal ballooning mode behaviour (i.e. a KBM-like behaviour). Interestingly, inspection of the mode frequency in this scan reveals an isolated mode transition occurring only at $\hat{s}=0$, to a mode that propagates in the electron diamagnetic direction. The dependence of the mode on $q$ (Panel (b) of Figure~\ref{fig:scan_sq}) is slightly more complicated but is also consistent with the behaviour of the ideal ballooning mode. As $q$ increases the stability boundary of the ideal ballooning mode moves towards higher $\hat{s}$ and lower $\beta^\prime$. Therefore, increasing $q$ will make access to the second stability region easier (which is why we see the mode stabilised with increasing $q$ from the reference value). At sufficiently low $q,$ the stability boundary gets pushed to higher $\beta^\prime$ enabling the equilibrium to lie in the first stability region (see \cite{davies2022} for a more careful discussion of first and second stability) which is consistent with the stabilisation of the KBM-like dominant mode. We note that the global MHD equilibrium is not varied consistently in these cases.} 

{These results in Section~\secref{sec:stabilising the hybrid KBM} motivate  the need to analyse the subdominant instability.}

\section{Subdominant MTM instability}
\label{sec:subdominant MTM instability}

We now turn our attention to the subdominant MTM instability, which may play an important transport role, especially if the hybrid mode is effectively stabilised by flow shear.
MTMs generate magnetic islands on rational surfaces that tear the confining flux surfaces and generate heat transport primarily through the electron channel~\cite{guttenfelder2011,giacomin2023a}. We note that local GK simulations have revealed MTMs as the dominant microinstabilities in the wavenumber range $k_y \rho_s<1$  locally at mid-radius (where $\beta_e \sim 5\%-10\%)$ in several  spherical tokamak plasmas (see \cite{kaye2021} and references therein). 

\begin{figure}
    \centering
    \subfloat[]{\includegraphics[height=0.23\textheight]{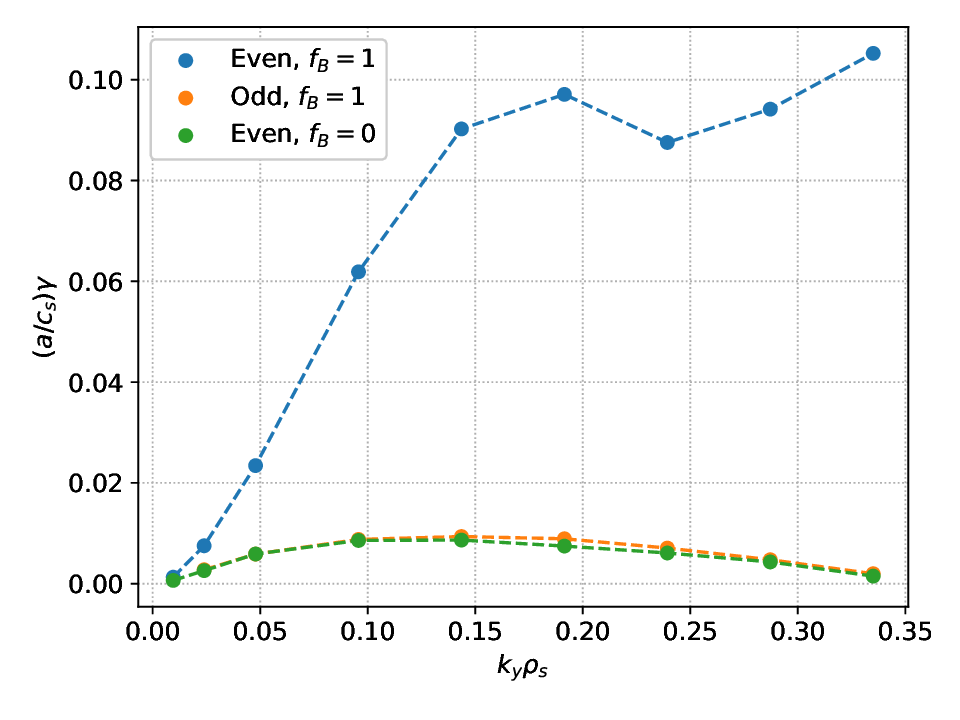}}\quad
    \subfloat[]{\includegraphics[height=0.23\textheight]{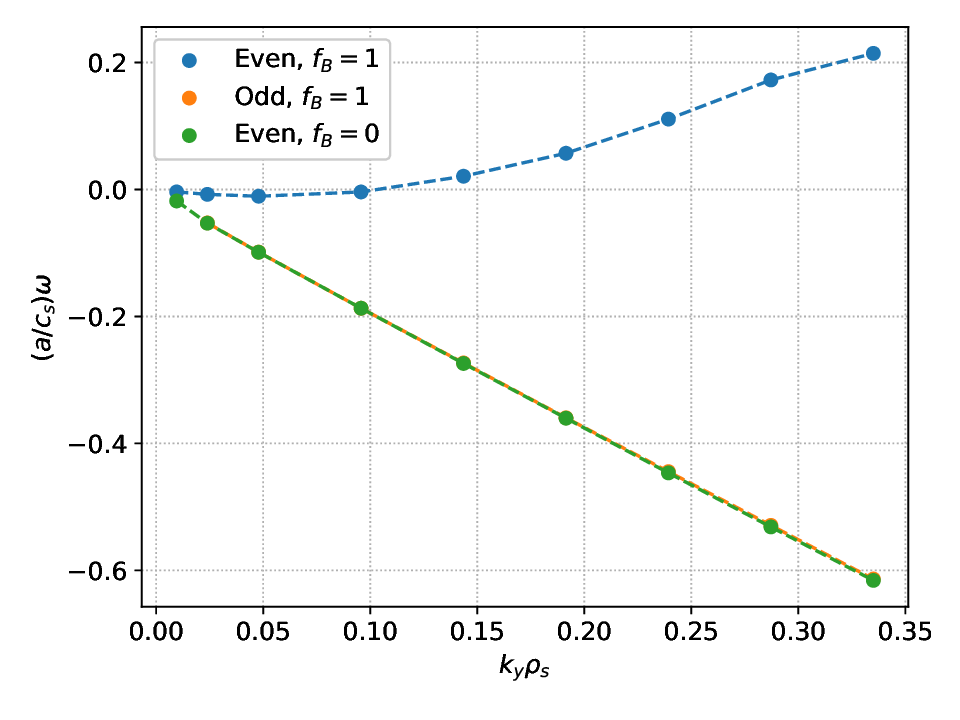}}
    \caption{Growth rate (a) and mode frequency (b) as functions of $k_y \rho_s$ at $\Psi_n = 0.49$ of STEP-EC-HD from GS2 linear simulations. The blue line refers to the nominal simulation with $\delta B_\parallel$ fluctuations ($f_B = 1$), the orange line to the simulation with tearing parity enforced in the distribution function and the green line to the simulation without $\delta B_\parallel$ fluctuations ($f_B = 0$). Only unstable modes are shown.}
    \label{fig:sub}
\end{figure}

GK studies for the high performance phase in conceptual ST reactors have also found MTMs dominant over an extended range of binormal scales and likely to have significant impacts on transport \cite{patel2021,dickinson2022}. 
The presence of the fastest growing tearing mode can be investigated by enforcing the tearing (odd) parity of the perturbed distribution function, exploiting the up-down symmetry of the Miller equilibrium. 
This test is carried out with GS2 at $\theta_0=0$ and the results are shown in Figure~\ref{fig:sub} (see orange curve). 
We thus see that there are in fact unstable modes with tearing parity (e.g., MTMs), but on this surface in STEP-EC-HD these are always subdominant to the \textit{hybrid}-KBM. 

Another way to obtain the MTM from an initial value solver as the fastest growing mode in our system specifically, without forcing the parity of the eigenmode, is to simply switch off compressive magnetic perturbations i.e., we exclude the $\delta B_{\parallel}$ contribution to the GK equation. Figure~\ref{fig:sub} shows that the simulation neglecting $\delta B_\parallel$ (green) is equivalent to the nominal simulation with a tearing parity initial distribution function (orange). Essentially, removing $\delta B\p$ fluctuations from the system stabilises the \textit{hybrid}-KBM, whilst having no impact on the MTM and thus leaving the MTM as the dominant mode.

\begin{figure}
    \centering
    \subfloat[]{\includegraphics[height=0.23\textheight]{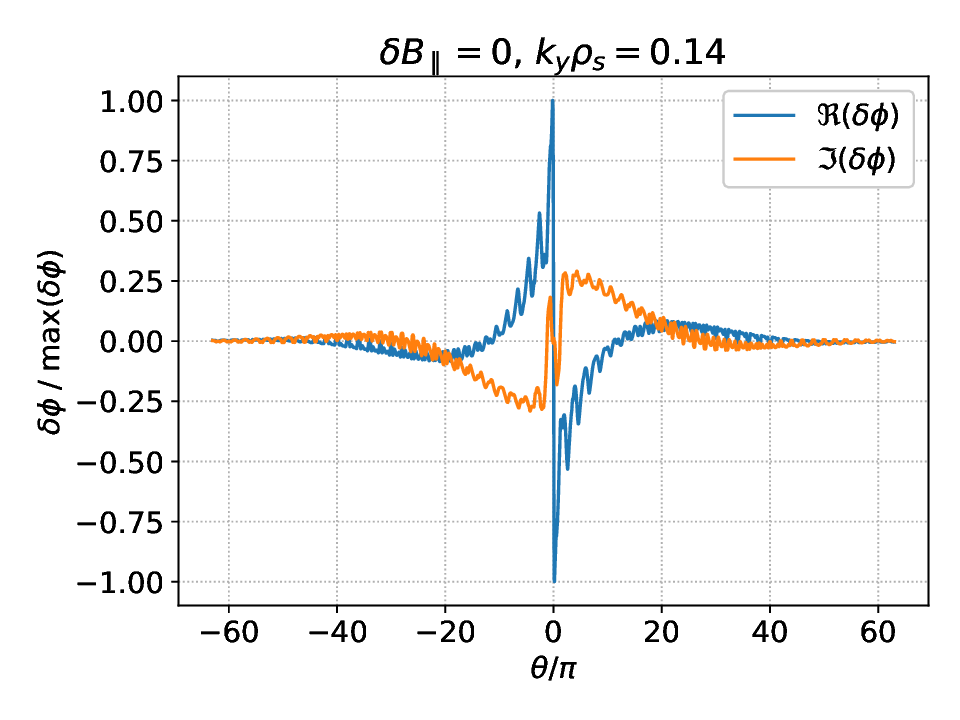}}\quad
    \subfloat[]{\includegraphics[height=0.23\textheight]{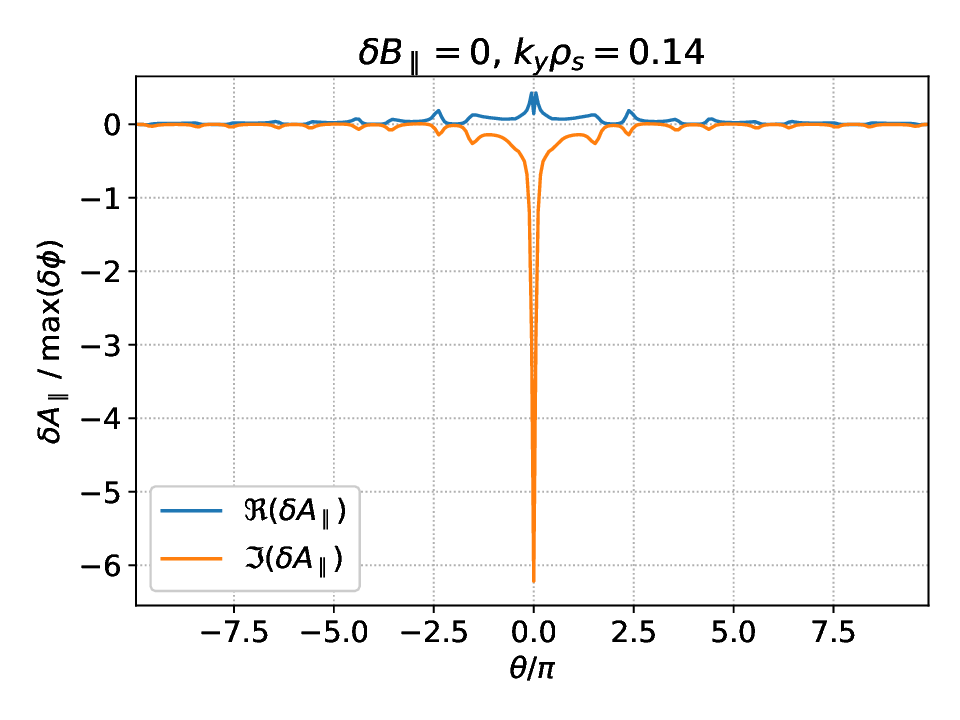}}
    \caption{Real and imaginary part of $\delta \phi/\max(\delta\phi)$ (a) and $\delta A_\parallel/\max(\delta\phi)$ (b) at $k_y\rho_s=0.14$ as functions of the ballooning angle $\theta$ from the GS2 linear simulation with $\delta B_\parallel=0$. In (b) the $\theta$ range is restricted to $[-10\pi, 10\pi]$ for ease of reading.}
    \label{fig:eig_sub}
\end{figure}

The eigenfunctions of $\phi$ and $A_\parallel$ corresponding to the MTM at $k_{y}\rho_{s} = 0.14$ are shown in Figure~\ref{fig:eig_sub}. We find that the $A_{\parallel}$ fluctuation is significantly larger than the electrostatic fluctuations close to the inboard midplane, as expected for MTMs. The electrostatic potential eigenfunction exhibits a clear multiscale structure (ion-scale in $k_y,$ electron scale in $k_x$), with a narrow oscillatory structure in $\theta$ overlaying a much broader oscillation. The $A_{\parallel}$ function is more strongly peaked about $\theta = 0$, with subsequent peaks occurring along the field line at $\theta\, \mathrm{mod}\, 2\pi = 0,$ the outboard midplane. Similar MTM eigenfunctions extended in ballooning angle have been seen in simulations of MAST \cite{applegate2007} and NSTX \cite{guttenfelder2011} discharges and BurST \cite{patel2021}. The extended nature of these modes, requiring a parallel domain $\theta \in [-70\pi, 70\pi],$ coupled with a very small growth rate, means that even linearly resolving the subdominant MTM can  become very computationally expensive. 

\begin{figure}
    \centering
    \subfloat[]{\includegraphics[height=0.23\textheight]{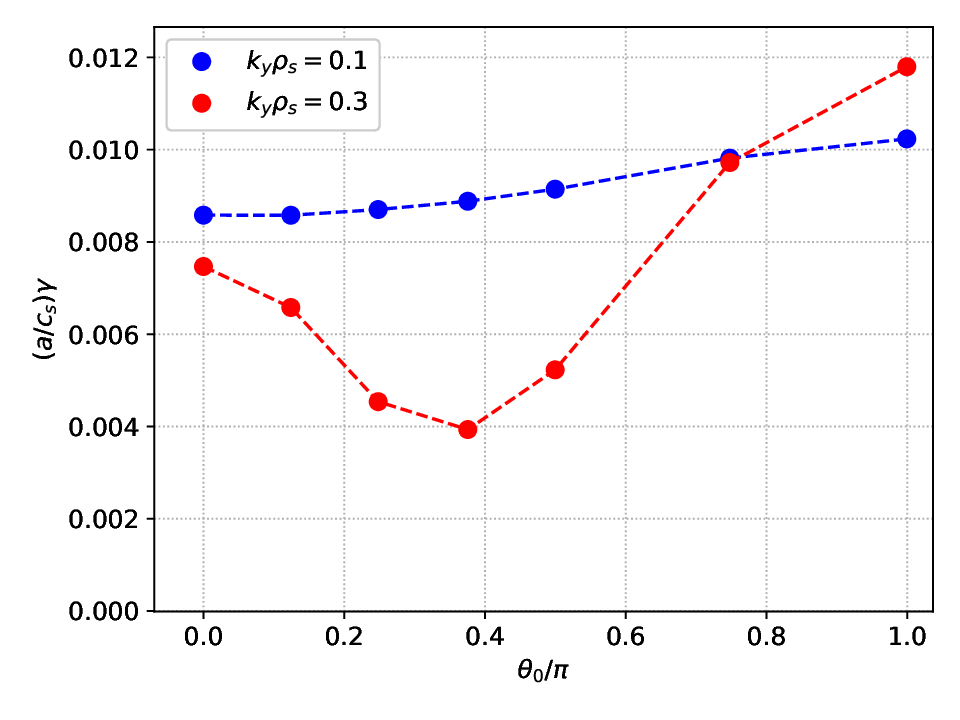}}\quad
    \subfloat[]{\includegraphics[height=0.23\textheight]{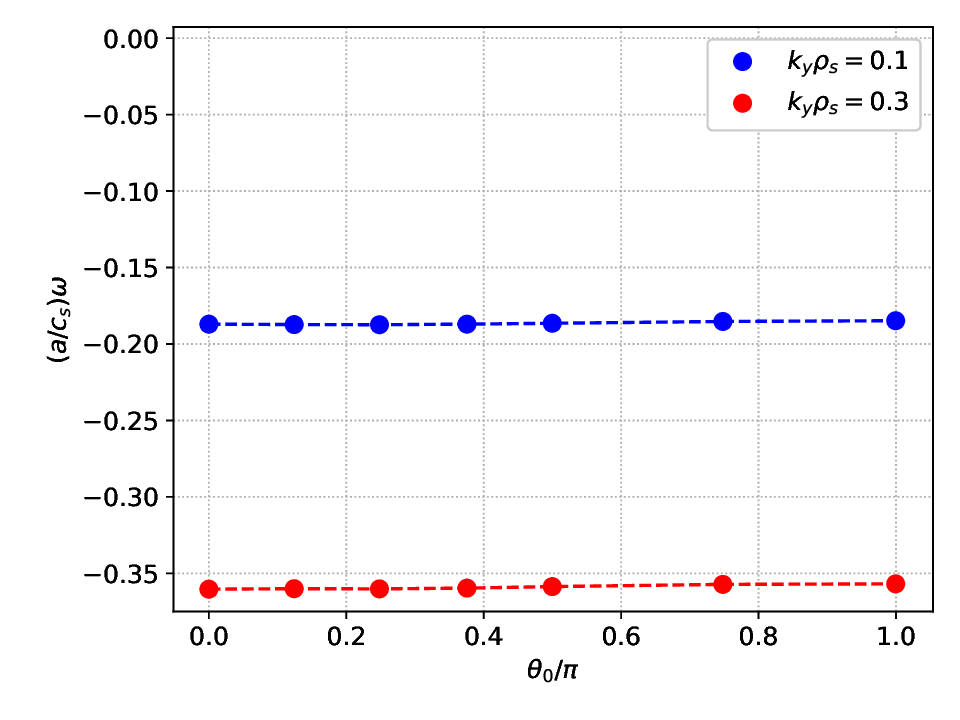}}
    \caption{Growth rate (a) and mode frequency (b) as functions of $\theta_0$ at $k_y\rho_s=0.1$ (blue line) and $k_y\rho_s=0.3$ (red line) from GS2 linear simulations with $\delta B_\parallel=0$.}
    \label{fig:theta0_sub}
\end{figure}

Nonlinear simulations involving the MTMs in Figure~\ref{fig:eig_sub} will be computationally challenging in these STEP plasmas, owing to the intrinsic multiscale character of the MTM in the radial direction (which is linked to its multiscale character in $\theta$ in ballooning space). Figure \ref{fig:theta0_sub} illustrates how the MTM growth rates (for modes at $k_y\rho_s=0.1$ and $k_y\rho_s=0.3$) depend on $\theta_0,$ showing that the MTM growth rate (particularly at $k_y\rho_s=0.1$) is much less sensitive than the \textit{hybrid}-KBM growth rate (see Figure~\ref{fig:theta0} of \secref{subsec:stabilising the hybrid KBM theta}) this therefore suggests that these MTMs should be much less susceptible than \textit{hybrid}-KBM modes to flow shear stabilisation.  We note that tokamak regimes exist where turbulent transport from MTMs is affected by flow shear stabilisation \cite{guttenfelder2011}.  Recent theoretical work has identified an important local equilibrium parameter that helps explain this \cite{Hardman2022}, and the relevance of this parameter in experiments and numerical simulations is explored in \cite{patel2023a}.  The insights gained here may be helpful in the future optimisation of STEP design points.   

\section{Code comparison on surfaces in STEP-EC-HD and STEP-EB-HD}
\label{sec:code_comparison}

Careful benchmarking is essential for ensuring the fidelity of GK simulations in next-generation reactor design, and to identify (and ideally rectify) issues that may arise in simulations using any single code (see e.g., the discussion of the numerical instability in paper II). Furthermore, this benchmarking also paves the way for the detailed nonlinear investigation of the companion article. In this section, we compare the results of CGYRO, GENE and GS2 linear GK simulations carried out at the radial surfaces corresponding to $q=3.0$ (STEP-EC-HD and STEP-EB-HD) and $q=3.5$ (STEP-EC-HD only) equilibria. As previously, we compare simulations and results (linear eigenvalues and eigenmodes) for both the \textit{hybrid}-KBM instability and the subdominant MTM instability. As before, the numerical resolutions used in these simulations are listed in Table~\ref{tab:resolution_spr45}, where again we resort to different resolutions for simulations of the \textit{hybrid}-KBM instability and simulations of the subdominant MTM instability. These simulations evolve three species (electron, deuterium and tritium) and include both perpendicular and parallel magnetic fluctuations, $\delta A_\parallel$ and $\delta B_\parallel$ for simulations of the \textit{hybrid}-KBM whilst including only $\delta \phi$ and $\delta A\p$ for MTM simulations. {In each code, we try to use the most advanced physics model available whilst also ensuring results are comparable by
adopting as similar approaches as possible.   We have therefore used the Sugama collision model \cite{sugama2009} in both CGYRO and GENE.\footnote{A more advanced exact Landau collision operator \cite{pan2020} available in GENE has not been used here.} The linearized Fokker-Planck collision model of \cite{barnes2009} is used in GS2.}

\begin{figure}
    \centering
    \subfloat[]{\includegraphics[height=0.23\textheight]{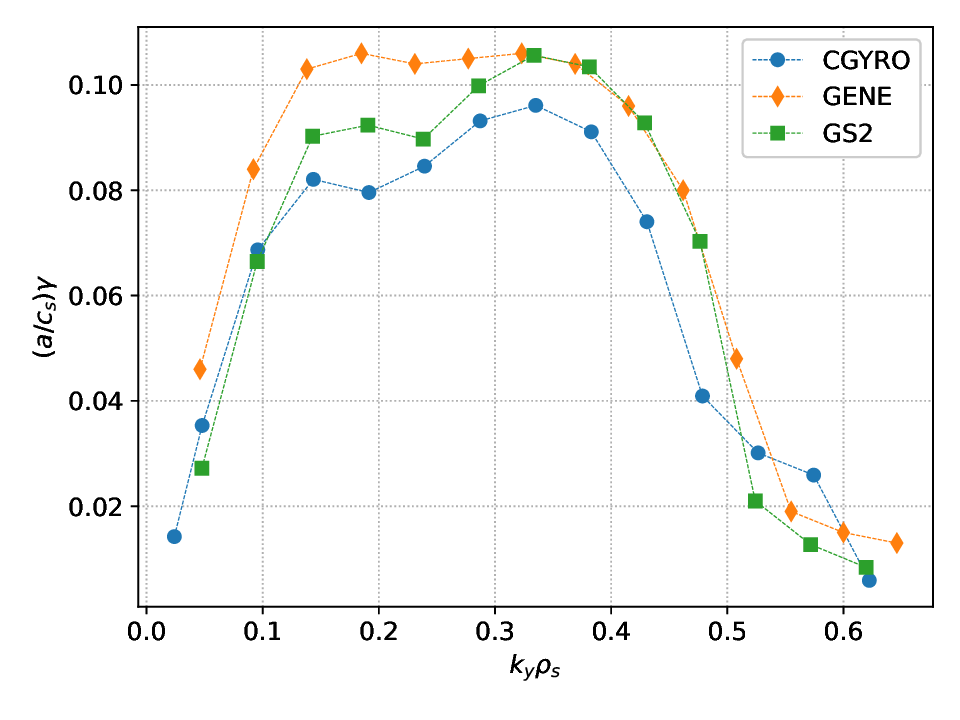}}\quad
    \subfloat[]{\includegraphics[height=0.23\textheight]{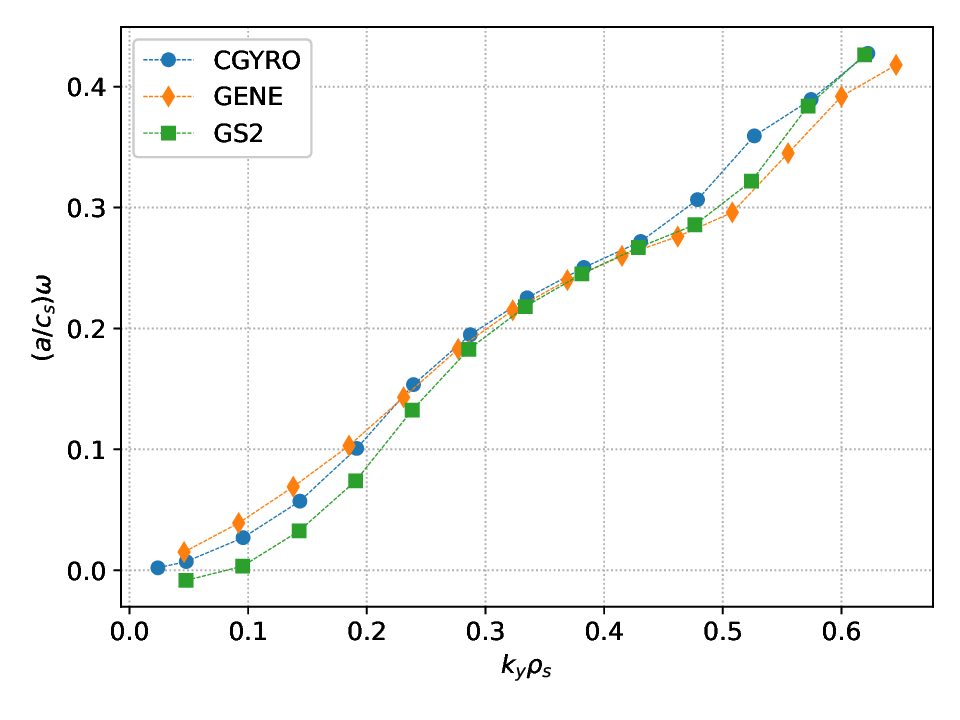}}
    \caption{Growth rate (a) and mode frequency (b) as functions of $k_y$ for the surface at $\Psi_n = 0.49$ ($q=3.5$) of STEP-EC-HD. Results from  CGYRO (blue line), GENE (orange line) and GS2 (green line) simulations with $\delta B_\parallel \ne 0$ (hybrid-KBM instability).}
    \label{fig:comp}
\end{figure}

Figure~\ref{fig:comp} compares the growth rate and the mode frequency at $\Psi_n=0.49$ of STEP-EC-HD, and a reasonable agreement is found between all three codes. We note that it is of no great surprise that there is some variation between the growth rates since; e.g., each code employs differing discretisations of the 5D space and schemes for parallel dissipation. 

\begin{figure}
    \centering
    \subfloat[]{\includegraphics[height=0.23\textheight]{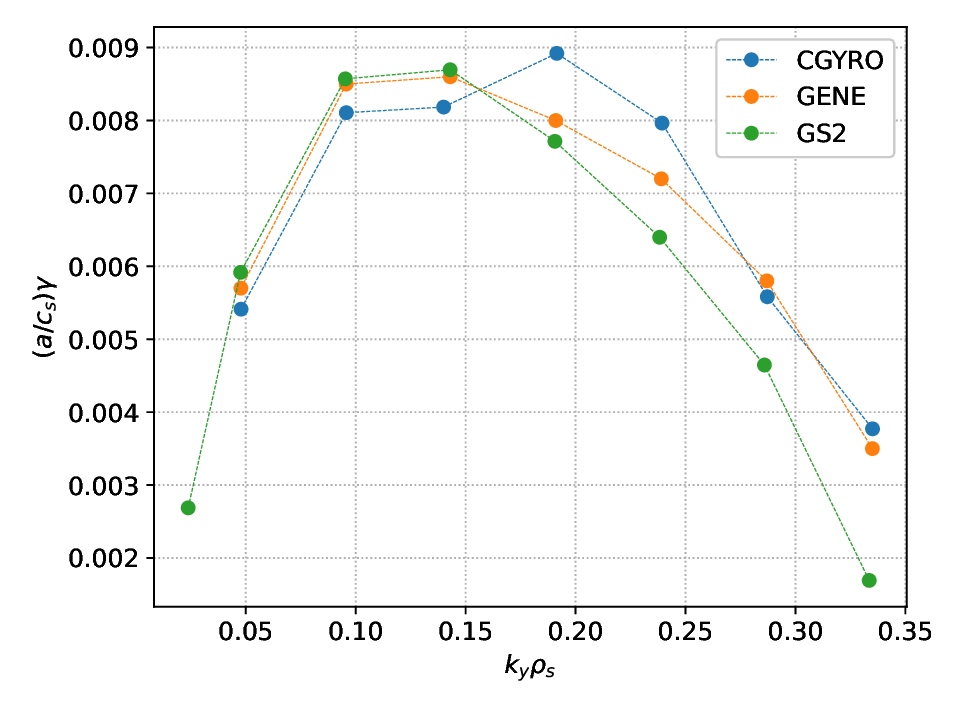}}\quad
    \subfloat[]{\includegraphics[height=0.23\textheight]{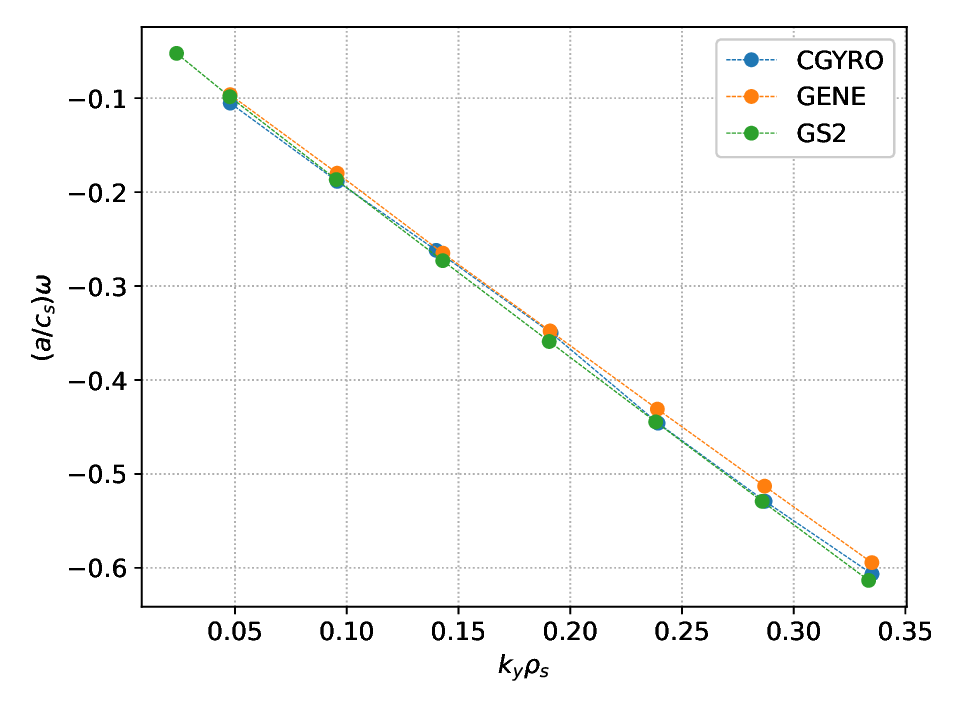}}
    \caption{Growth rate (a) and mode frequency (b) as functions of $k_y$ for the surface at $\Psi_n = 0.49$ ($q=3.5$) of STEP-EC-HD. Results from  CGYRO (blue line), GENE (orange line) and GS2 (green line) simulations with $\delta B_\parallel = 0$ (MTM instability).}
    \label{fig:comp_sub}
\end{figure}

\begin{figure}
    \centering
    \subfloat[\textit{Hybrid}-KBM]{\includegraphics[height=0.23\textheight]{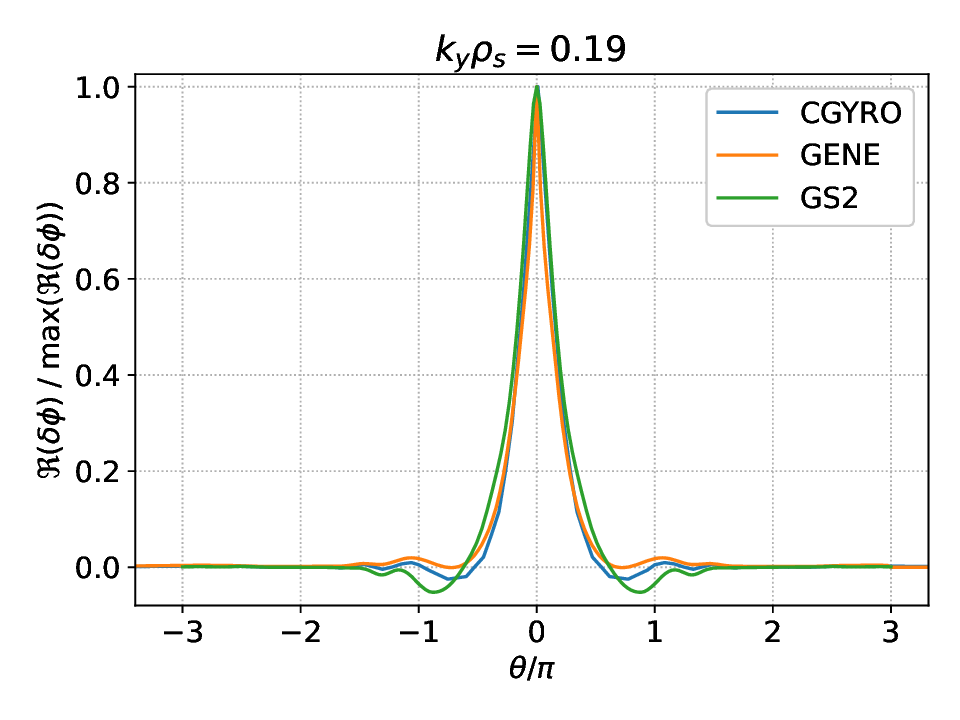}}\quad
    \subfloat[\textit{Hybrid}-KBM]{\includegraphics[height=0.23\textheight]{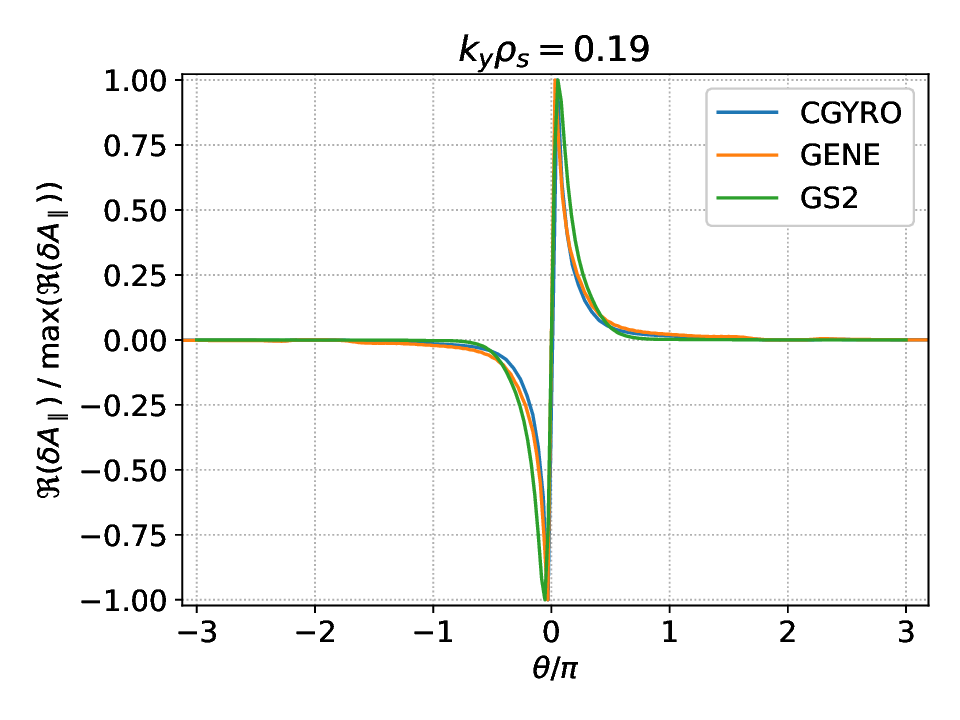}}\\
    \subfloat[MTM]{\includegraphics[height=0.23\textheight]{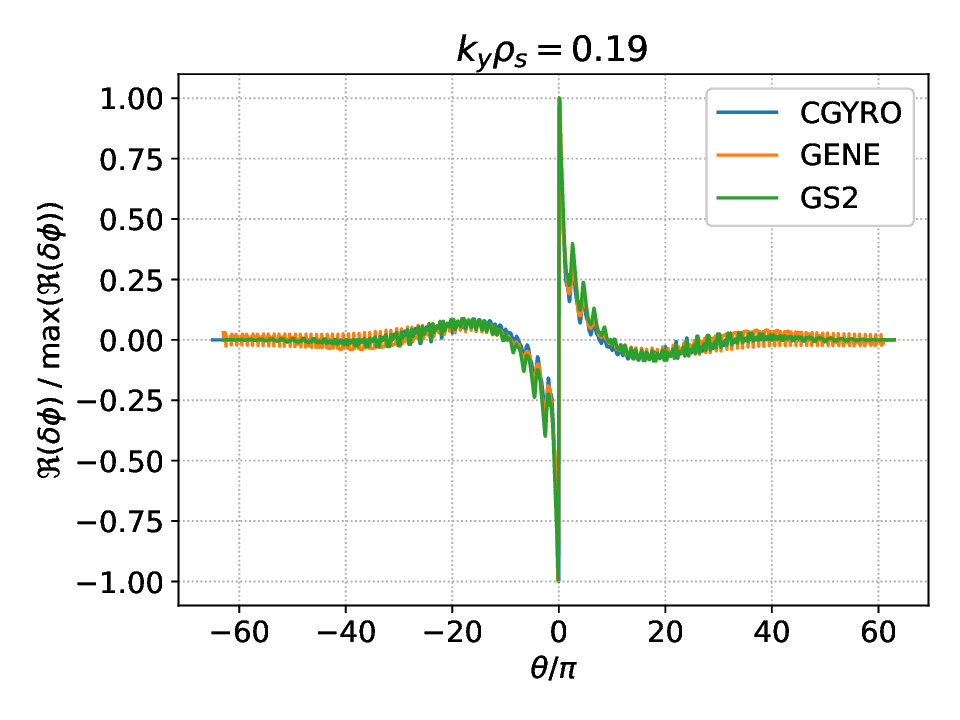}}\quad
    \subfloat[MTM]{\includegraphics[height=0.23\textheight]{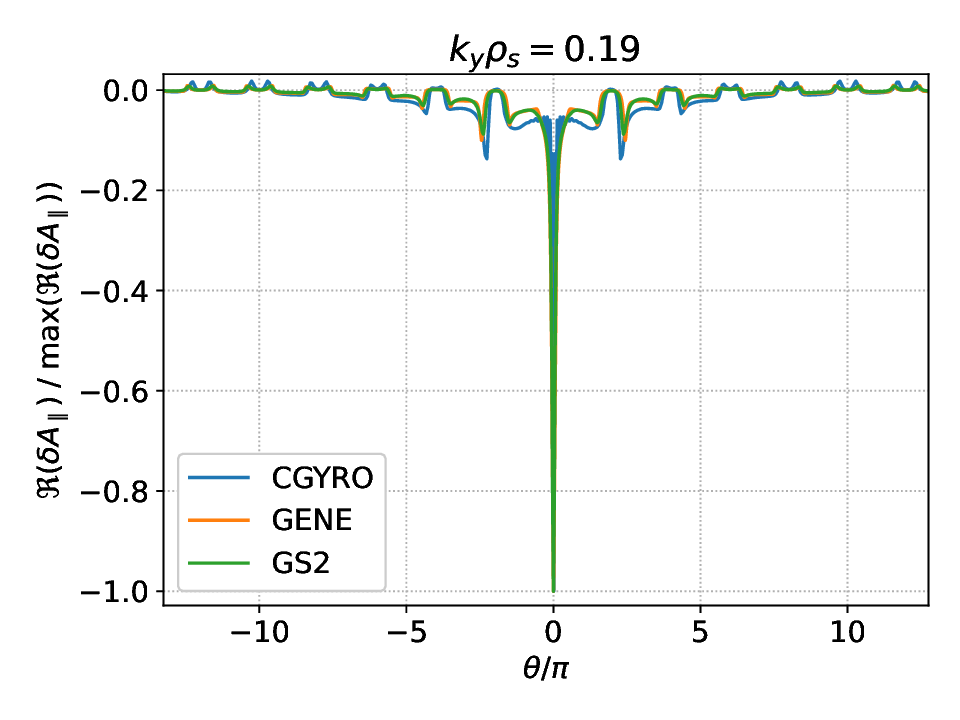}}
    \caption{Parallel mode structure $\Re(\delta \phi)$ [(a) and (c)] and $\Re(\delta A_\parallel)$ [(b) and (d)] at $k_y\rho_s = 0.19$ from CGYRO (blue line), GENE (orange line) and GS2 (green line) linear simulations of the hybrid-KBM (top row) and MTM (bottom row) instability. Results at $\Psi_n=0.49$ of STEP-EC-HD.}
    \label{fig:eig_comp}
\end{figure}

The comparison for the subdominant MTM instability is shown in Figure~\ref{fig:comp_sub}. Retrieving a good agreement here is much more challenging, since the growth rates are relatively small and therefore more sensitive to the different numerical implementations and dissipation employed in the three codes (see \cite{gs2,gene,candy2016} for code-specific details). In addition, it was found that numerical convergence in these simulations required a very high pitch-angle resolution (CGYRO) and a very high $\theta$ resolution (GENE) in order to capture the parallel structure of these very extended modes (\ref{app: numerical resolution convergence}). The resolutions used in these simulations are as listed in Table \ref{tab:resolution_spr45}.  
We also note that the maximum growth rate differs by less than 20~\% when a lower resolution is considered, thus motivating the lower numerical resolution used in some of the nonlinear simulations of Paper (II). As shown in Figure~\ref{fig:eig_comp}, we can see that all three codes show a good agreement on the eigenfunctions for both the dominant and subdominant modes. 


\begin{figure}
    \centering
    \subfloat[]{\includegraphics[height=0.23\textheight]{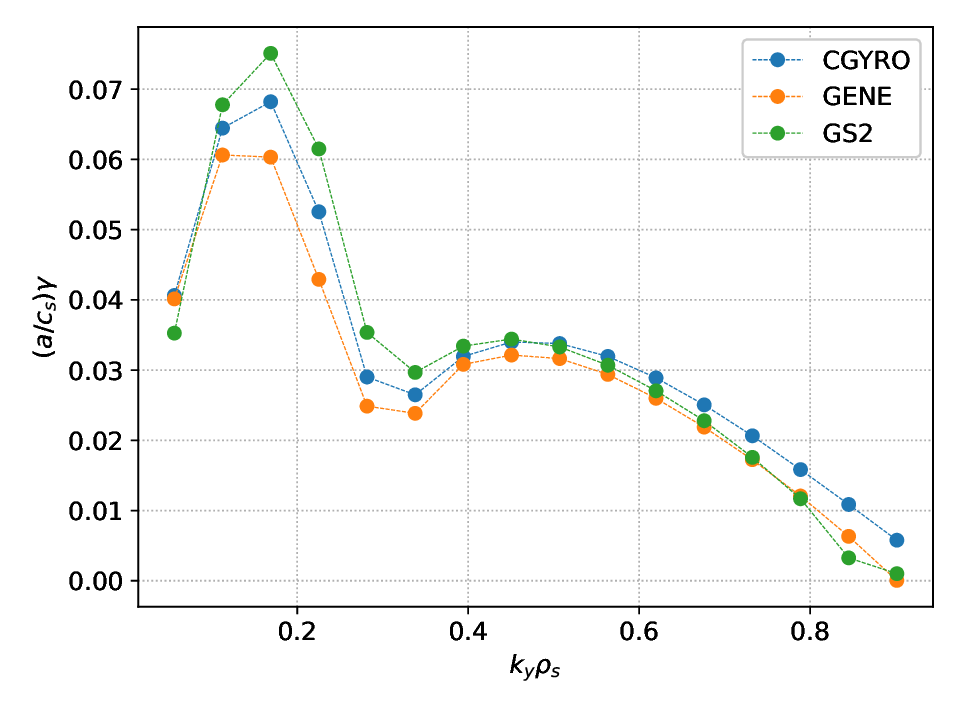}}\quad
    \subfloat[]{\includegraphics[height=0.23\textheight]{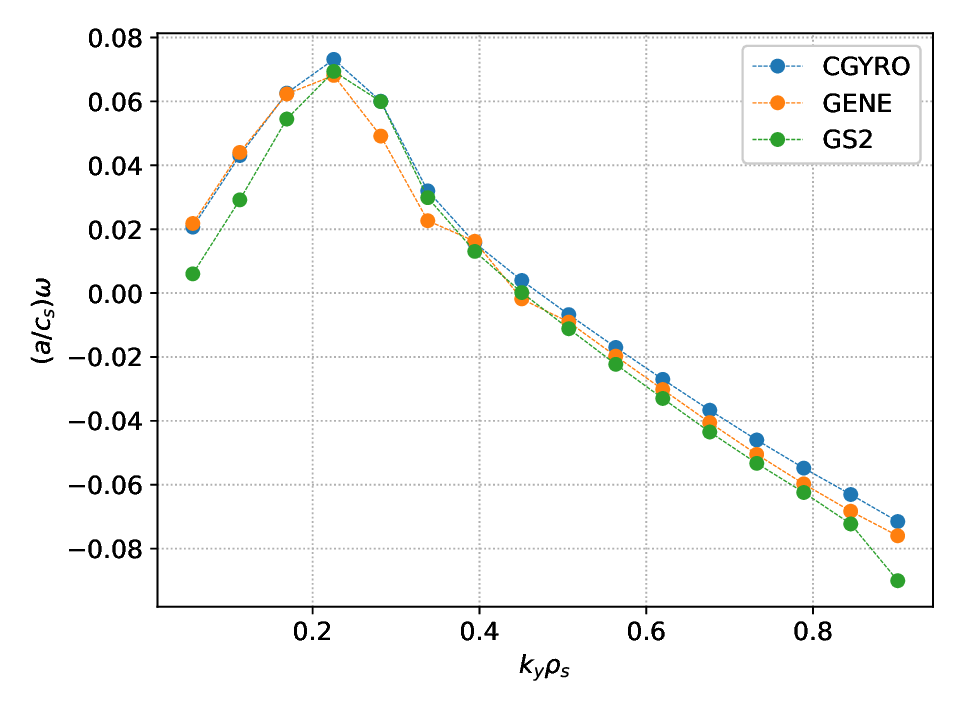}}
    \caption{Growth rate (a) and mode frequency (b) as functions of $k_y$ for the surface at $\Psi_n = 0.36$ ($q=3.0$) of STEP-EC-HD. Results from CGYRO (blue line), GENE (orange line) and GS2 (green line) linear simulations with $\delta B_\parallel \ne 0$ (hybrid-KBM instability).}
    \label{fig:psin036_comp}
\end{figure}

\begin{figure}
    \centering
    \subfloat[]{\includegraphics[height=0.23\textheight]{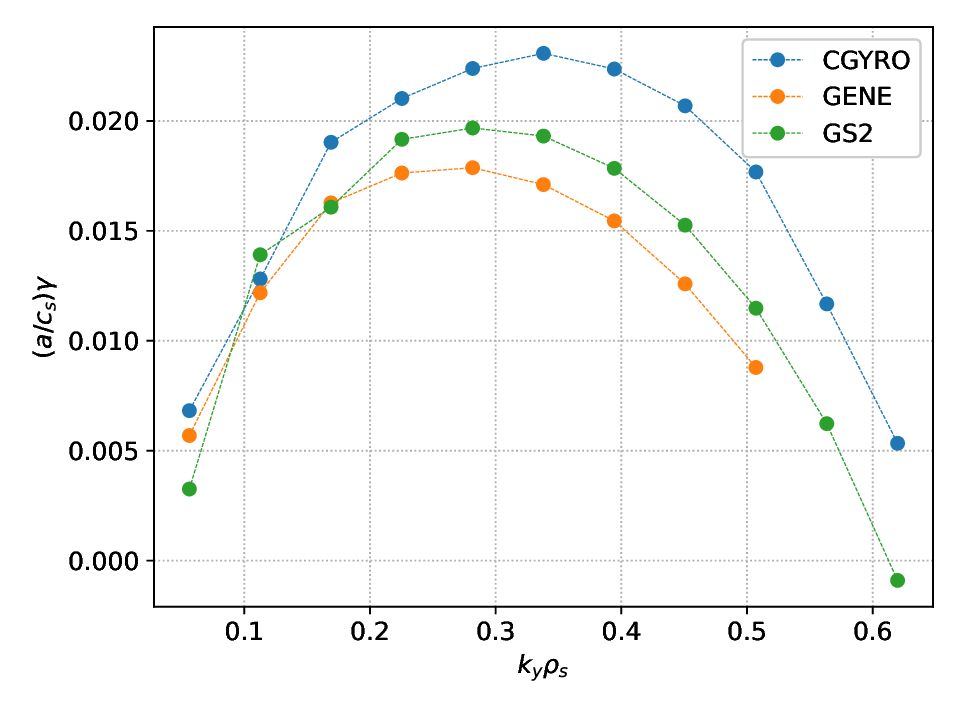}}\quad
    \subfloat[]{\includegraphics[height=0.23\textheight]{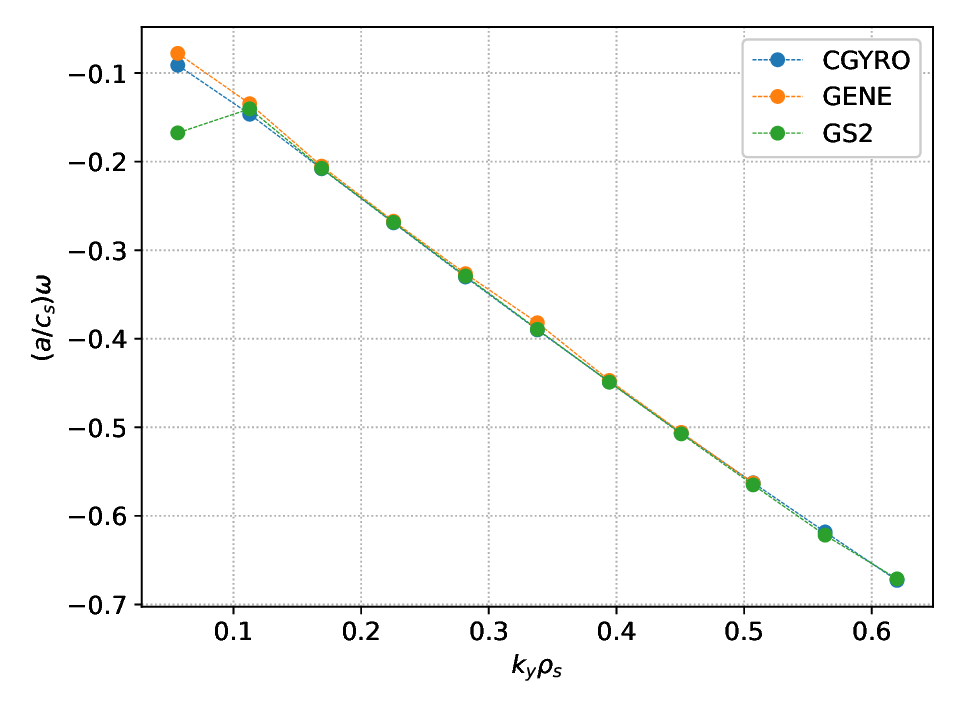}}
    \caption{Growth rate (a) and mode frequency (b) as functions of $k_y$ for the surface at $\Psi_n = 0.36$ ($q=3.0$) of STEP-EC-HD. Results from CGYRO (blue line), GENE (orange line) and GS2 (green line) linear simulations with $\delta B_\parallel = 0$ (MTM instability).}
    \label{fig:psin036_sub}
\end{figure}

\begin{figure}
    \centering
    \subfloat[]{\includegraphics[height=0.22\textheight]{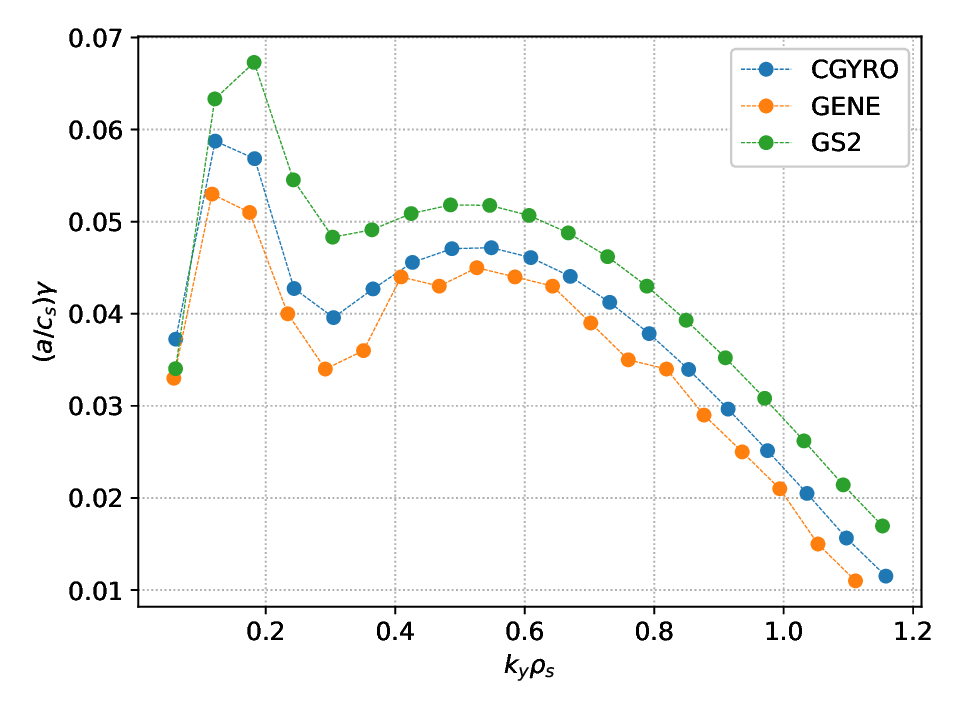}}\quad
    \subfloat[]{\includegraphics[height=0.22\textheight]{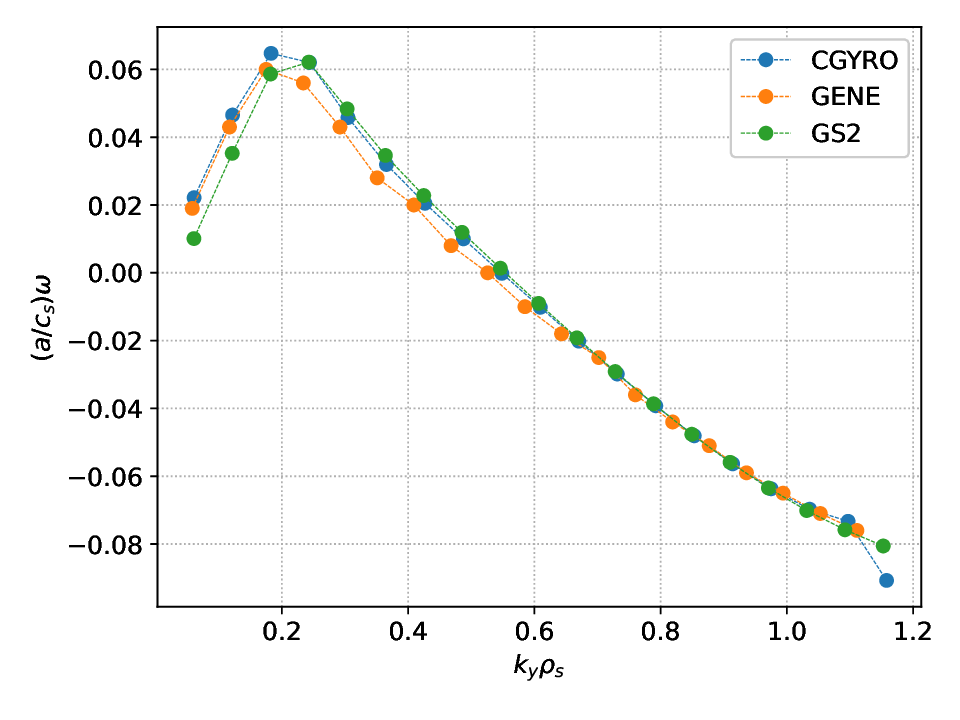}}
    \caption{Growth rate (a) and mode frequency (b) as functions of $k_y$ for the surface at $\Psi_n = 0.35$ ($q=3.5$) of STEP-EB-HD. Results from CGYRO (blue line), GENE (orange line) and GS2 (green line) linear simulations with $\delta B_\parallel \ne 0$ (hybrid-KBM instability).}
    \label{fig:spr46}
\end{figure}

\begin{figure}
    \centering
    \subfloat[]{\includegraphics[height=0.22\textheight]{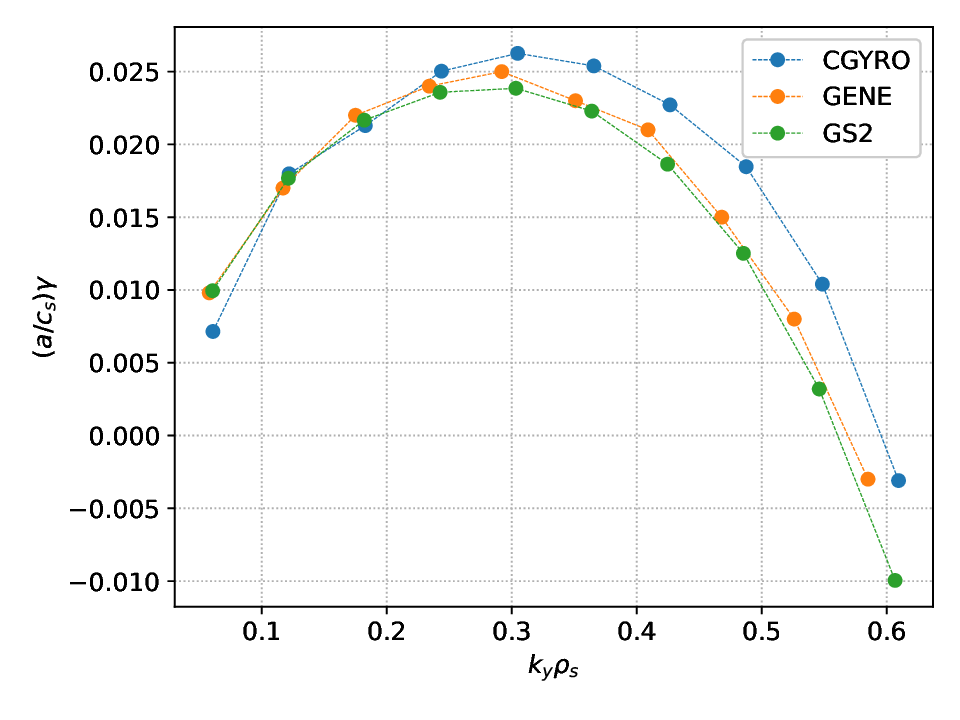}}\quad
    \subfloat[]{\includegraphics[height=0.22\textheight]{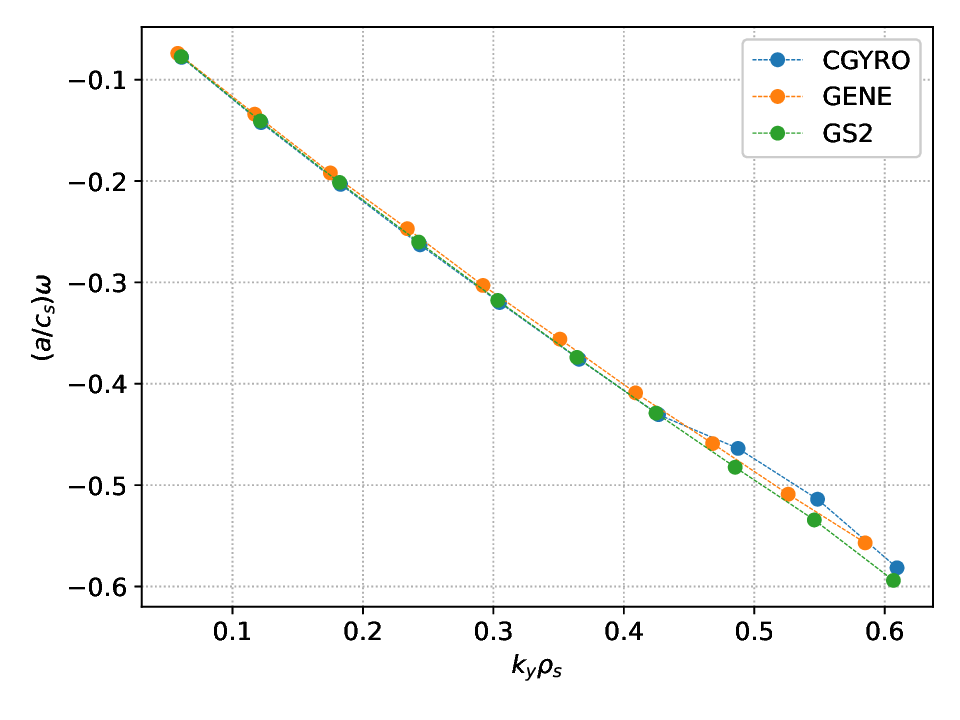}}
    \caption{Growth rate (a) and mode frequency (b) as functions of $k_y$ for the surface at $\Psi_n = 0.35$ ($q=3.5$) of STEP-EB-HD. Results from CGYRO (blue line), GENE (orange line) and GS2 (green line) linear simulations with $\delta B_\parallel = 0$ (MTM instability).}
    \label{fig:spr46_sub}
\end{figure}

The three code comparison is also carried out on the $q = 3.0$ flux surface of STEP-EC-HD; the dominant instability is shown in  Figure~\ref{fig:psin036_comp} and a comparison for the subdominant MTM instability is shown in Figure~\ref{fig:psin036_sub}; and also for the $q=3.5$ surface of STEP-EB-HD; the dominant instability is shown in  Figure~\ref{fig:spr46} and a comparison for the subdominant MTM instability is shown in Figure~\ref{fig:spr46_sub}.

We conclude by noting that there is a good agreement between the three codes in all the considered cases.

\section{Conclusions}
\label{sec:conclusions} 

In this paper, we have presented the results of local linear microinstability studies of the thermal plasma on a range of flux surfaces from the core to the pedestal top in the two preferred STEP flat-top operating points. We find that the linear spectra is dominated by a hybrid mode, sharing features of the KBM, ITG, and TEM instability, at the ion Larmor scale, with weakly-growing subdominant MTMs present at similar scales. The local equilibria examined here ($q = 3.0,\ 3.5,\ 4.0,\ 5.0$ in STEP-EC-HD and $q=3.5$ in STEP-EB-HD) were found to be completely stable to electron scale modes. 

A summary of the dominant microinstabilities from some of our simulations is given in Table \ref{tab:linear results} alongside the results from some similar conceptual designs of burning ST plasmas, namely the TDoTP high $q_{0}$ equilibrium taken from \cite{tdotp}; and an earlier prototype for a burning ST reactor BurST taken from \cite{patel2021}. Shown here is a summary for only the $q=3.5$ surface in each equilibrium.\footnote[1]{Note that data was only available for the $q=4.3$ flux surface in BurST. However, we remark that the stability properties of the $q=4$ and $q=5$ surfaces of STEP-EC-HD are qualitatively identical to those of the $q=3.5$ surface of STEP-EC-HD. As such, we believe that the comparison between the $q=3.5$ surfaces of STEP-EC-HD and STEP-EB-HD to the $q=4.3$ surface of BurST is still relevant for the purposes of broad understanding.}
\begin{table}[h]
\small
\begin{center}
    \begin{tabular}{ccccccl}
\toprule 
Design & \, $q$ \, &\, $\Psi_{N}$ \,\,& \,$\beta$\, & \, $\beta^{\prime}$ \, & \, $a/L_{Te}$ \, &   Dominant modes at ion Larmor scale $k_{y}\rho_{i} \ll 1$ \\
    &                          &           &                           &              &                    &             Dominant modes at intermediate scale $k_{y}\rho_{i} \gtrsim 1$  \\
    &                          &           &                           &              &                    &             Dominant modes at electron Larmor scale $k_{y}\rho_{i} \gg 1$  \\
    &                          &           &                           &              &                    &             Subdominant modes at any scale. \\
\midrule
\rowcolor{Gray} STEP-EC-HD & 3.5 & 0.49 & 0.09 & -0.48 & 1.58 &  {Hybrid KBM/TEM/ITG at ion scale $k_{y}$}\\
\rowcolor{Gray}     &      &      &      &             &       &  {No purely intermediate scale instability }
\\
\rowcolor{Gray}     &      &      &      &             &       &  {No purely electron scale instability }\\
\rowcolor{Gray}     &      &      &      &             &       &  {MTM at ion scale $k_{y}$ but electron scale $k_{x}$ }\\
\rowcolor{Gray} STEP-EB-HD & 3.5 & 0.35 & 0.11 & -0.40 & 1.40 &  {Hybrid KBM/TEM/ITG at ion scale $k_{y}$}\\
\rowcolor{Gray}     &      &      &      &             &       &  {No purely intermediate scale instability }
\\
\rowcolor{Gray}     &      &      &      &             &       &  {No purely electron scale instability }\\
\rowcolor{Gray}     &      &      &      &             &       &  {MTM at ion scale $k_{y}$ but electron scale $k_{x}$ }\\
\rowcolor{black!0} TDotP-high-$q_{0}$ & 3.5 & 0.5 & 0.18 & -1.19 & 3.35 &  {Hybrid KBM/ITG at ion scale $k_{y}$}\\
\rowcolor{black!0}     &      &      &      &             &       &  {Collisionless MTM at $k_{y}\rho_{i} \sim 4$}
\\
\rowcolor{black!0}     &      &      &      &             &       &  {No purely electron scale instability }\\
\rowcolor{black!0}     &      &      &      &             &       &  {MTM at ion scale $k_{y}$ but electron scale $k_{x}$ }\\
\rowcolor{black!0} BurST & 4.3 & 0.5 & 0.15 & -0.99 & 2.77 &  {KBM at ion scale $k_{y},$ MTM at $k_{y}\rho_{i} \ll 1$}\\
\rowcolor{black!0}     &      &      &      &             &       &  {Collisionless MTM at $k_{y}\rho_{i} \sim 4$}
\\
\rowcolor{black!0}     &      &      &      &             &       &  {No purely electron scale instability}\\
\rowcolor{black!0}     &      &      &      &             &       &  {MTM at ion scale $k_{y}$ but electron scale $k_{x}$ }\\
\bottomrule
\end{tabular}
\end{center}
\caption{Summary of the nature of the dominant microinstabilities found on core flux surfaces at $q=3.5$ in the two preferred STEP flat-top operating points (STEP-EC-HD and STEP-EB-HD). For comparison, results are also shown for the TDoTP high $q_{0}$ case ($q=3.5$) \cite{tdotp}; and for BurST ($q=4.3$) \cite{patel2021}.}
\label{tab:linear results}
\end{table}

We remark that
\begin{itemize}
    \item The dominant mode shares properties of a hybrid KBM/TEM/ITG. 

 \begin{itemize}
     \item The mode is electromagnetic (KBM-like) but can be tracked consistently back to the electrostatic limit.
     \item The mode has many features typical of the KBM: it generally propagates in the ion-diamagnetic direction, has eigenfunctions which are strongly peaked in ballooning space and the mode has twisting parity (KBM-like), but the mode is unstable in a regime below the ideal $n=\infty$ MHD limit.
     \item The mode is driven by the pressure gradient (KBM-like) but it is more sensitive to the electron temperature gradient than the ion temperature gradient. 
    \item The mode is sensitive to trapped electron dynamics.
    \item The mode is sensitive to $\theta_0$.
    \item The mode growth rate is smaller at larger values of $\beta$ and $\beta^\prime$.
    \item The mode growth rate is less sensitive to $\beta$ and $\beta^\prime$ at low $\beta$ values.
    \item The mode requires access to $\delta B_{\parallel}$ drive in order to be unstable.
\end{itemize}
    
    \item There is no unstable branch of collisionless MTMs at intermediate scales $(k_y\rho_s \sim 4$) unlike in \cite{tdotp} and \cite{patel2021}
    \item A collisional MTM is unstable but is always subdominant to the KBM-like instability.
\end{itemize}

{It is important to remark that for the equilibria examined in this work, the confinement is assumed and this is a substantial caveat. In a natural extension to the linear work presented in this paper, a parallel companion article \cite{giacomin2023b} explores whether or not these assumed plasma profiles can be sustained by the available heating and fuelling, i.e. whether the turbulence driven by the instabilities studied in this paper are compatible with the assumptions in the design of STEP-EC-HD and STEP-EB-HD. Linear analysis presented here has uncovered different routes to stabilising the \textit{hybrid}-KBM which could be very useful in future exploration of optimised operating points.} We have shown that the \textit{hybrid}-KBM can be stabilised by increasing $\beta'$. Due to the strong sensitivity on $\theta_0$, the mode might be suppressed by $\mathrm{E}\times \mathrm{B}$ flow shear, and this is explored in \cite{giacomin2023b} by means of nonlinear turbulent simulations.

A detailed three-code comparison involving GS2, CGYRO, and GENE was performed, and we found reasonable agreement between the three codes for a range of different plasma parameters. The result of this benchmark increases our confidence in the fidelity of GK modelling of electromagnetic instabilities, as well as highlighting the need for care in the handling of parallel dissipation in order to avoid encountering numerical instabilities in these challenging computations (see discussion in \cite{giacomin2023b}). This linear stability analysis paves the way for detailed nonlinear turbulence studies undertaken in the companion article \cite{giacomin2023b}. 

\ack
We are indebted to E. Belli, J. Candy, B. Chapman, D. Hatch, P. Ivanov and M. Hardman for helpful discussions and suggestions at various stages of this project. The authors would also like to thank the GENE team – most notably T. G{\"o}rler and D. Told - for their help and support. The first author is grateful to The Institute for Fusion Studies (IFS), Austin TX, for its splendid hospitality during a stimulating and productive visit. 
This work has been founded by the Engineering and Physical Sciences Research Council (grant numbers EP/R034737/1 and EP/W006839/1).
Simulations have been performed on the Viking research computing cluster at the University of York and on the Marconi supercomputer from the National Supercomputing Consortium CINECA, under the project STETGMTM. Part of this work was performed using resources provided by the Cambridge Service for Data Driven Discovery (CSD3) operated by the University of Cambridge Research Computing Service (\url{www.csd3.cam.ac.uk}), provided by Dell EMC and Intel using Tier-2 funding from the Engineering and Physical Sciences Research Council (capital grant EP/T022159/1), and DiRAC funding from the Science and Technology Facilities Council (\url{www.dirac.ac.uk}). To obtain further information on the data and models underlying this paper please contact PublicationsManager@ukaea.uk.

\appendix
\section{Numerical resolution convergence}
\label{app: numerical resolution convergence} 

Here we discuss the numerical resolution convergence studies in GS2 linear simulations for the dominant \textit{hybrid}-KBM instability at the $q=3.5$ flux surface of STEP-EC-HD. Similar resolution convergence scans were performed also with CGYRO and GENE at all the radial surfaces considered in this work, and for both the dominant and subdominant modes. These detailed convergence tests are also used to inform the resolutions used in the nonlinear simulations which are the focus of Paper (II).

\begin{figure}
    \centering
    \subfloat[]{\includegraphics[height=0.22\textheight]{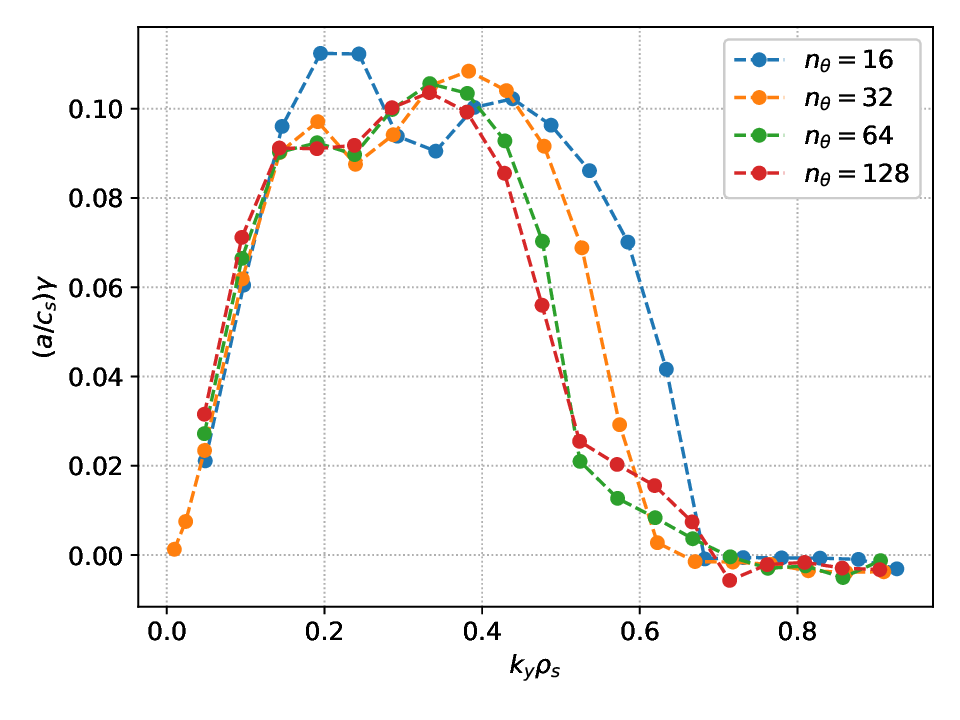}}\quad
    \subfloat[]{\includegraphics[height=0.22\textheight]{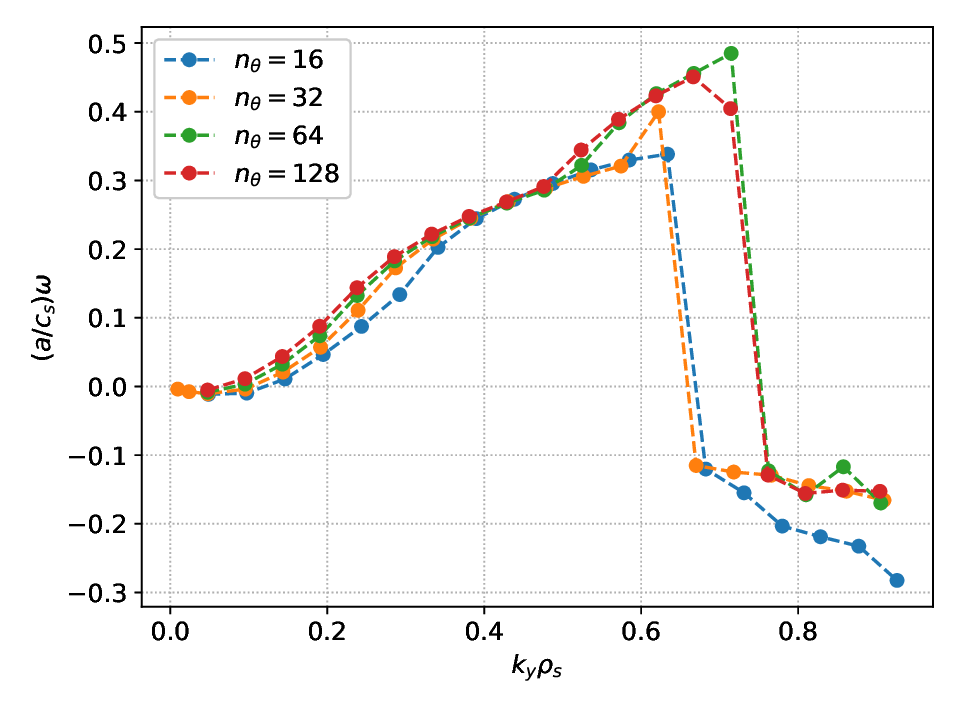}}
    \caption{Growth rate (a) and mode frequency (b) as functions of $k_y$ from GS2 linear simulations at $\Psi_n=0.49$ of STEP-EC-HD with different parallel grid resolution.}
    \label{fig:ntheta}
\end{figure}

Figure~\ref{fig:ntheta} shows the growth rate and mode frequency at different parallel grid resolutions. Convergence is achieved for $n_\theta \geq 32 $ at low mode numbers, while a higher resolution ($n_\theta \geq 64$) is required at $k_y\rho_s>0.4.$

\begin{figure}
    \centering
    \subfloat[]{\includegraphics[height=0.22\textheight]{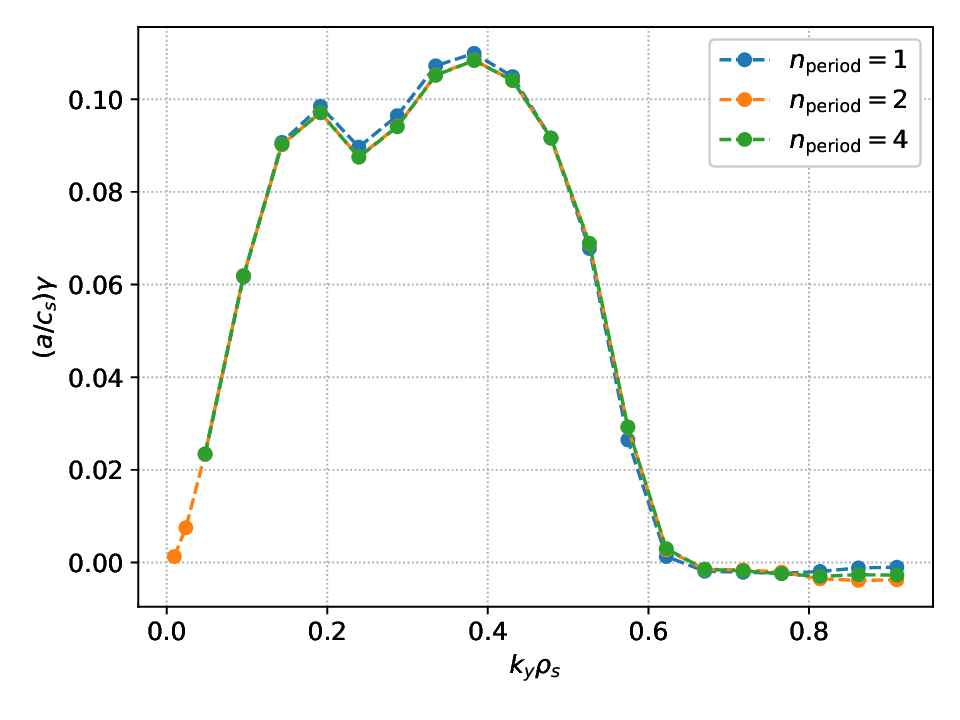}}\quad
    \subfloat[]{\includegraphics[height=0.22\textheight]{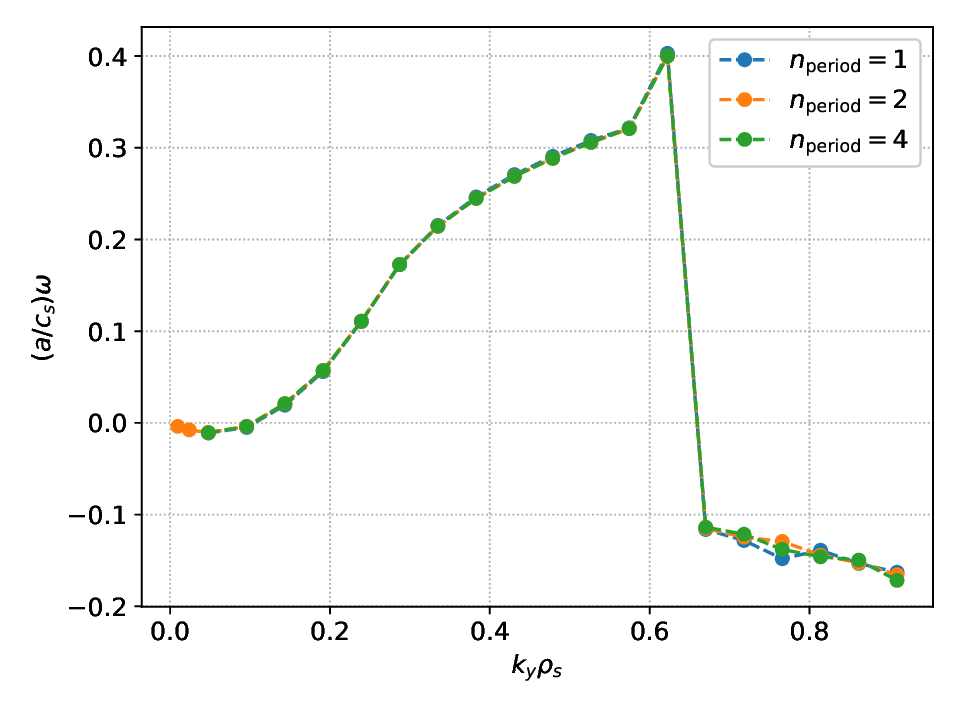}}
    \caption{Growth rate (a) and mode frequency (b) as functions of $k_y$ from GS2 linear simulations at $\Psi_n=0.49$ of STEP-EC-HD with different values of \emph{\texttt{nperiod}}.}
    \label{fig:nperiod}
\end{figure}

Convergence with respect to  the grid extent in ballooning space (which is equivalent to the radial grid resolution in the flux tube) is controlled in GS2 by the parameter \texttt{nperiod}, and is investigated in Figure~\ref{fig:nperiod}. Growth rate is only slightly affected by this parameter, as expected since for the \textit{hybrid}-KBM both $\delta \phi$ and $\delta A_\parallel$ are very localised around $\theta = 0$ (see Figure~\ref{fig:eig_kbm_02}). It should of course be noted once again that the MTM has much more stringent conditions on the radial and parallel grid resolutions (see Table \ref{tab:resolution_spr45}). 

\begin{figure}
    \centering
    \subfloat[]{\includegraphics[height=0.22\textheight]{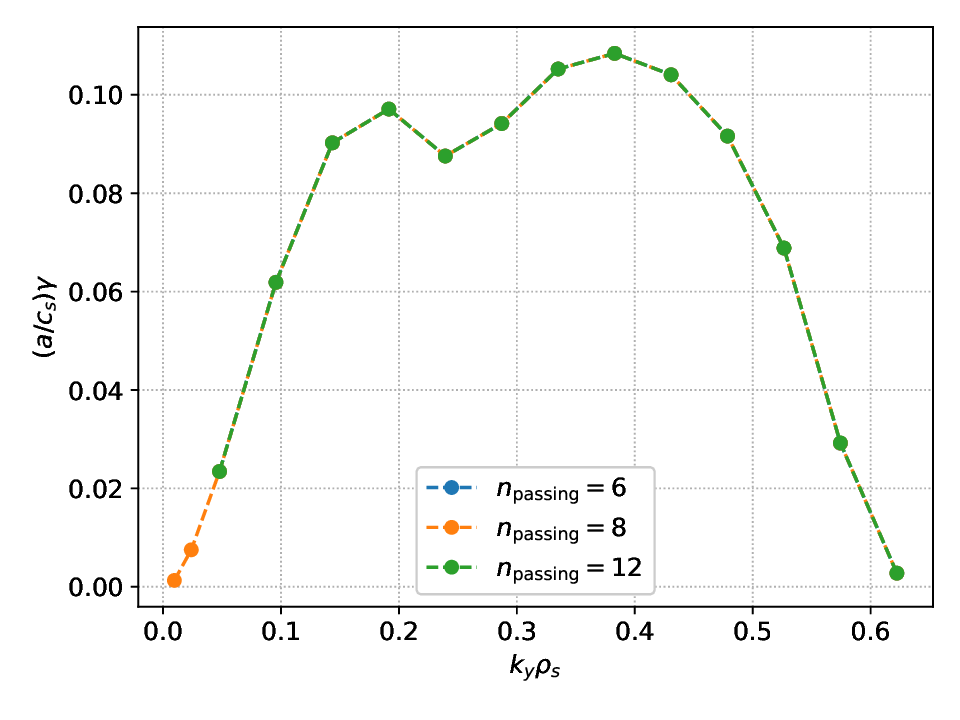}}\quad
    \subfloat[]{\includegraphics[height=0.22\textheight]{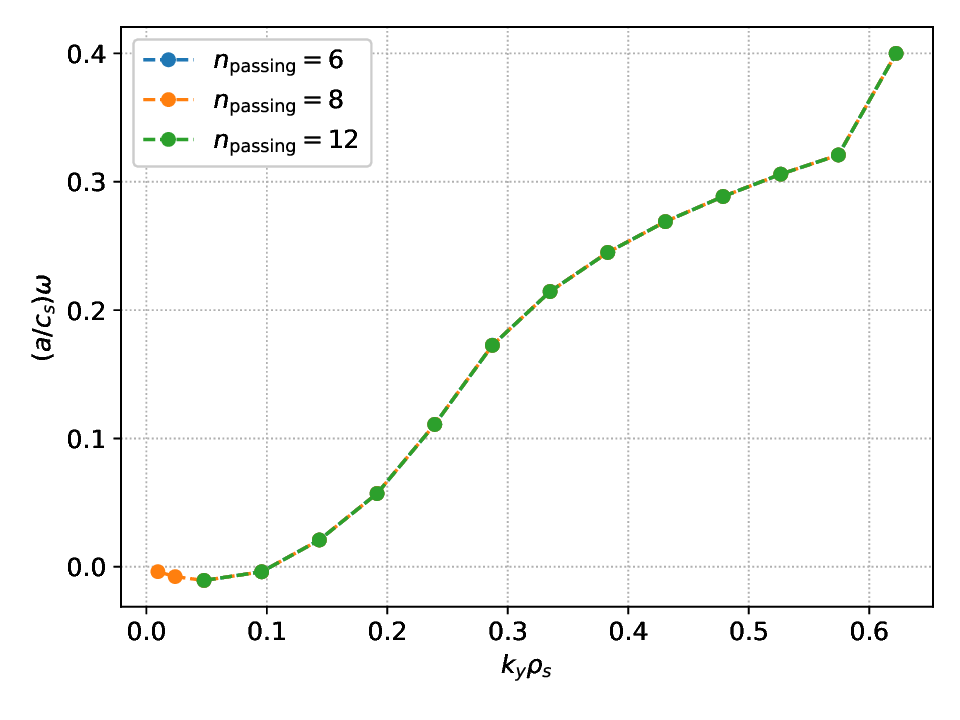}}
    \caption{Growth rate (a) and mode frequency (b) as functions of $k_y$ from GS2 linear simulations at $\Psi_n=0.49$ of STEP-EC-HD with different values of \emph{\texttt{npassing}}.}
    \label{fig:nl}
\end{figure}

\begin{figure}
    \centering
    \subfloat[]{\includegraphics[height=0.22\textheight]{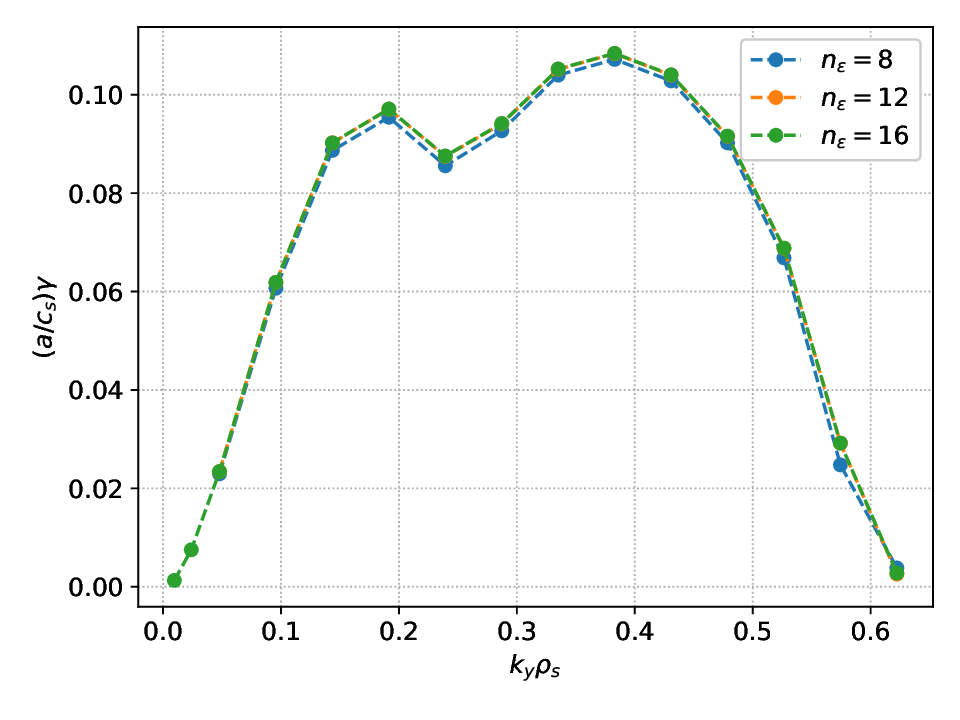}}\quad
    \subfloat[]{\includegraphics[height=0.22\textheight]{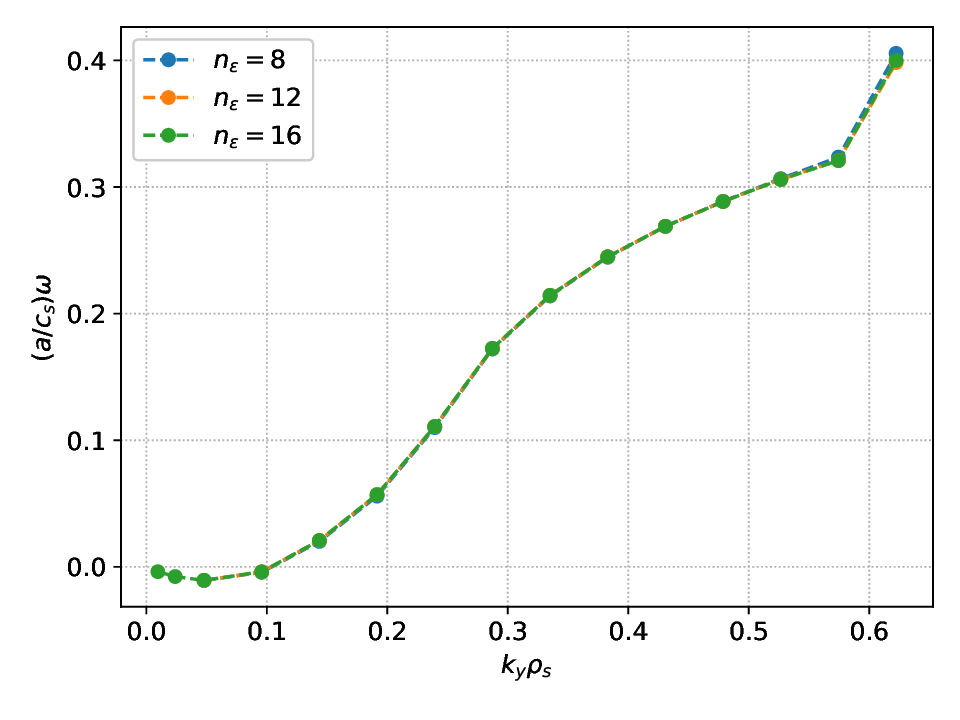}}
    \caption{Growth rate (a) and mode frequency (b) as functions of $k_y$ from GS2 linear simulations at $\Psi_n=0.49$ of STEP-EC-HD with different values of $n_\epsilon$.}
    \label{fig:negrid}
\end{figure}

Velocity space resolution convergence is tested in Figures~\ref{fig:nl}~and~\ref{fig:negrid}, where the number of passing pitch-angles and the number of energy grid points are varied. Over these grid parameters there is a weak dependence of the growth rate and mode frequency on the velocity space resolution over the entire $k_y$ spectrum.  

\section{Dependence on the collision model}
\label{app:collision_model}

Here we briefly discuss the dependence of our simulation results on the collision model. In particular, we observe that the linear spectrum is more sensitive to the plasma composition (the number of species evolved) when the Sugama collision operator~\cite{sugama2009}, a sophisticated approximation to the full
linearized, gyro-averaged Fokker-Planck operator, is used instead of the full linearized Fokker-Planck collision operator which is used in all of our GS2 simulations. 

\begin{figure}
    \centering
    \subfloat[]{\includegraphics[height=0.22\textheight]{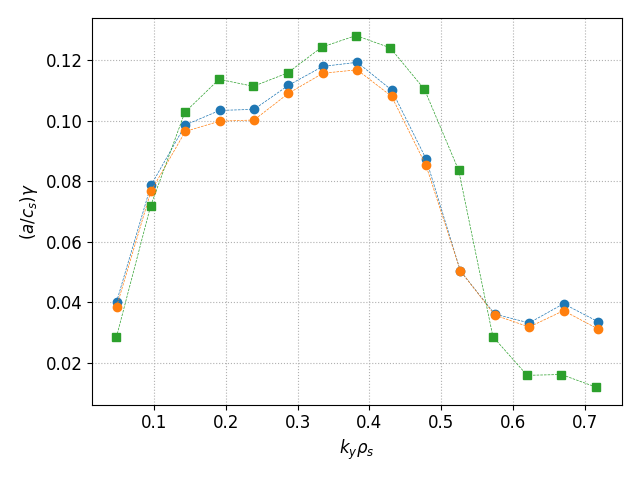}}\quad
    \subfloat[]{\includegraphics[height=0.22\textheight]{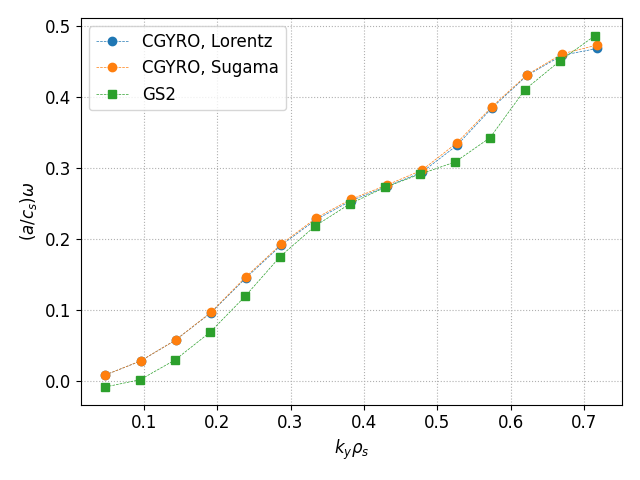}}
    \caption{Comparison of the growth rate (a) and mode frequency (b) as functions of $k_y$ from CGYRO simulations with two different collision operators (Lorentz and Sugama) and two species. Also shown is the growth rate and mode frequency from the GS2 simulation with two species.}
    \label{fig:coll_2spec}
\end{figure}

\begin{figure}
    \centering
    \subfloat[]{\includegraphics[height=0.22\textheight]{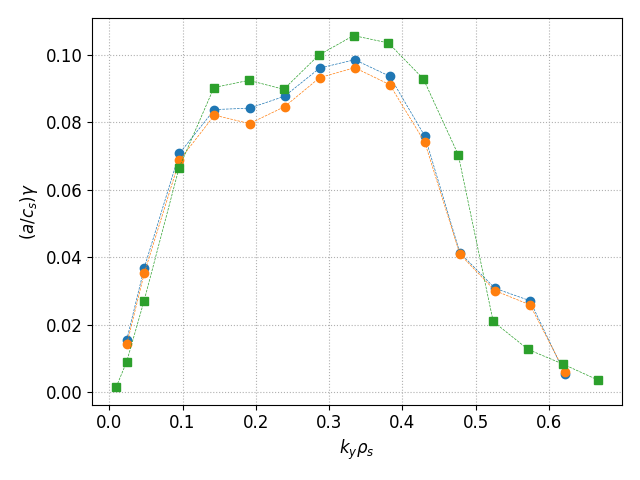}}\quad
    \subfloat[]{\includegraphics[height=0.22\textheight]{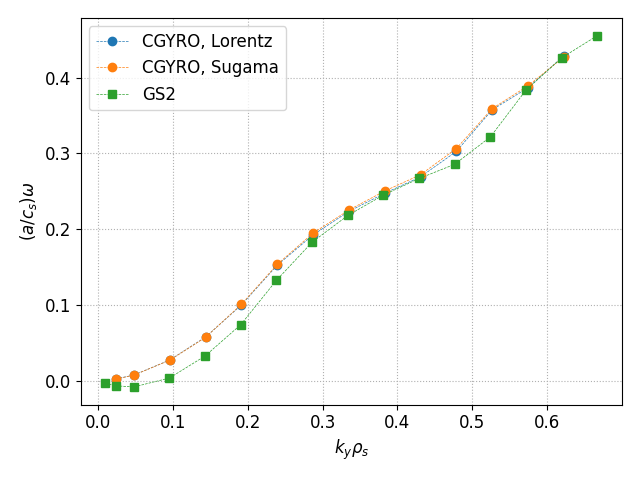}}
    \caption{Comparison of the growth rate (a) and mode frequency (b) as functions of $k_y$ from CGYRO simulations with three different collision operators (Lorentz and Sugama) and three species. Also shown is the growth rate and mode frequency from the GS2 simulation with three species.}
    \label{fig:coll_3spec}
\end{figure}

We first compare the growth rate and mode frequency values from CGYRO and GS2 simulations carried out by evolving two species (see Figure~\ref{fig:coll_2spec}) and three species (see Figure~\ref{fig:coll_3spec}) and considering different collision operator models: CGYRO simulations are performed by using the Lorentz and the Sugama collision operators, while GS2 uses the linearized Fokker-Planck collision operator. In all of these simulations, we note that there is only a very weak dependence on the type of collision model employed.

\begin{figure}
    \centering
    \subfloat[]{\includegraphics[height=0.22\textheight]{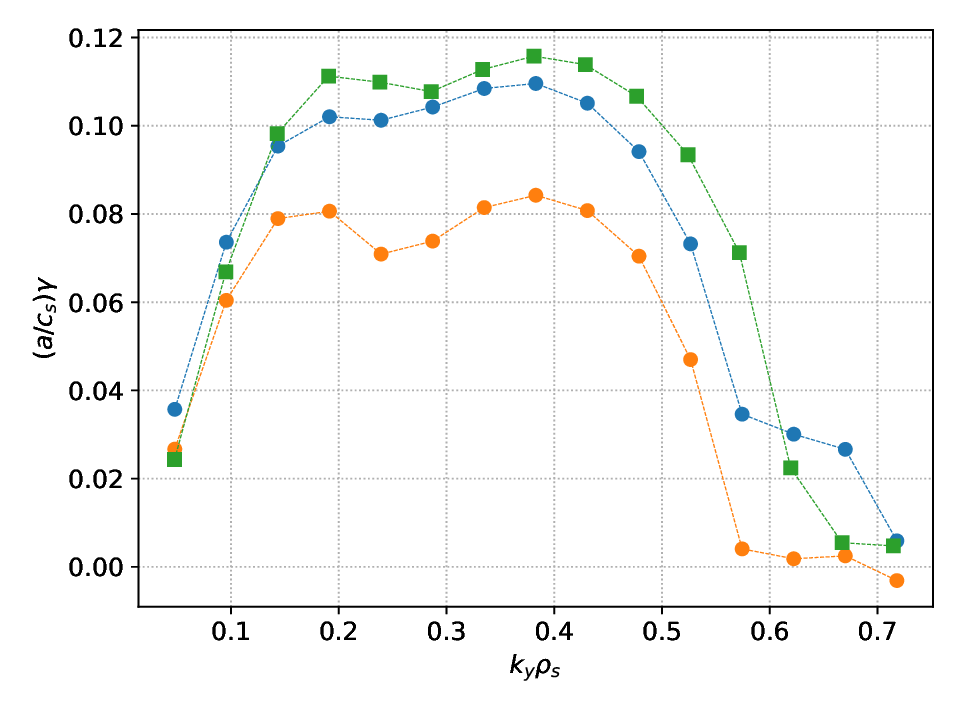}}\quad
    \subfloat[]{\includegraphics[height=0.22\textheight]{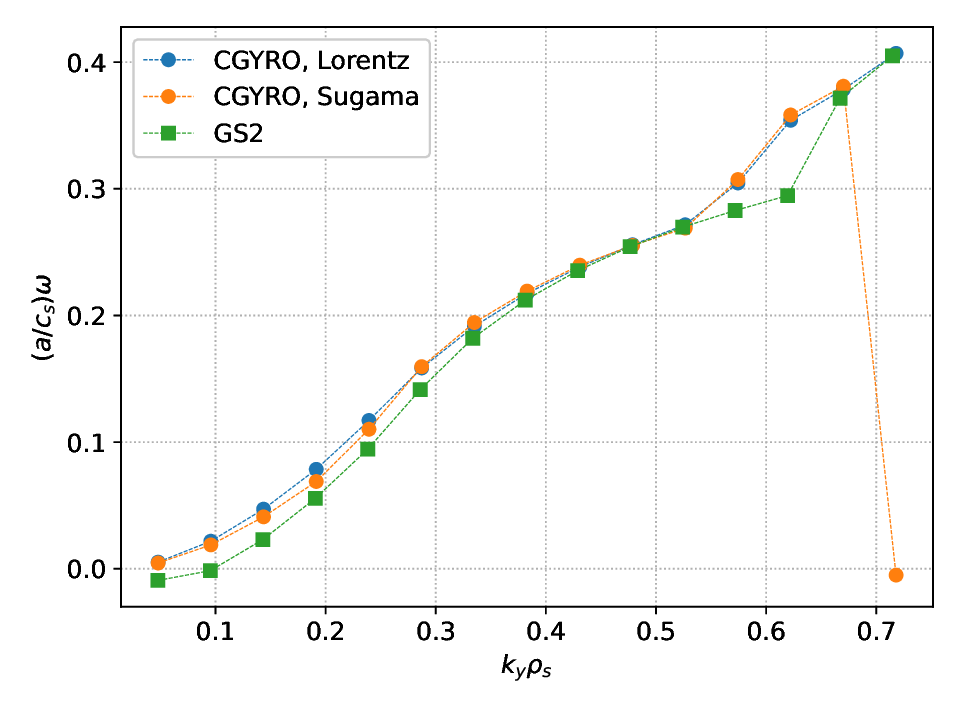}}
    \caption{Comparison of the growth rate (a) and mode frequency (b) as functions of $k_y$ from CGYRO simulations with two different collision operators (Lorentz and Sugama) and five species. Also shown is the growth rate and mode frequency from the GS2 simulation with five species.}
    \label{fig:coll_6spec}
\end{figure}

However, Figure~\ref{fig:coll_6spec} shows that growth rate values are more sensitive to the collision model when five species are considered. Interestingly, although there a relatively good agreement observed between CGYRO and GS2 simulations results when the Lorentz operator is used in CGYRO, we find that the growth rate values are smaller (by approximately 20\%) when the Sugama collision operator is used in CGYRO. It is important to note however that there is no qualitative difference in the main instability. We will also see in Paper (II) that reducing the linear growth rates by such a small amount makes no significant difference to the transport properties on the surface considered.

\section*{References}
\bibliographystyle{unsrt}
\bibliography{bibliography}

\end{document}